\documentclass[acmtog]{acmart}

\acmJournal{TOG}

\usepackage{version}

\usepackage{mathtools} %
\usepackage{textcomp}
\DeclareMathOperator{\Tr}{Tr}
\usepackage{siunitx}
\usepackage{stmaryrd}
\usepackage{scalerel}
\usepackage{bm}

\usepackage{ifthen}
\usepackage{float} %
\usepackage{hyperref}
\hypersetup{
    colorlinks = true,
}
\usepackage[capitalize,noabbrev]{cleveref}

\usepackage{csquotes}
\usepackage{parskip}
\usepackage{calc}
\usepackage{longtable}

\usepackage{algorithm}
\usepackage{algorithmicx}
\usepackage{algpseudocode}
\usepackage{listingsutf8}

\usepackage{array}
\usepackage{booktabs}
\usepackage{multirow}

\usepackage{float}
\usepackage{wrapfig} %
\usepackage{graphicx}
\usepackage[percent]{overpic}
\usepackage{subcaption}
\usepackage{supertabular}

\usepackage{natbib}

\usepackage{appendix}

\usepackage[shortlabels]{enumitem}
\newcommand{\pd}{\partial}
\newcommand{\dt}{\Delta t\,}
\newcommand{\pde}{\mathbf{h}}

\newcommand{\fh}{\mathcal{H}}
\newcommand{\hht}{\hat{\pde}}
\newcommand{\cL}{\mathcal{L}}

\newcommand{\fprm}{q}  %
\newcommand{\fprms}{{\bar{\fprm}}} %
\newcommand{\prm}{\mathbf{\fprm}}  %
\newcommand{\prms}{{\mathbf{\fprms}}}  %
\newcommand{\adj}{p}  %
\newcommand{\vadj}{\mathbf{p}}  %
\newcommand{\lmT}{\vadj^T} 
\newcommand{\dpar}{\partial_\prm} %
\newcommand{\dparms}{\partial_\prms} %
\newcommand{\fdpar}{d_\prm} %
\newcommand{\dpp}{\partial_\prm}  %
\newcommand{\dppm}{\partial_{\prm^m}}  %
\newcommand{\Omegar}{\Omega_{\mathrm{ref}}} %
\newcommand{\vva}{\mathbf{a}}
\newcommand{\vu}{\mathbf{u}} %
\newcommand{\dpu}{\partial_\vu} %
\newcommand{\dpui}{\partial_{\vu^i}} %
\newcommand{\dpuim}{\partial_{\vu^{i-1}}} %
\newcommand{\dpuz}{\partial_{\vu^{0}}} %
\newcommand{\dpun}{\partial_{\vu^{N}}} %
\newcommand{\dpunm}{\partial_{\vu^{N-1}}} %
\renewcommand{\vv}{\mathbf{v}} %
\newcommand{\vf}{\mathbf{f}} %
\newcommand{\vw}{\mathbf{w}}  %
\newcommand{\vx}{\mathbf{x}}  %
\newcommand{\vpsi}{\bm{\psi}}  %
\newcommand{\vmu}{\bm{\mu}}  %
\newcommand{\vnu}{\bm{\nu}}  %
\newcommand{\ubas}{\phi} %
\newcommand{\vbas}{\xi} %
\newcommand{\ubasref}{\hat{\phi}} %
\newcommand{\vbasref}{\hat{\xi}} %
\newcommand{\vd}{\mathbf{d}} %
\newcommand{\vg}{\mathbf{g}} %
\newcommand{\loc}[2]{\mathrm{loc}_{#1}(#2)} 
\newcommand{\Loc}[1]{\mathrm{Loc}_{#1}} 
 
\newcommand{\ic}{\mathbf{g}} %
\newcommand{\obj}{J}  %
\newcommand{\objt}{J} %
\newcommand{\htobj}{\hat{\obj}} %
\newcommand{\per}{\theta} %
\newcommand{\vper}{\bm{\theta}}
\newcommand{\stress}{\sigma}

\newcommand{\friccoeff}{\gamma}

\definecolor{DaviColor}{rgb}{0.56,0.34,0.62}
\definecolor{DenisColor}{rgb}{0.55,0.35,0.05}
\definecolor{ArviColor}{rgb}{0.99,0.37,0.07}
\definecolor{ZizhouColor}{rgb}{0.1, 0.5, 0.5}
\definecolor{DanieleColor}{rgb}{0.3411764706,0.02352941176,0.5490196078}
\definecolor{ZachColor}{rgb}{0.6274509804,0.2039215686,0.4470588235}
\definecolor{TeseoColor}{rgb}{0.1,0.8,0.1}

\definecolor{forestgreen}{rgb}{0.13,0.54,0.13}
\definecolor{darkblue}{rgb}{0,0,.5}
\hypersetup{
	unicode=true,
	colorlinks=true,
	citecolor=forestgreen, %
	linkcolor=darkblue, %
	urlcolor=darkblue, %
}

\renewcommand{\paragraph}[1]{{\bfseries #1.}}
\renewcommand*{\paragraph}[1]{{\bfseries #1.}}

\crefname{algocf}{alg.}{algs.}
\Crefname{algocf}{Algorithm}{Algorithms}

\crefname{appsec}{Appendix}{Appendices}

\let\originalleft\left
\let\originalright\right
\renewcommand{\left}{\mathopen{}\mathclose\bgroup\originalleft}
\renewcommand{\right}{\aftergroup\egroup\originalright}

\renewcommand{\geq}{\geqslant}
\renewcommand{\leq}{\leqslant}

\renewcommand{\vec}[1]{{\bm{{#1}}}}

\renewcommand\vec[1]{\ensuremath{\mathbf #1}}

\DeclareFontFamily{U}{mathx}{\hyphenchar\font45}
\DeclareFontShape{U}{mathx}{m}{n}{<-> mathx10}{}
\DeclareSymbolFont{mathx}{U}{mathx}{m}{n}

\renewcommand{\vec}[1]{{\textbf{#1}}}

\def\u{\vec{u}}

\def\q{\vec{q}}
\def\vf{\vec{f}}

\def\stress{\sigma}
\def\R{\, \mathbb{R}}

\newcommand{\RNum}[1]{\uppercase\expandafter{\romannumeral #1\relax}}

\begin{document}

\setcopyright{acmlicensed}
\acmJournal{TOG}
\acmYear{2024} \acmVolume{1} \acmNumber{1} \acmArticle{1} \acmMonth{1}\acmDOI{10.1145/3657648}

\title{Differentiable solver for time-dependent deformation problems with contact}

\ccsdesc[500]{Computing methodologies~Physical simulation}

\keywords{Differentiable Simulation, Finite Element Method, Elastodynamics, Frictional Contact}

\author{Zizhou Huang}
\affiliation{%
\institution{New York University}
\country{United States}
}
\email{zizhou@nyu.edu}
\authornote{Joint first authors with equal contributions}

\author{Davi Colli Tozoni}
\affiliation{%
\institution{NTop and New York University}
\country{United States}
}
\email{davi.tozoni@nyu.edu}
\authornotemark[1]

\author{Arvi Gjoka}
\affiliation{%
\institution{New York University}
\country{United States}
}
\email{arvi.gjoka@nyu.edu}

\author{Zachary Ferguson}
\affiliation{%
\institution{Massachusetts Institute of Technology and New York University}
\country{United States}
}
\email{zfergus@mit.edu}

\author{Teseo Schneider}
\affiliation{%
\institution{University of Victoria}
\country{Canada}
}
\email{teseo@uvic.ca}

\author{Daniele Panozzo}
\affiliation{%
\institution{New York University}
\country{United States}
}
\email{panozzo@nyu.edu}

\author{Denis Zorin}
\affiliation{%
\institution{New York University}
\country{United States}
}
\email{dzorin@cs.nyu.edu}

\begin{abstract}
We introduce a general differentiable solver for time-dependent deformation problems with contact and friction. Our approach uses a finite element discretization with a high-order time integrator coupled with the recently proposed incremental potential contact method for handling contact and friction forces to solve ODE- and PDE-constrained optimization problems on scenes with complex geometry. It supports static and dynamic problems and differentiation with respect to all physical parameters involved in the physical problem description, which include shape, material parameters, friction parameters, and initial conditions. Our analytically derived adjoint formulation is efficient, with a small overhead (typically less than 10\% for nonlinear problems) over the forward simulation, and shares many similarities with the forward problem, allowing the reuse of large parts of existing forward simulator code.

We implement our approach on top of the open-source PolyFEM library and demonstrate the applicability of our solver to shape design, initial condition optimization, and material estimation on both simulated results and physical validations.

\end{abstract}

\begin{teaserfigure}\centering\footnotesize
\includegraphics[width=0.3\linewidth]{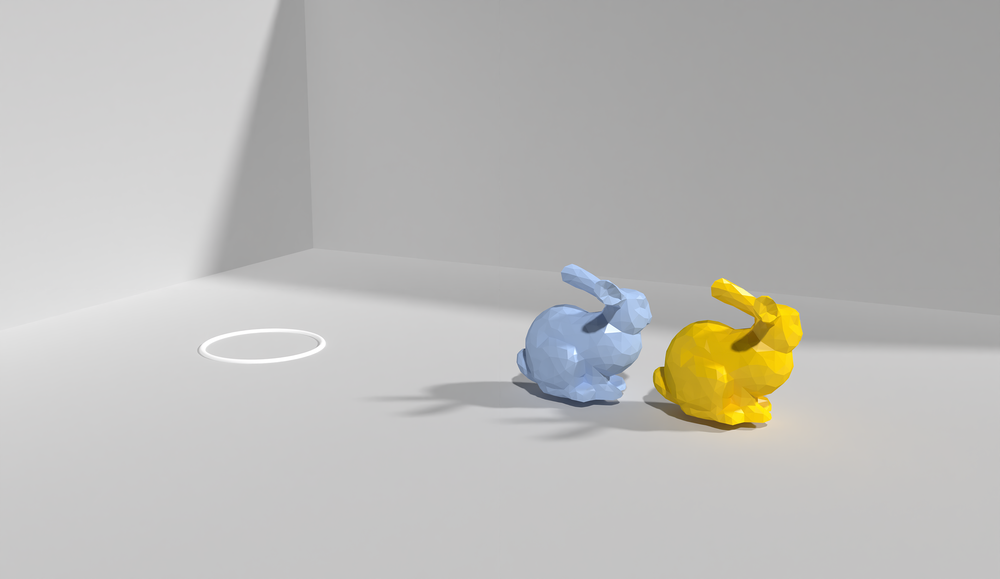}\hfill
\includegraphics[width=0.3\linewidth]{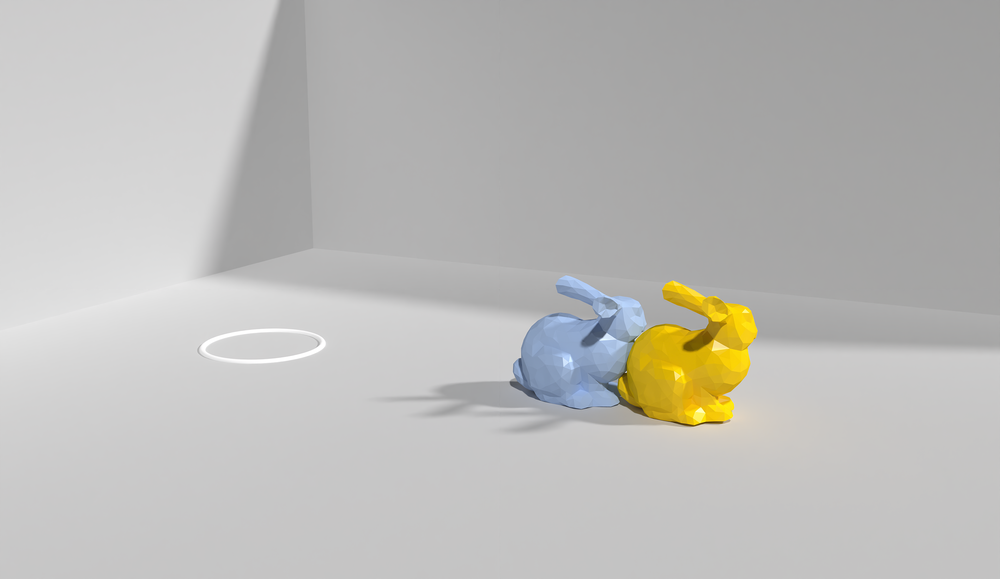}\hfill
\includegraphics[width=0.3\linewidth]{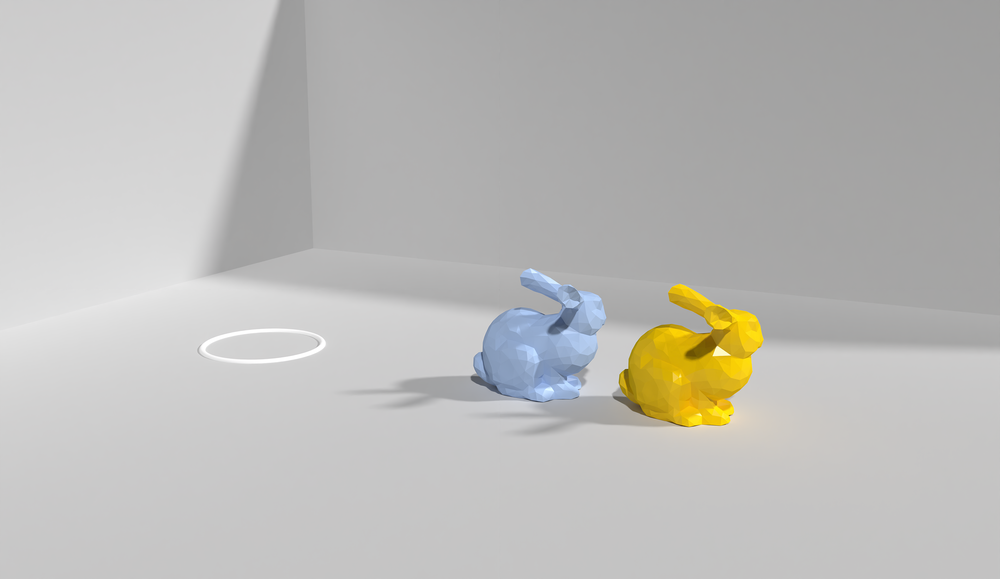}\\
\vspace{1em}
\includegraphics[width=0.3\linewidth]{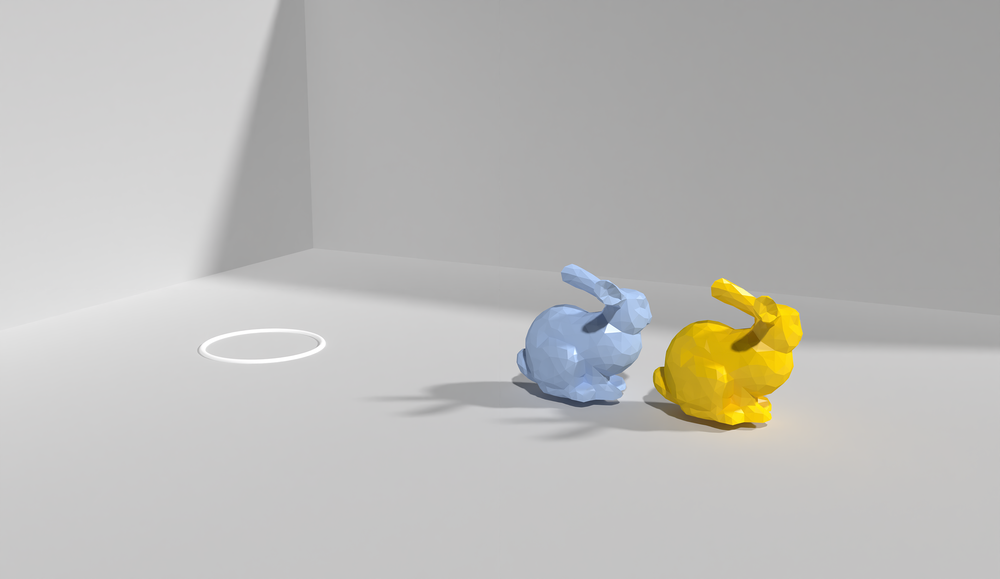}\hfill
\includegraphics[width=0.3\linewidth]{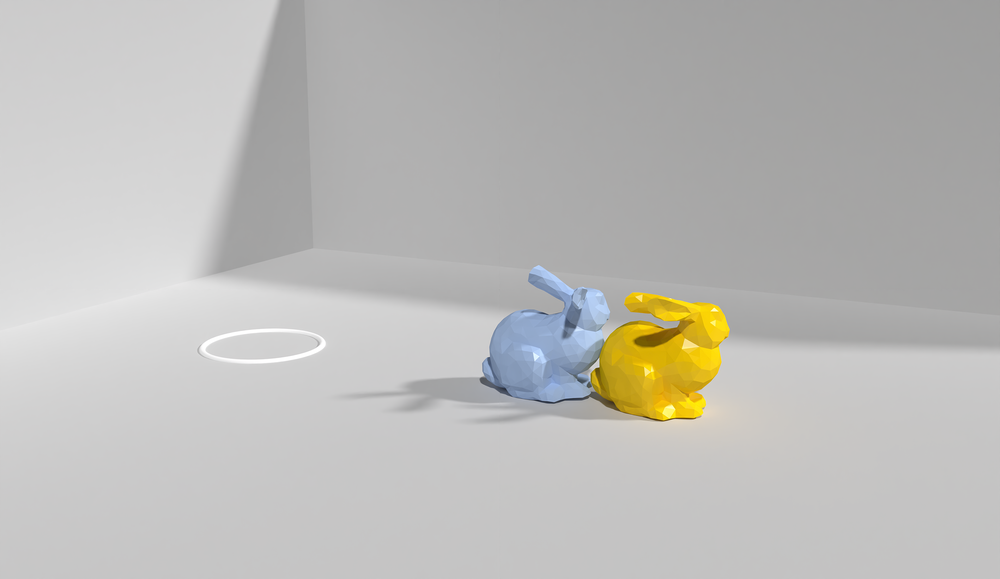}\hfill
\includegraphics[width=0.3\linewidth]{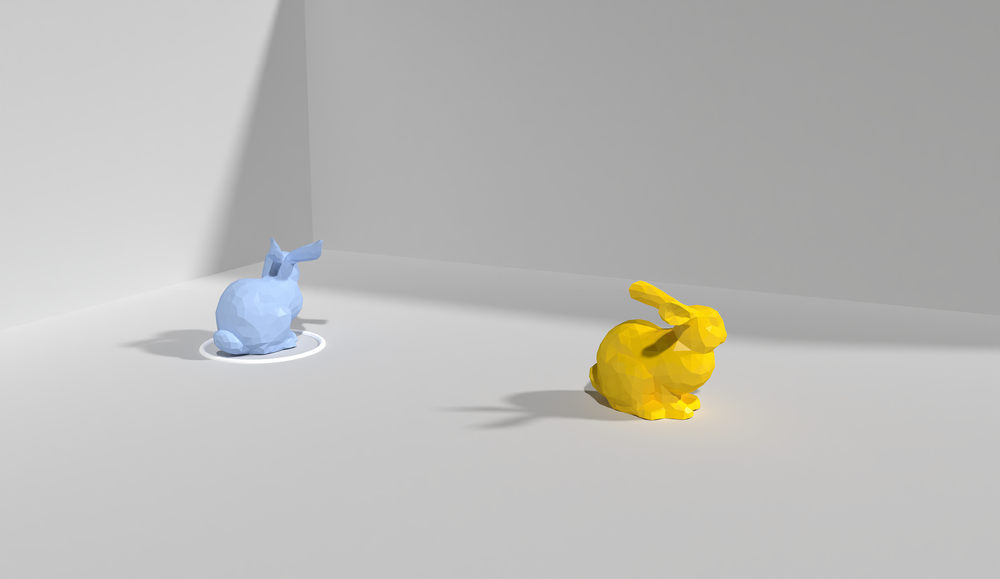}\\
\parbox{0.3\linewidth}{\centering $t = 0$}\hfill
\parbox{0.3\linewidth}{\centering $t = 0.15$}\hfill
\parbox{0.3\linewidth}{\centering $t = 1.5$}\hfill
\caption{The direction and magnitude of the initial velocity of the yellow bunny is optimized to push, after contact, the blue bunny into the white circle marker. Top row is the initial configuration, and bottom row is our optimized result. This scene involves a elastodynamic simulation with a non-linear material model with contact and friction forces.}
\label{fig:bunny-pool}
\Description{}
\end{teaserfigure}

\maketitle

\newenvironment{explicitexpr}{}{}
\excludeversion{derivation}
\excludeversion{explicitexpr}

\section{Introduction}
\label{sec:introduction}

ODE- and PDE-constrained optimization problems, i.e. the minimization of a functional depending on the state of a physical system modeled using a set of (partial) differential equations, appear in many application areas: optimized design in engineering and architecture, metamaterial design in material science, inverse problems in biomedical applications, controllable physically-based modeling in computer graphics,  control policy optimization, and physical parameter estimation in robotics.

A common family of PDE-constrained optimization problems in graphics, robotics, and engineering involves static or time-dependent elastic deforming objects interacting with each other via contact and friction forces. A significant number of approaches have been proposed to tackle PDE-constrained optimization problems of this type (Section \ref{sec:related}). 

However, these approaches often make application-specific assumptions aimed at simplifying the differentiable simulator, often sacrificing generality (e.g., handling contact only with simple rigid obstacles or differentiating with respect to material parameters only),  robustness (e.g., using a contact model that requires per-scene parameter tuning to prevent failure), accuracy (e.g., using approximate spatial discretizations, or non-physical material and friction models), or scalability (e.g., restricting the number of system parameters with respect to which it can be optimized).

Building on and integrating a broad range of previous work on PDE-constrained optimization, including shape optimization, material property estimation, and trajectory control, we develop a differentiable solver that eliminates or reduces these shortcomings. 
Our solver has the following characteristics:

\begin{enumerate}
\item Maximally \emph{general differentiability}: we support differentiation with respect to all physical parameters (Section \ref{sec:general-forces}) involved in the physical problem description: shape, material parameters, friction parameters, and initial conditions. The user can pick an arbitrary subset of these parameters to use in objective functions (Section \ref{sec:objectives}), differently from previous works which limit this selection (Table \ref{tab:characteristics}).
\item Our contact/friction formulation builds upon the recently proposed Incremental Potential Contact (Section \ref{sec:contact}) approach \cite{Li2020IPC}. 
 Our differentiable simulator supports complex geometry, is automatic and robust (with only two main parameters controlling the accuracy of the spatial and temporal discretizations), and guarantees physically valid configurations at all timesteps, without intersections nor inverted elements. Many previous works instead use a restricted set of contact scenarios (Table \ref{tab:characteristics}).
 \item We use discretizations of arbitrary order (Section \ref{sec:results}), both in space and time with general non-linear elasticity material models, ensuring \emph{accuracy}. Many competing works instead rely on linear time and spatial discretization and often use simplified material models, leading to lower accuracy solutions (Section \ref{sec:related}).
 \item Our formulation supports both static and dynamic problems in a unified framework (Section \ref{sec:adjointderivatives}).
\item Our differentiation approach is \emph{low cost.} The computation of the derivatives for one PDE-constrained optimization step is at most as expensive as a forward evaluation of the underlying forward simulation of the physical systems (Section \ref{sec:summary-grad},
Table~\ref{Table:stats}), and, for nonlinear problems, we observe that the differentiability adds at most 10\% to the cost. 
\end{enumerate}

While individually most of these features appeared in previous works  in some form,
they have never been combined in a unified formulation and algorithm for accurately solving inverse problems in elastodynamics with contact.  The foundation of our approach is the \emph{adjoint method}, which we systematically apply to obtain derivatives with respect to all parameters in a unified and general way, while achieving high efficiency.  We discuss our design choices and compare to alternatives in Section~\ref{sec:related:compute-grad}.

We demonstrate the effectiveness of our approach on a set of examples involving multiple objectives and optimizing for the shape, material parameters, friction parameters, and initial conditions.
\section{Related Work}
\label{sec:related}

We summarize the most relevant simulation frameworks, primarily focusing on those supporting differentiable simulations of elastic deformable objects.

For the works closer to our targeted applications, we provide an explicit breakdown of which subset of the characteristics of our solver they support (Table \ref{tab:characteristics}). We also highlight the generality of our solver by explicitly identifying which solvers cannot reproduce the examples in our paper (Table \ref{tab:figures}). While implementing additional derivatives with respect to parameters already present in one of these codes is easy in some cases, other features are harder to add, e.g., contact between soft bodies or self-collisions. The reasons why specific solvers cannot handle certain problems are included in the caption of Table \ref{tab:figures}. We note that prior works in Tables~\ref{tab:characteristics}, \ref{tab:figures} can solve problems that our method cannot handle, e.g. the application in visuomotor control tasks in~\cite{gradsim} is not included in this work.

\begin{table*}[htb]
\caption{The table columns correspond to five comparison characteristics: 
(1) High-order space and time discretization, (2) supported optimization parameters (3) support for complex contacts between arbitrary surfaces, including self-collision, (4) support for static and dynamic simulations, and (5) method for the derivatives computation. No existing differentiable solver supports all features of our solver simultaneously; in particular, most do not support differentiating with respect to the shape of the domain. 
}
\footnotesize
\begin{tabular}{l l l l l l}
Method      & (1) HO & (2) Parameters                  & (3) Collisions                                              & (4) Static and Dynamic & (5) Differentiation \\ \hline

Elastic Texture \cite{Panetta2015} & Yes & Shape & No support & Static-Only & Adjoint \\
CB-Assemblies \cite{Tozoni2021} & Yes & Shape & Static and Prescribed & Static-Only & Adjoint \\
ADD \cite{geilinger2020add} & No & Material, Initial & Only planes or SDF, no self-collisions & Dynamic-Only & Adjoint \\
GradSim \cite{gradsim} & No & Material, Initial & Only planes, no self-collisions & Dynamic-Only & Code transformation \\
DiSECt \cite{heiden2021disect} & No & Material & Only planes or SDF, no friction & Dynamic-Only & Code transformation/autodiff \\
NeuralSim \cite{heiden2020neuralsim}& No & Material, Initial & Only rigid-bodies & Dynamic-Only & Code transformation/autodiff \\
DiffPD \cite{du2021diffpd} & No & Material, Initial & Only planes or SDF & Dynamic-Only & Adjoint \\
\textbf{Ours} & \textbf{Yes} & \textbf{Shape, Material, Initial} & \textbf{No restrictions} & \textbf{Static and Dynamic} & \textbf{Adjoint} \\
\end{tabular}
\label{tab:characteristics}
\end{table*}

\begin{table*}[htb]
\caption{To clarify the differences between our approach and other differentiable simulators, we show which simulators support the features needed for each experiment in our paper. The figure captions provide more details for each experiment; most significantly, almost no other simulators support shape optimization (Fig \ref{fig:bridge-shape}--\ref{fig:shock-protect}), and the ones that do lack support for dynamic.
From left to right: Fig \ref{fig:bunny-pool} and \ref{fig:ball-h} require contact handling between soft bodies; Fig \ref{fig:bridge-shape}--\ref{fig:shock-protect} require shape optimization; Fig \ref{fig:sine}--\ref{fig:bridge-mat} require material distribution optimization; and Fig \ref{fig:physical-microstructure} and \ref{fig:bunny} require self-collision handling. 
}
\begin{tabular}{l c c c c c c c c c c c c}
Method & Fig\ref{fig:bunny-pool} & Fig\ref{fig:bridge-shape}--\ref{fig:hanger3D} & Fig\ref{fig:bouncing-ball}--\ref{fig:shock-protect} & Fig\ref{fig:bouncing-puzzle} & Fig\ref{fig:sine}--\ref{fig:bridge-mat} & Fig\ref{fig:physical-cube} & Fig\ref{fig:physical-microstructure} & Fig\ref{fig:kangaroo} & Fig\ref{fig:bunny} & Fig\ref{fig:ball-h} & Fig\ref{fig:bounce-ball} \\ \hline
Elastic Texture \cite{Panetta2015}& & & & & Y & & & & & & \\
CB-Assemblies \cite{Tozoni2021}& & Y & & & & & & & & & \\
ADD \cite{geilinger2020add}& & & & Y & & Y & & Y & & & Y \\
GradSim \cite{gradsim}& & &  & Y &  & Y & & Y & &  & Y \\
DiSECt \cite{heiden2021disect}& & &  & &  & Y & & Y & & & Y \\
NeuralSim \cite{heiden2020neuralsim}& & &  &  &  &  &  &  &  &  &  &  \\
DiffPD \cite{du2021diffpd}& & &  & Y &  & Y & & Y & Y & & Y \\
\textbf{Ours}& \textbf{Y} & \textbf{Y}& \textbf{Y}  & \textbf{Y} & \textbf{Y}  & \textbf{Y}  & \textbf{Y} & \textbf{Y}  & \textbf{Y} & \textbf{Y} & \textbf{Y} \\
\end{tabular}
\label{tab:figures}
\end{table*}

\paragraph{Differentiable deformable object simulators}
Numerous  differentiable elastic body simulators have been developed for applications in
optimal design of shapes~\cite{Panetta2015,panetta2017,Tozoni2020, elasticshell2018}, actuators~\cite{Skouras2013,Chen2020,Maloisel2021}, sensors~\cite{Tapia2020}, material characterization~\cite{Hahn2019,Schumacher2020}, and robotic control~\cite{Bern2019,Hoshyari2019}. Differentiable simulators are also developed for fluid simulations in~\cite{McNamara:2004:fluid, pmlr-v87-schenck18a, 2023SPHRigid}.
These simulators broadly fit into three categories:
(i) those employing analytic derivatives computed using sensitivity analysis;
(ii) those using automatic differentiation  libraries~\cite{hu2019difftaichi,heiden2020neuralsim} based on overloading, or algorithmic differentiation
(iii) neural surrogate models replacing the entire simulation with a differentiable neural network~\cite{chang2016compositional,Zhang2016,Baque2018,Bern2020}.

Our method belongs to the first category: analytic sensitivity analysis generally requires manual differentiation of the physics equations, but allows one to reuse existing solvers most easily; direct differentiation is feasible only if the number of parameters is very small; a large number of parameters requires construction of the adjoint equations for specific functionals~\cite{liang2019differentiable,qiao2020scalable,Bern2019,rojas2021differentiable,du2021diffpd,li2022diffcloth, elasticshell2018}, and is more efficient than all other approaches.  One exception is Dolphin-Adjoint \cite{Mitusch2019}, which  automatically and robustly derives adjoint models for models written in the finite element software FEniCS~\cite{AlnaesEtal2015}. Automatic differentiation methods are most general but require existing simulators to be rewritten using data structures required for gradients and Hessians, and typically incur a significant performance penalty. Surrogate models, though enabling dramatic speedups in some cases, require huge training sets and long training times for even simple design spaces~\cite{Gavriil2020}, and currently are unsuitable for high-precision applications~\cite{Bacher2021}. 
Code transformation and auto-differentiation, e.g. in simulators 
such as \cite{gradsim} and \cite{heiden2021disect},  based on technology developed in NVIDIA Warp~\cite{nvidiawarp}, while potentially allowing one to reuse existing codes, typically places a few limitations on what the code may contain.
To the best of our knowledge, none of the existing simulators support robust handling of contact and friction for complex geometries, and they only support a subset of the design parameters compared to the more general formulation of this paper.

We provide direct comparisons of our solver, \cite{du2021diffpd}, and \cite{gradsim} in \cref{sec:res:comparisons}.

\paragraph{Differentiable Simulations with Contact}
Differentiable simulators incorporating various contact models have recently been developed for rigid~\cite{heiden2020neuralsim} and soft bodies~\cite{liang2019differentiable,qiao2020scalable,geilinger2020add, gradsim, heiden2021disect}. These contact models often require per-scene parameter tuning if complex contact scenarios are present, which makes these methods hard to use in optimization, especially shape optimization.

Our approach uses the recently proposed Incremental Potential Contact (IPC) formulation \cite{Li2020IPC}, replacing the traditional zero-gap assumption~\cite{wriggers1995finite,Kikuchi88,Stewart01,Brogliato99,belytschko2000nonlinear,bridson2002robust, otaduy2009implicit,harmon2008RTSC,harmon2009asynchronous,gilles20111hybrid,verschoor2019efficient} with a smooth version ensuring a (small) non-zero separation between objects at every frame of the simulation. This approach was designed with the explicit goal of guaranteeing \emph{robustness} and its smooth formulations of contact and friction avoids the need for handling non-smooth constraints.

\citet{Stupkiewicz2010} is one of the few papers that demonstrate sensitivity analysis of elastic problems with contact with respect to a range of parameters, including shape and material properties. This method, tested on a limited set of regular-grid problems, uses direct differentiation requiring a solve per parameter, and does not use a robust contact model.

We compare our solver, \cite{du2021diffpd}, and \cite{gradsim} in Section \ref{sec:res:comparisons} in  scenes involving both contact and self-contact.

\paragraph{Shape and topology optimization with contact}  Historically shape optimization was primarily considered separately, e.g., for physical parameter or initial condition estimation, primarily in static settings, often with additional assumptions on bodies involved in contact.

Some previous works  in this area have considered the specific case of optimization in the presence of contact between a soft body with fixed rigid surfaces~\cite{Beremlijski2014,Haslinger1986b,Herskovits2000}. Other works, like ours, have studied the interaction of two or more deformable bodies in contact~\cite{Desmorat2007,Stupkiewicz2010,Maury2017,Tozoni2021}.  Most papers do not consider friction or use a simplified model (compared to the standard Coulomb formulation) as discussed by \citet{Maury2017}.
 
Most closely related to our approach, \citet{Maury2017} presented a level set discretization technique where contact and friction were modeled with penalty terms,  using smooth approximations to the problem. Using a similar contact and friction model, \citet{Tozoni2021} designed a shape optimization technique that focused on reducing stress of static assemblies that are held together by contact and friction. Both these works followed the mathematical model of contact presented by \citet{Eck2005}, which allows for interpenetration and assumed that contact zones are fixed.

For the specific use case of avoiding sag due to gravity forces, \citet{sagfree} proposes a global/local approach to optimize the rest shape and initial displacement of input geometries to avoid the deformation introduced by gravity forces. This work uses the IPC contact model in some simulation examples, but does not use IPC in their optimization procedure.

Our approach supports dynamic simulation, allows contact zones to change with both optimization parameter changes and in the course of the simulation, and supports contact and self-contact  between arbitrary deformable objects.

\paragraph{Meshfree methods}
A number of differentiable simulation methods use meshfree discretizations. Especially for shape optimization, methods like XFEM~\cite{Schumacher2018,Hafner2019}
and MPM~\cite{hu2019chainqueen} that do not maintain conforming meshes are often considered to circumvent remeshing-induced discontinuities~\cite{Bacher2021}.
However, these methods sacrifice accuracy~\cite{de2019material}, particularly for stress minimization problems~\cite{Sharma2018}.
Our approach computes accurate displacement and stressed by using a finite element method framework using high-order elements, coupled with dynamic remeshing to compensate for the distortion introduced by large deformations.

\subsection{Choice of approach to computing gradients}
\label{sec:related:compute-grad}

A broad variety of approaches to differentiable simulation exist in the literature on optimal control, shape optimization, and inverse problems (see, e.g.,  \cite{VANKEULEN20053213} for a systematic overview); in this section, we briefly discuss the motivation for the design choices in our algorithm.  Our choice is significantly influenced by the features of our problem setting: 
\begin{itemize}
\item High dimension: e.g., shape and variable material property optimization may require thousands of parameters. 
\item Complex linear solvers: we aim to accurately solve highly nonlinear, time-dependent or static, stiff problems, requiring complex linear solvers for large sparse linear systems in the inner loop of nonlinear solvers.  
\item  Contact: resolving contacts requires additional complex algorithms for continuous collision detection, in the nonlinear solver line searches.
\item Large shape changes and deformations: shape differentiation often leads to large shape changes, which may require remeshing.
\end{itemize}

\paragraph{Choice of the overall approach}
The two most general approaches are, in a sense, opposite extremes, but neither is a good fit for our setting.  

\emph{Finite difference} methods can be used with any black-box solver, but require an extra solve for each parameter, so it is not suitable for high-dimensional problems or even problems of moderate dimension (Table~\ref{Table:finite_diff} compares the efficiency of our method and finite differences).

\emph{Code differentiation}  \cite{naumann2012art,griewank2008evaluating,bischof2000computing,margossian2019review} 
 through overloading operators, or using a specialized language, has two fundamental problems, making it unsuitable for 
complex nonlinear codes with contact: it requires rewriting all of the simulation code, including supporting numerical libraries, e.g., sparse linear solvers and contact handling, and even more significantly is likely to produce unnecessarily inefficient code (fully differentiable sparse linear matrix inversion is going to be slow, and differentiating through a nonlinear solve is unnecessary, as we see below). 
While automatic code transformation in principle may eliminate the need to rewrite the code, and there is promising work \cite{autodiff,Moses:2022} in this direction, we are unaware of fully automated tools capable of handling large software systems, and the concerns about the efficiency of the resulting code remain.

Existing differentiable solvers following this route use explicit time integration and/or a few iterations of an iterative linear solver. Both these options are unsuitable for applications requiring robustness and accuracy, limiting their applicability. For more details, we refer to Appendix E.3 of \cite{hu2019difftaichi}, where the authors discuss that it is not realistic to differentiate stably a complex linear solver (the paper refers to a multigrid solver, but it is even more true for a sparse direct solver), so they use 10 Jacobi iterations to approximate the linear solve in the smoke simulation.

We opt for the approach based on \emph{adjoint equations}, well established in scientific computing and optimal control, as described in Section~\ref{sec:overview}. It is widely considered the most efficient approach to computing sensitivities, with the cost of a single additional linear solve per time step, and reusing important parts of the forward solver, at the expense of requiring derivations specific to a particular time-stepping algorithm.
It allows us to implement efficient differentiability for solvers with all the features listed above. 

\paragraph{Fixed vs. changing discretization} The adjoint method is particularly simple to apply to a purely algebraic problem, in which both the objective and PDE are discretized once, and then the problem is treated as a purely algebraic finite-dimensional optimization problem with PDE acting as a constraint.  However, in our context, as shape optimization may change the domain, we cannot view the optimization problem as purely algebraic, 
as the discretization may change at every optimization step: both the forward and adjoint systems are rebuilt starting from a new discretization.

\paragraph{Discretize-then-optimize vs optimize-then-discretize}
In the context of adjoint methods, we need to choose between the "discretize-then-optimize" approach and "optimize-then-discretize" \cite{VANKEULEN20053213}.  In the first approach, the original PDE and objectives are converted to a discrete form which is then differentiated with respect to discrete optimization parameters. In the second approach, a PDE for the sensitivities is derived, and this PDE is discretized. The difference between these approaches is relatively small for differentiation with respect to material parameters, but more significant for shape derivatives. In this context, "optimize-then-discretize" is the most common approach: its convergence theory is better established, and directly follows from the discretization convergence. On a more practical side, it leads to a simpler form of adjoint equations for the shape derivatives formulated in the physical domain, enabling better reuse of the forward solver code. We refer to Section 2.3 in \cite{ALLAIRE20211} for additional discussion and to  Appendix~\ref{app:optimizediscretize} for an example illustrating the differences for the Poisson problem.  We emphasize that both approaches, for a suitable choice of discretization, lead to the same discrete solution; however, discretize-then-optimize in the context of shape optimization obscures the essential fact that the system matrix need not be recomputed. We use a specific discretization that ensures that the computed gradients are consistent with differentiating the discretized objective, as this simplifies the implementation of optimization algorithms.

\cite{dokken2020automatic} uses the "discretize-then-optimize" approach to support shape derivatives in FEniCS, which has its own DSL. This approach allows one to support a broad range of PDEs but at the expense of higher complexity and significant additional performance overhead.

\paragraph{Constructing adjoint equation components: AD vs analytic approach}
The adjoint method requires partial derivatives of the objective for the right-hand side of the adjoint system, and the stiffness and mass matrices for the adjoint PDE itself, which are similar to or coincide with those for the forward PDE. These can be computed using an AD method (note that the code to be differentiated is a straightforward algebraic computation, not a complex algorithm like a linear solver) or in closed form. 

 This can be done by transforming the code of assembling force vectors and computing objectives to an AD framework or applying code transformation to these parts of the code. However, AD leads to less reuse compared with the analytic case and higher computational complexity. We briefly compare these options in Section~\ref{sec:results}.  We opt for doing extra analytical work to derive all derivatives explicitly, but the approaches can be combined - one can add additional forces or objectives using AD.   

\section{Overview}
\label{sec:problem-setup}
\label{sec:overview}
In Sections~\ref{sec:adjointderivatives}-\ref{sec:all-objectives}, we provide a self-contained description of our method. While this contains a mix of known and new material, we aim to present all components of the method in a unified and systematic notation to ensure reproducibility.

\renewcommand{\arraystretch}{1.4}
\begin{table*}[htb]
\caption{Notation.}
 \tabletail{\hline}
\begin{tabular}[t]{|p{0.6in}|p{2.5in}|}
\hline
\multicolumn{2}{|l|}{\textbf{Domains and bases}}\\ \hline
$D_d$ & Domain dimension, $2$ or $3$.\\ 
$D_s$ & Solution dimension, $1$, $2$ or $3$.\\
$\Omegar$ &\textbf{Reference domain} $\Omegar \subset \mathbb{R}^{D_d}$ consists of copies of identical reference  elements $\hat{K}_j$, $j=1\ldots n_K$ identified along edges.\\ 
 $\hat{x}_\ell$ and  $\hat{z}_\ell$
  & \textbf{Nodes}  are points in $\Omegar$  used to define bases, $\ell= 1 \ldots n^x_N$, and   $1 \ldots n^z_N$ respectively.
\emph{The set of nodes $\hat{z}_\ell$ does not include nodes with Dirichlet boundary conditions; the set of nodes $x_i$ does include these nodes. }\\
$\ubasref^\ell$ and $\vbasref^\ell$ & \textbf{FE basis functions}  are scalar basis functions defined on $\Omegar$; $\vbasref^\ell$ correspond to nodes  $\hat{x}_i$, and is used for geometric maps (we use p.w. linear basis);  $\ubasref^\ell$ correspond to $\hat{z}_\ell$ and used for all other quantities (arbitrary order Lagrangian).
\\
$\fprms$, $\fprms^j(y^j)$, $\prms$,  $x_\ell$ &\textbf{Geometric map} $\fprms$ embedding a reference element in space,  is defined on each $\hat{K}_j$ in $\Omegar$ with local coordinates $y^j$ as 
$
\fprms^j(y^j) =  \sum_{\ell}  x_\ell \vbasref^\ell(y^j),
$
where $x_\ell \in \mathbb{R}^{D_d}$ are the positions  of the nodes of the element $j$ forming the vector $\prms$. %
Concatenation of these  maps yields the global geometric  map 
$\fprms: \Omegar \rightarrow \mathbb{R}^{D_d}$.
\\
 $\Omega_\prms$ & \textbf{Physical domain} is the domain on which the PDE is solved,  parametrized by  $\prms$,  $\Omega_\prms = \fprms(\Omegar)$. 
 $
$
The global coordinate on $\Omega_\prms$ is $x = x^\prms \in \mathbb{R}^{D_d}$.
\\
$\ubas^\ell(x)$,$\vbas^\ell(x)$ & \textbf{FE bases on $\Omega_\prms$.}  The bases $\ubasref^\ell$ and $\vbasref^\ell$ can be pushed forward to the 
 domain $\Omega_\prms$ via $\ubas(x) = \ubasref\circ \fprms^{-1}(x)$ and  $\vbas(x) = \vbasref \circ \fprms^{-1}(x)$. 
 \\
 $\Omega_{\prms + \vper t}$&  \textbf{Perturbed domain} obtained using  a perturbation direction $\vper$  in $\prms$. Perturbation $\per(x) \in \mathbb{R}^{D_d}$ is: 
 $\per = \per(x) = \sum_\ell \per_\ell \, \vbasref^\ell \circ (\fprms^j)^{-1} (x) = \sum_\ell \per_\ell \, \vbas^\ell(x)$.
\\
\hline
\end{tabular}
\hfill
\begin{tabular}[t]{|p{0.5in}|p{2.6in}|}
\hline
\multicolumn{2}{|l|}{\textbf{Functions on physical domain} $\Omega_\prms$}\\ \hline
$u_\prms(x)$,$\vu$ & \textbf{PDE solution defined on $\Omega_\prms$} with values in $\mathbb{R}^{D_s}$.
We denote the vector of coefficients of  $u$ in the FE basis 
$\ubas$ by $\vu$. 
$
u(x) = \sum_\ell u_\ell \ubasref^\ell \circ \fprms^{-1} (x) = \sum_\ell u_\ell \ubas^\ell(x).
$
\\
 $w(x)$, $\psi(x)$,$\vw$,$\vpsi$ & \textbf{Test functions} (scalar)
defined similarly to $u(x)$ in the same basis and  vectors of their coefficients are  $\vw$ and $\vpsi$.
\\
$\adj(x)$,$\vadj$ &\textbf{Adjoint solution} is the solution of the adjoint equation and the vector of its coefficients, with values in $\mathbb{R}^{D_s}$.
\\
 $q^m(x)$, 
$\prm^m$ & \textbf{$m$-th optimization parameter}  
 $ q^m(x) = \sum_{\ell=1}^{n^m_\fprm} \fprm^m_{\ell} \zeta^\ell(x)$
 with a basis $\zeta^\ell$ with values in $\mathbb{R}^{D^m_{\fprm}}$
 parameters can be material properties, boundary conditions etc, defined on all or parts of $\Omega_\fprms$. 
 For the geometry map $\fprms$,  $\vbas$ on $\mathrm{Dom}(\fprms) = \Omegar$, and $\zeta^\ell = \xi^\ell$.
\\
&\rule{0pt}{5ex}\\
\hline
\multicolumn{2}{|l|}{\textbf{PDE and derivatives}}\\
\hline
$\pde(\vu,\prm) = 0$  &\textbf{Discretized  form of the PDE,}
i.e., a system of $n_u$ algebraic equations with components
of $\vu$ as unknowns. 
\\
$\obj(\vu,\prm)$ &\textbf{Discretized form of the objective}.
\\
 $\dpar a(\vu,\prm)$ &\textbf{Derivative of  a (possibly) vector quantity $a$ with respect to a vector of optimization parameters}, not including dependence through $u$. The vector is the vector of coefficients of one of $\fprm^m$ or $\fprms$.
 If the dimension of $a$ is $n_a$,  then $\dpar \vva$ is a matrix of size $n_a \times D^m_\fprm n^m_\fprm$.
\\
$\dpu a(\vu,\prm)$ &\textbf{Derivative of  a quantity $a$ with respect to the the PDE solution $u$}; it is a vector of length $ D_s n_u$.   
\\
$\fdpar a(\vu,\prm)$ &\textbf{Full derivative of $a$ with respect to $\prm$, including through the dependence on $u$.}
\\ 
$\nabla a(v)$, $\nabla_i a(v,w)$ &\textbf{Derivatives of $a$ with respect to arguments $v,w \in \mathbb{R}^{D}$}.
\\
\hline
\end{tabular}
\label{tab:notation}
\end{table*}

\paragraph{Typographical conventions} 
We use lower-case italic for functions $a(z)$ and variables $z$,  with both $z$ and $a$ in $\mathbb{R}^D$, where $D=2,3$. Boldface lower-case letters ($\mathbf{a}$) are used for vectors of coefficients of a FEM (or any other) discretization of a function.
For a vector or matrix quantity, superscripts are used to index whole vectors or matrices: e.g., $\mathbf{p}^i$ may denote  $\mathbf{p}$ at time step $i$. Subscripts are used for the indices of components of a vector, e.g., 
$a(z) = \sum_{\ell=1}^n a_\ell \phi^\ell(z)$
means that the function $a(z):  \mathbb{R}\rightarrow \mathbb{R}^D$ is a linear combination of basis functions $\phi^\ell$, with coefficients $a_\ell$ which are components of $\mathbf{a}$.
If $a(z)$ has values in $\mathbb{R}^D$, its coefficients in a scalar basis $\phi^\ell$ are $D$-dimensional, 
Then $\mathbf{a}$ is a vector of length $D\cdot n$, with $D$ coordinates of  each component of $a_\ell$ in sequential entries. 

\paragraph{General problem form}  We solve static and dynamic optimization problems of the form 
\begin{equation}
    \min_q \obj(u, x, q),\; \mbox{such that,  $\fh(u,x,q) = 0$}
    \label{eq:static-general}
\end{equation}
and
\begin{equation}
    \begin{split}
   &\hspace{50pt}\min_q J(\vu,\prm) = \min_q\int_{t=0}^T \objt(\vu,t,\prm) \;\\
   &\mbox{{\small such that} $\rho \ddot{u}  = \fh(u,x,t,q)$ {\small on} $\Omega_\prms$},\;
     u(0) = g^u(q), \dot{u}(0) = g^v(q),
    \end{split}
    \label{eq:dynamic-general}
\end{equation}
where $\obj$ is an objective, possibly including constraints in penalty form, $u(x,t)$  is the displacement
of a material point $x$ satisfying a static or dynamic physics equation, and $g^u$ and $g^v$ are the initial conditions for the displacements and velocities. In this work, we consider nonlinear elastic deformation, contact, friction, and damping forces. 
We assume the density $\rho$ to 
be constant in time.  The optimization parameter functions $q = (\prms, \prm^1, \ldots, \prm^m)$ include all parameters of the system: material properties (elastic, friction, and damping), object shape, and initial and boundary conditions. 
The first of these, $\fprms$ plays a special role: it determines the shape of the domain  $\Omega_\prms$ on which the PDE is defined;  it is a function on  a reference domain $\Omegar$ defining its deformation.
Parameters $q^i$ may be global constants, or dependent on the points of the  reference domain, or pairs of points (as it is the case for the friction coefficient).   

This problem statement is similar to \cite{geilinger2020add} and other works on differentiable simulators; however, our goal is to support full differentiability, including shape, in a systematic way (see Table~\ref{tab:characteristics} for details) which affects the adjoint formulation and requires deriving expressions for a number of gradients of forces and functionals. 

\paragraph{Discrete problem} 
We postpone the exact description of the discrete problem to Section~\ref{sec:physicalmodel}.  The discretized static problem obtained using FEM discretization has the general form:  
\begin{equation}
\min_\prm \obj(\vu,\prm),\; \mbox{s.t., $\pde(\vu,\prm) = 0$},
\label{eq:discrete-static}
\end{equation}
where $\vu$ is the vector of FE basis coefficients of $u$ and 
$\prm$ is the concatenation of the vectors of coefficients of 
$\prms, \prm^1, \ldots, \prm^m$.

The dynamic discretized problem with BDF of order $m$ discretization in time has the general form:
\begin{equation}
\begin{split}
&\min_\prm J(\vu,\prm) = \min_\prm \sum_{i=0}^N w_i \objt_{i}(\vu^i,\prm) \\ %
&\vu^i + \sum_{j=1}^{\min(i,m)}\alpha^i_j \vu^{i-j} = \beta_i \dt \vv^{i}\\
&M \bigg(\vv^i + \sum_{j=1}^{\min(i,m)}\alpha^i_j \vv^{i-j}\bigg) = \beta_i \dt \pde^i(\vu^i, \vu^{i-1},\prm) = \hht^i,
\end{split}
\label{eq:discrete-dynamic}
\end{equation}
where $M$  is the mass matrix.
The higher-order BDF schemes  need to be initialized with lower-order steps; more specifically, 
$\alpha^i_j$ is $j$-th coefficient of BDF$i$, for $1 \leq i < m$, 
and $j$-th coefficient of BDF$m$ otherwise. 
In the formulation above,  $ \pde(\vu,\prm)$  does not depend on  velocities $\vv$.  If the dependence on velocities is needed, as for damping forces,  we discretize in time, and handle it as dependence on $\vu$ at  different time steps.  

\paragraph{Overview of the method} We aim to present a complete, largely self-contained formulation, to ensure reproducibility as well as support easy addition of new types of forces.  This requires restating briefly some of the known facts and formulas using our notation; we identify parts that are not present in previous work.

We first assume the discretized form of the problem \eqref{eq:discrete-static} and \eqref{eq:discrete-dynamic}, and derive consistent adjoint equations for the static and dynamic cases.  

Each force and objective can be added to this general framework by deriving a set of matrices and vectors needed to compute partial force and objective derivatives. 

We then proceed by computing these quantities analytically for the set of forces involved in our formulation, and a broad selection of functionals, including most used in the previous work both on differentiable dynamic simulation and shape optimization.  We compute these in a form that allows for easy remeshing of $\Omegar$ and
$\Omega_\prms$, which is necessary for the large changes in physical domain introduced by shape optimization.

\section{Adjoint-based objective derivatives}
\label{sec:adjointderivatives}
The derivatives of the objective  $\obj$ with respect to optimization parameters can be computed efficiently using the classic adjoint method. 
While the basic principles of derivation are well-known, we show how these are applied in the context of our problem. The general form of our equations is similar to \cite{geilinger2020add}, which in turn is based on \cite{Hahn2019} for the specific case of BDF2 time-stepping and material parameter differentiation.  We derive the abstract form of the adjoint system for a general form of BDF time-stepping, and importantly we ensure that the dynamic adjoint solution is \emph{consistent}, i.e., yields identical, rather than approximately identical, results to direct differentiation, as well as consider variable mass matrix needed for shape derivatives.

\subsection{Static case}

With the adjoint method, the gradient with respect to \emph{any} number of parameters can be obtained by solving a single additional linear PDE (the adjoint PDE), and then evaluating an expression depending on this unknown. 
The adjoint PDE is obtained by considering the Lagrangian

\begin{align}
    \cL &= \obj(\vu,\prm) & \mbox{\{objective term\}} \\
    &+ \vadj^T \, \pde(\vu,\prm) &\mbox{\{physical constraint term\}}
\end{align}
and differentiating it with respect to the parameters $\prm$:
\begin{align}
    d_\prm\cL  &= \partial_\prm\obj  + \partial_\vu J \, d_\prm \vu  + \vadj^T \, \partial_\prm \pde \  + \vadj^T \, \partial_\vu \pde \, d_\prm \vu.
    \label{eq:lagrangian-diff}
\end{align}
$d_\prm\cL\,$ is expensive to compute if the 
dimension of $\prm$ is large; a direct computation involves computing $d_{q_m} \vu$ (how solution changes according to parameter $q_m$) for every optimized parameter $q_m$ in $\prm$, which means solving $\lvert q \rvert$ different linear PDEs.
Isolating all terms multiplying $d_\prm \vu$:
\begin{align}
    d_\prm\cL  &= \partial_\prm\obj  +  \vadj^T \, \partial_\prm \pde + \left( \partial_\vu J  + \vadj^T \, \partial_\vu \pde \right) \, d_\prm \vu.
    \label{eq:lagrangian-diff-deltau}
\end{align}
We can then eliminate the last term by choosing the \emph{adjoint variable} $\vadj$ such that it solves \emph{the adjoint problem}:
\begin{align}
    \vadj^T \, \partial_\vu \pde = - \partial_\vu J.
    \label{eq:adjoint-static}
\end{align}

Then, by plugging the solution $\vadj$ of the adjoint PDE into the Lagrangian, we obtain the final shape derivative:
\begin{align}
    d_\prm J  = d_\prm\cL(\vadj) &= \partial_\prm\obj + \vadj^T \, \partial_\prm \pde.
    \label{eq:static-deriv-final}
\end{align}

\paragraph{Combining contributions from different forces and objectives together} 
Our discretized equation has the form %
\[
\pde(\vu,\prm) = \sum_k \pde^k(\vu,\prm) = 
0,\]
where $\pde^k$ is a contribution from each type of force 
(elasticity forces, contact forces, etc). 
Similarly, the objective $\obj$ is a sum of contributions 
from several objective components or constraints in penalty form: 
\[
\obj(\vu,\prm) = \sum_\ell \obj^\ell(\vu,\prm).
\]
Thus, the adjoint system and the full parametric derivative 
have the following form, respectively:
\begin{equation}\label{eq:static-adjoint}
    \begin{split}
    \vadj^T\left( \sum_k  \dpu \pde^k \right)   & = -\sum_\ell \dpu \obj^\ell, \\
    \fdpar \obj   &= \sum_\ell \dpar \obj^\ell  +       \sum_k \vadj^T \dpar \pde^k.
    \end{split}
\end{equation}

Thus, for each force, we need  $\dpu \pde^k$ and $\dpar \pde^k$ and each objective component, $\dpu \obj^k$ and $\dpar \obj^\ell$.

\subsection{Dynamic case}
\label{sec:dynamicadj}

\paragraph{Discrete time-dependent Lagrangian} 
We write the time-dependent Lagrangian $\cL$ for the functional $J$ viewing  the equations for $\dot{\vv}$ and $\dot{\vu}$ as constraints
with Lagrange multipliers $\vadj$ and $\vmu$. 

Similar to the static case,  we expand the derivative $\fdpar \cL$, and isolate the terms containing $\fdpar \vu$ and $\fdpar \vv$. By setting the sum of each of these two sets of terms to zero, we obtain two adjoint equations. 

Our Lagrangian consists of three parts, corresponding to the objective ($J$), physics constraints ($\cL_c$), and initial condition constraints ($\cL_{in}$):
\[
\cL(\vu,\vv,\vadj,\vmu,\prm) = J(\vu,\prm) + \cL_c(\vu,\vv,\vadj,\vmu,\prm) + \cL_{in}
(\vu^0,\vv^0,\vadj^0,\vmu^0,\prm),
\]
where
\[
\cL_{in} = \vadj_0^T (\vv^0 - \ic^v) + \vmu_0^T(\vu^0 - \ic^u),
\]
and
\[
\small
\cL_c = \sum_{i=1}^N \lmT_i M \bigg(\vv^i + \sum_{j=1}^{\min(i,m)}\alpha^i_j\vv^{i-j} - \hht^i \bigg) + \vmu^T_i \bigg( \vu^i + \sum_{j=1}^{\min(i,m)}\alpha^i_j \vu^{i-j} - \beta_i \dt\vv^{i}\bigg).
\]
\paragraph{Adjoint equations}
As shown in the Appendix, this leads to the following adjoint equations:
\begin{equation}
\begin{aligned}
& \bigg( \vadj_i   + \sum_{j=1}^{\min(m,N-i)} \alpha^{i+j}_j \vadj_{i+j}\bigg)   = \beta_i \dt \vnu_i\\
&M^T \left( \vnu_i    + 
 \sum_{j=1}^{\min(m,N-i)}
 \alpha^{i+j}_j \vnu_{i+j}\right)
 = \\&(\dpui \hht^i)^T \vadj_i  + (\dpui \hht^{i+1})^T \vadj_{i+1} - (\dpu \htobj^{i})^T,
\end{aligned}
\label{eq:time-adjoint}
\end{equation}
where we introduce a new variable $\vnu$ satisfying $\vmu = M^T \vnu$.

Note that this system is very similar to the forward time-stepping, with the following differences: it proceeds backward, from $\vnu_{i+1}$ to $\vnu_i$; there is a single \emph{linear} solve per time step, rather 
than a nonlinear solve as for the forward system; for higher-order time stepping the first few steps in the forward system are lower-order BDF steps; however, this is \emph{not} the case for the adjoint system: to maintain consistency, we derive the initial low-order steps from the forward system. If BDF2 is used for the forward time-stepping, the resulting scheme is different from the standard BDF2 scheme used in  \cite{Hahn2019} for the adjoint system. If the system were discretized inconsistently as in \cite{Hahn2019}, a sufficiently small time step is needed to maintain accuracy of the gradient that would ensure that the discrete energy decreases along the gradient direction.

By introducing $\vadj_{N+1},\ \vnu_{N+1}$, the initial condition can be simplified as
\begin{equation}
\begin{aligned}
&\vadj_{N+1} = 0,\\
&\vnu_{N+1} = 0.
\end{aligned}
\label{eq:adjoint-init}
\end{equation}

The first (last in the adjoint solve) values need to be treated separately, as shown in Appendix \ref{appsec:adjoint-bdf}:%
\begin{equation}
\vmu_0 = -(\dpu \htobj^{0})^T - \sum_{j=1}^m \alpha^{j}_j   \vmu_{j} + \vadj_1^T \dpuz \hht^{1},\;
\vadj_0 =  -\sum_{j=1}^m \alpha^{j}_j M^T \vadj_{j}. 
\label{eq:zero-adjoint}
\end{equation}

\paragraph{Computing the derivative of $J$ from the forward and adjoint solutions}  From the adjoint variables,  we can compute
$\fdpar J = \fdpar\cL$:

\begin{equation}
\begin{aligned}
\fdpar J &= -\vadj_0^T \dpp \ic^v - \vmu_0^T\dpp \ic^u\\
&+ \sum_{i=0}^{N}  \dpp \htobj^{i} \\
&+ \sum_{i=1}^{N} -\lmT_i \dpp \hht^i 
+ \beta_i \dt \vnu_i^T \fdpar M
\vv^i \\
&+\sum_{j=1}^m
\alpha^{j}_j \lmT_{j} \fdpar M \vv^0.
 \end{aligned}
 \label{eq:deriv-from-adjoint}
\end{equation}

Partial derivatives $\dpp \hht, \dpu \hht$ and $\dpp \htobj_i$, $\dpu \htobj_i$
are exactly the same as used in the construction of the system for 
static adjoint and computation of the functional. The differences, specific 
to time discretization, are:  
\begin{itemize}
\item Mass matrix derivative $\fdpar M$. See Appendix (section \ref{appsec:mass-matrix-deriv}).%
\item Partial derivatives  of the initial conditions with respect to parameters  $\dpp \ic^v$ and $\dpp \ic^u$, for positions and velocities. See Section \ref{appsec:initcond-derivatives} in the Appendix. %
Typically, a 3D position and velocity for the whole object (or angular velocity for the object rotating as a rigid body) are used as parameters, so these are trivial to compute. 
\end{itemize}

\subsection{Summary of the parametric gradient computation}  
\label{sec:summary-grad}
Computing the derivative $\fdpar J$ requires the following 
components
\begin{itemize}
    \item Derivatives $\dpu J_i$,  $\dpu \pde_i$, $\dpp J_i$ and 
    $\dpp \pde_i$ for each time step $i$.
   See Sections \ref{sec:general-forces} to \ref{sec:specific-objectives} for corresponding formulas.
    \item For the dynamic problems, $\dpp \ic^u$ and $\dpp \ic^v$, derivatives of the initial conditions. See Section~\ref{appsec:initcond-derivatives} in the Appendix. %
\end{itemize}

To compute the parametric derivative of $J$, the steps are as follows: 
\begin{enumerate}[1.]
    \item Solve the forward system \eqref{eq:discrete-static} or \eqref{eq:discrete-dynamic}, and store 
    the resulting solutions $\vu$ for the static problem; for the dynamic problem, we store $\vu^i, \vv^i$, $i=0\ldots N$ at every step. 
    \item Initialize adjoint variables $\vadj_{N+1},\vnu_{N+1}$ as shown in  
    \eqref{eq:adjoint-init}. 
    \item  For the static problem, solve the adjoint system \eqref{eq:adjoint-static}. For the dynamic problem, perform backward time stepping using  %
     ~\eqref{eq:time-adjoint}. 
    \item At every step of the dynamic solve, evaluate derivative of the 
    mass matrix $\fdpar M$, if applicable, and use formulas 
    \eqref{eq:deriv-from-adjoint} to update $\fdpar J$.
\end{enumerate}

\section{Optimization algorithm}
\label{sec:method}

We provide a high-level summary of our method in Algorithm \ref{alg:the_alg}, its major components are:
\begin{itemize}
     \item \textsc{ForwardSolve} solves the nonlinear elasticity system, retaining all solution steps for time-dependent problems;
     \item \textsc{Objective} computes the objective function given the solution and parameters;
     \item \textsc{AdjointSolve} solves the adjoint system \eqref{eq:time-adjoint} stepping backward in time and using the solutions of the forward problem; 
     \item \textsc{DiscreteDerivative} computes gradients given displacements and adjoint variables;
     \item \textsc{LineSearch} is the standard Wolfe-Armijo line search, with additional prevention of element inversion and contact~\cite{Li2020IPC};
     \item \textsc{Remesh} performs remeshing of $\Omegar$ and $\Omega_\prm$ to improve the mesh quality before restarting optimization;
     \item  \textsc{Converged} is the outer iteration stopping criterion.
\end{itemize}

We omit the pseudo-code for the forward solve as it closely follows that of~\cite{Li2020IPC} with only a few notable changes: (1) we use an area weighting inside the barrier potential for convergence (see Section~\ref{sec:contact}), (2) we use a fixed barrier stiffness $\kappa$ as changing it adaptively throughout the simulation would require computing its gradient through the update, and (3) to speed up convergence, we only project the Hessian to positive semi-definite in the Newton update if the unprojected direction is not a descent direction. 

The inner loop works on a fixed mesh for $\Omegar$, and is close to the standard L-BFGS algorithm with two additional features, essential for handling shape derivatives and large deformations: (1) we check for any inversions of tetrahedra and contacts resulting from changes to the shape of the domain $\Omega_\prms$ as a result of changing shape parameters and (2) after each update of the boundary vertices, we call the \textsc{SLIM} smoothing algorithm \cite{rabinovich2017scalable}, with boundary vertices $p$ fixed, to move the interior vertices to improve mesh quality.

Unlike previous work we support remeshing. If the mesh quality $Q$ is smaller than a tolerance $\delta_\text{remesh}$, the domain is remeshed. If the gradient w.r.t. $\prm$ is smaller than a tolerance $\delta_\text{grad}$ or the step size is smaller than a tolerance $\delta_x$, the optimization is stopped.

\begin{algorithm}[htb!]
    \begin{algorithmic}
    \caption{Optimization algorithm overview}
    \label{alg:the_alg}
    \Function{Gradient}{$\prm$}
         \State $\vu \gets$ \Call{ForwardSolve}{$\prm$}
         \State $\vadj \gets$ \Call{AdjointSolve}{\mbox{\textsc{Objective}}, $\vu$, $\prm$}
         \State $\vg \gets$  \Call{DiscreteDerivative}{\mbox{\textsc{Objective}}, $\vu$, $\vadj$, $\prm$}
         \State \Return $\vg$
    \EndFunction
    \Statex
       \Function{ParameterOptimization}{}
       \State $\prm \gets$ initial parameter values
           \State $oi \gets 0$ \Comment{Optimization iteration count}
           \Repeat
            
            \State $\vg \gets$ \Call{Gradient}{$\prm$}
            \State $\vd \gets$ \Call{LBFGSDirection}{$\vg$, $\prm$}
            \State $s \gets$ \Call{LineSearch}{$\vd$}
            \State $\prm \gets \prm + s\vd$
            \If{$Q < \delta_\text{remesh}$}
            \State $\prm \gets$ \Call{Remesh}{$\Omega_\prm$}
            \EndIf
            \State $oi \gets oi + 1$
            \Until{$oi = oi_{\max}$ or $\|\vg\| < \delta_\text{grad}$ or $\|s\vd\| < \delta_{x}$}
  \EndFunction
    \end{algorithmic}
\end{algorithm}

\section{Physical model and discretization}
\label{sec:physicalmodel}
In this section, we summarize the physical model we use. The model is similar to the one used in \cite{Li2020IPC}, with some minor modifications to the friction and contact formulation (Section~\ref{sec:contact}), most significantly, addition of damping.

To discretized the model we use arbitrary-order Lagrangian elements and 
arbitrary-order BDF time stepping (our experiments are with schemes of order 1 and 2). 

The forces, which contribute to the PDE and need to be included in the adjoint equations and corresponding parametric gradient terms are:
\begin{itemize}
\item geometrically non-linear elasticity (with linear and Neo-Hookean constitutive laws as options);
\item contact forces in smoothed IPC formulation; 
\item friction forces also in smoothed IPC formulation; 
\item strain-rate proportional viscous damping for elastic objects;
\item external forces such as gravity or surface loads. 
\end{itemize}

The right-hand side of the system of equations we solve on $i$-th time step 
of \eqref{eq:time-adjoint} can be written as
\[ 
\begin{split}
    &\pde^e(\vu^i;\lambda(x),\mu(x)) + 
    \pde^c(\vu^i) + \pde^f(\vu^i,\vu^{i-1};\mu(x,y))\\
    & + \pde^d(\vu^i,\vu^{i-1};\alpha(x),\beta(x)), %
\end{split}
\]
where $\pde^e$ is the discrete elastic PDE term, $\pde^c$ and $\pde^f$ define contact and friction forces, and $\pde^d$ defines damping.
 In greater detail, all these forces are defined in the next section, along with $\dpu \pde$ and $\dpar \pde$ for each one of them.

The physical parameters $\prm$ of the model with respect to which it can be differentiated include:
\begin{itemize}
    \item (possibly spatially variant) 
Lame coefficients for elasticity $\lambda(x), \mu(x)$;
    \item friction coefficient between pairs of points $\friccoeff(x,y)$ (we consider it fixed for each pair of objects, to reduce the number of variables involved); 
    \item damping coefficients $\alpha(x),\beta(x)$.   
\end{itemize}

\begin{figure}
    \centering
    \includegraphics[width=\columnwidth]{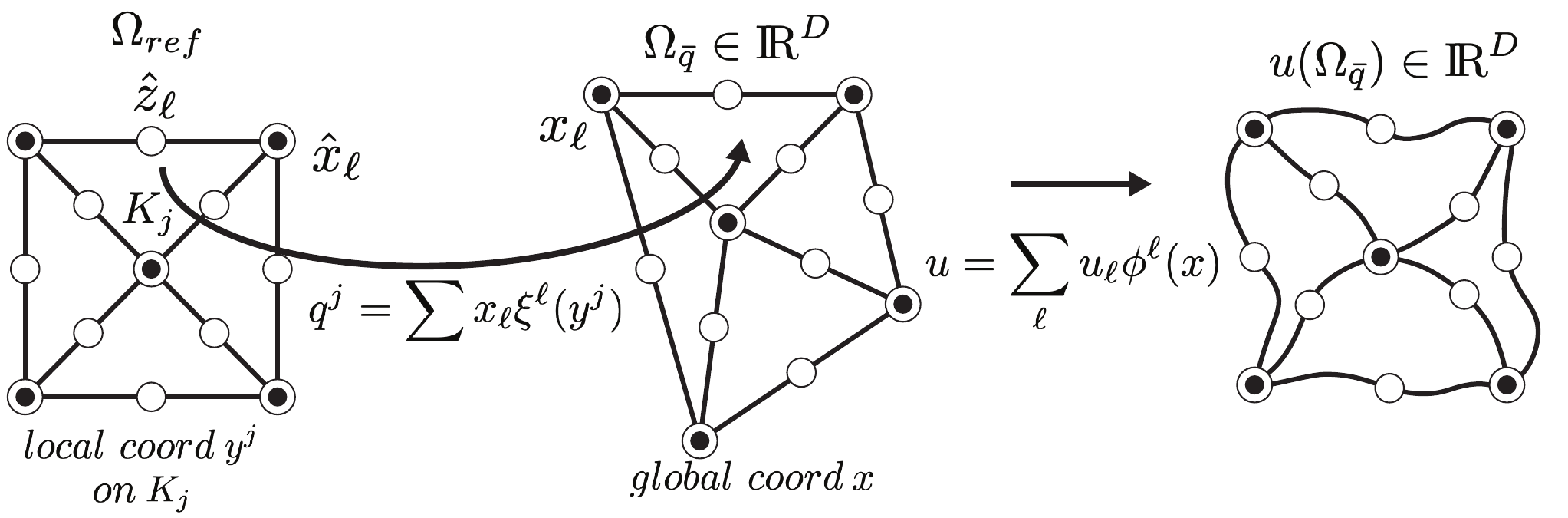}
    \caption{Notation for domains and maps we use, see Table~\protect\ref{tab:notation}.}
    \label{fig:notation_maps}
    \Description{}
\end{figure}

\paragraph{Domains} A critical aspect of the formulation at the foundation of our solver is the distinction between 
\emph{reference domain} $\Omegar$,  and (undeformed) \emph{physical domain} $\Omega_\prms$, where $\prms$ denotes parameters defining the shape (Figure~\ref{fig:notation_maps}). The physics equations $\fh(u,x,t,q)$ and the solution $u(x,t)$ is defined on $\Omega_\prms$ most naturally, but this domain may be changed by optimization. 
The optimization parameters $q$ are defined on $\Omegar$.  This distinction is present in previous work on shape optimization (e.g., \cite{Tozoni2021}) but not in the more general setting of dynamic differentiable simulation. 

\begin{figure}
    \centering
    \includegraphics[width=0.5\columnwidth]{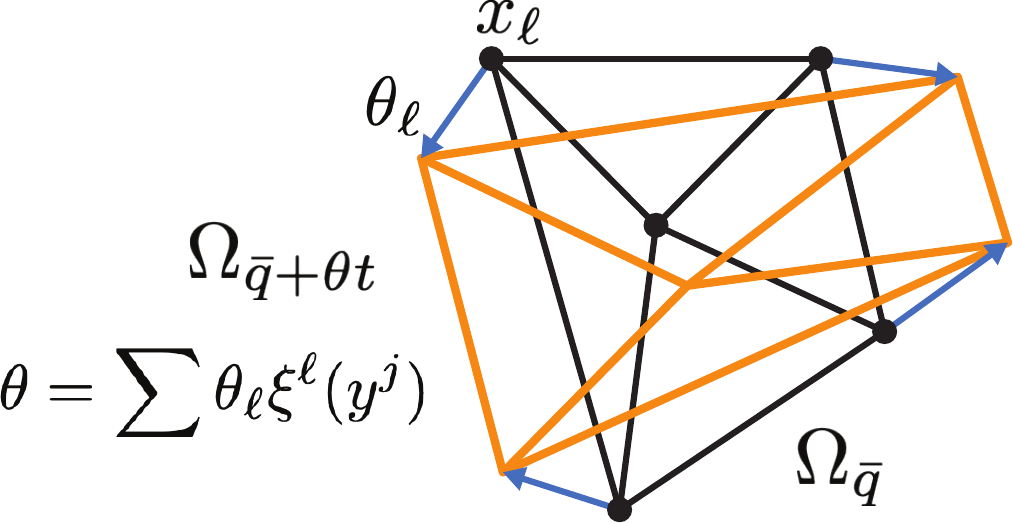}
    \caption{Domain perturbation $\theta$, see Table~\protect\ref{tab:notation}.}
    \label{fig:notation_deform}
    \Description{}
\end{figure}

\section{Example: Poisson equation}
\label{sec:poisson}
To explain the principles of how individual derivatives for forces and target functionals are computed, we use a simple example. For more complex forces in our problem formulation, we state the final result in this paper, and we refer to the Appendix for the derivation.

Consider a variable-coefficient Poisson equation $\nabla \cdot (c(x) \nabla u) = f$ and zero Neumann boundary conditions on a domain $\Omega_\fprms$ that can be changed by the optimization. We take as the optimization objective the squared gradient of the solution on the domain. Then 
\begin{itemize}
\item the optimization parameters are $q = [ \fprms, c]$;
\item The PDE in weak form is \[\fh(u,q,w) = \fh(u,\fprms,c,w) =  \int_{\Omega_\fprms} c \nabla u \nabla w  - f w\, dx.\]
\item The objective is 
 \[J(u,\fprms) = \int_{\Omega_\fprms} \| \nabla u\|^2 dx.\]
\end{itemize}

Discretizing in FE basis, with basis functions $\ubasref^\ell$  (e.g., quadratic) used for $u = \sum_\ell u_\ell \ubasref^\ell $ and $c = \sum_\ell c_\ell \ubasref^\ell$,
and basis $\vbasref^\ell$ used for the geometric map $\prms = \sum_\ell x_\ell \vbasref^\ell$, we obtain the following.
 (Note that both our basis $\vbasref$ and $\ubasref$ are
 defined on the fixed triangulated domain $\Omegar$.) 

\begin{itemize}
    \item $\prm = [\prms, \mathbf{c}] = [x_1\ldots x_{n_N}, c_1 \ldots c_{n_N} ]$, where $x_\ell \in \mathbb{R}^2$ are vertices of the physical domain $\Omega_{\prms}$, which we optimize, and $c_\ell$ are the coefficients of $c$ in FE basis. 
    \item The PDE discretization is performed on the physical domain $\Omega_\prms$, and has the form
     $\pde(\vu,\prm) = S(\prm) \vu - M(\prm) \vf$. 
   The entry $(m,\ell)$ of the matrix $S(\prm)$ are obtained by substituting $\u=\phi^m\circ g^{-1}$ and $w=\phi^\ell\circ g^{-1}$, and the discrete expression for $c$ into the expression below; entries of $M(\prm)$ are obtained in a similar way  
    \begin{equation} S(u,w) =\int_{\Omega_\prms} c(\prms^{-1}(x)) \nabla u\cdot  \nabla w \, dx;\quad M(v,w) = \int_{\Omega_\prms} v w \, dx.
        \label{eq:poisson-form}
    \end{equation}  
    \item  The discrete objective is $J(\vu,\prm) = \vu^T T(\prm) \vu $, with entries of $T(\prm)$ also obtained by substituting pairs of basis functions 
    into the bilinar form
    \begin{equation}
        T(u,w)  = \int_{\Omega_\prms} \nabla u \cdot \nabla w \, dx.
        \label{eq:dirichlet-form}
    \end{equation}
\end{itemize}

Computing derivatives of $S$, $\partial_{c_k} S(u,w)$, %
with respect to $c$ is straightforward, as the dependence on the coefficients of $c$ is linear.  Computation 
of shape derivatives is more complex, as the integration domain and the gradient operator $\nabla$ with respect to physical domain variables are affected by the change of shape parameters.  

\paragraph{Direct approach}
The direct approach is to perform a change of variables in \eqref{eq:poisson-form} and \eqref{eq:dirichlet-form} and to the domain $\Omegar$, and differentiate with respect to $x_\ell$; e.g.,\eqref{eq:poisson-form} becomes
\[
    S(u,w) = \int_{\Omegar} c(y) \, \nabla_y \hat{u} ^T (\nabla_y \fprms )^{-1} (\nabla_y \fprms)^{-T} \nabla_y \hat{w} \,\, \det \nabla \fprms \,\, dy,
\]
where $\hat{u}$ and $\hat{w}$ denote compositions $u \circ \fprms^{-1}$.   These expressions are highly nonlinear in $x_\ell$ and the final expressions for 
$\partial_{x_\ell} S(u,w)$ needed for $\partial_{\prms} \pde$ are unwieldy, especially for more complex forces like nonlinear elasticity and friction.

\paragraph{Shape derivative approach}
Instead, we use \emph{shape derivative calculus} commonly used in shape optimization to obtain the derivatives with respect to the shape parameters directly on the physical domain $\Omega_\prms$ (for the parameters not affecting domain shape the  approaches using $\Omegar$ and $\Omega_\prms$ are identical). 

To compute $\dpar h$, or $\dpar \obj$, we consider the perturbed domain  $\Omega_{\prms+\vper \epsilon}$, where $\vper$ is a 
vector field, and compute the full derivative as limit of 
\[\frac{1}{\epsilon}\left(h(u_{\prms+\vper \epsilon}, \prms+\vper \epsilon)
- h(u_{\prms}, \prms) \right),\]
as $\epsilon \rightarrow 0$.
In the resulting expression, the terms not containing the change $\delta u(x)$ correspond to $\dpar \obj \vper$, and the 
terms containing derivatives of $u(x)$ are transformed  to $\dpu \obj \vpsi $ by substituting  $\psi$ instead of $\delta u(x)$.

\section{Parametric derivatives of forces}
\label{sec:general-forces}

In this section, we derive expressions for $\dpu\pde$ and $\dpar\pde$ for specific forces needed for the adjoint equations and the final derivative formula respectively. 

For each force, we obtain expressions of the forms $B$ and $A$ below, 
from which the matrices for corresponding derivatives can be obtained using:
\begin{equation}
B^k(\adj,\per) = \vadj^T \dpar \pde^k\, \vper,\;
A^k(\adj,\psi) = \vadj^T \dpu \pde^k \vpsi,
\label{eq:BA-def}
\end{equation}
with $\per$ going over basis vectors for this parameter type,
$\adj$ going over adjoint variable components, and 
$\psi$ over the test function basis vectors for the adjoint; i.e., two matrices  of size $D_s n_N^z\times D_s n_N^z$.

While nonlinear elasticity derivatives with respect to material parameters and initial conditions were used in \cite{geilinger2020add} and \cite{Hahn2019}, and static-problem shape derivatives for a different (static, allowing interpenetration) contact and friction model were obtained in \cite{Tozoni2021}, we present expressions for all force-related derivatives with respect to all parameters (material, shape, initial conditions) in a unified way, simplifying adding additional forces, building whenever possible on a general form described in Section~\ref{sec:general-volume}. 

\subsection{Volume forces}
\label{sec:general-volume}
Many forces in continuum mechanics  have the general weak form
\begin{equation}
\fh^{v}(u,w,\fprm) = \int_{\Omega_\prms} f^v( \nabla u,\fprm ): \nabla w\, dx,
\label{eq:vol-force}
\end{equation}
where $u$ is the displacement vector, with the components of the vector $\pde^v(\vu)$ obtained as $\fh^v(u,\phi^\ell,\fprm)$, for all basis functions $\phi^\ell$, and the column denotes tensor contraction. 
In our case, elastic forces, irrespective of the constitutive law used, belongs to this category.  

In these expressions $f^v(\nabla u, \fprm)$ is a tensor of
dimension $D_d \times D_s$; e.g., for elasticity, $D_d = D_s$, and this expression  is the stress tensor, as a function of $\nabla u$. %

If the force is associated with a volume  energy density $W^v(\nabla u,\fprm)$, associated forces have the form above, 
specifically,  $f^v(\nabla u,\fprm) = \nabla_1 W^v$. (Here, $\nabla_1$ means the gradient with respect to the first parameter, which in this case is $\nabla u$).  For a surface energy density $W^s(u,\fprm)$, the formulas are similar, but the integrals are over the surface. 

We also formulate damping forces in a similar way, as explained in more detail below, except at each timestep  $W^v$ depends on displacements  $u^i$ and $u^{i-j}$, $j=1\ldots m$ at the current and $m$ previous steps, where $m$ is the order of approximation of velocity used in damping (we use $m=1$). 
The formulas for $A^{v}$ and $B^{v}$ in this case are obtained 
in exactly the same way as for the dependence on $u^i$ only, separately 
for  $u^i$ and $u^{i-1}$, corresponding to $\partial_{\vu_i}\pde^i$ and $\partial_{\vu_{i-1}}\pde^i$ respectively.

To obtain matrices $A^v$ and $B^v$ corresponding to  $\dpu \pde^v $ and $\dpar \pde^v $ 
\eqref{eq:BA-def},  we split  $\dpar \pde^f $ into   $\partial_{\prms} \pde^f$ and $\partial_{\prm^1} \pde^f $, the shape and non-shape parameter derivatives, assuming  $f$ depends on a single volume   vector of parameters $\fprm = \prm^1$ (e.g., Lame constants). 
We treat these two types of parameters separately, as $\prms$ affects the domain of integration but not the integrand, and conversely, $\fprm$ affects the integrand but not the domain. 

\paragraph{Shape derivatives}
For the shape derivative contribution, we obtain the following forms (the derivation and explicit form of matrix entries can be found in supplementary material). 
\begin{equation}
\begin{split}
B^v(\per,p) = &\int_{\Omega_\prms} - f(\nabla u) \nabla \per^T : \nabla p
\\&- (\nabla_1 f(\nabla u) : (\nabla u \nabla \per)) : \nabla p + \left( f(\nabla u) : \nabla p \,  \right) \nabla \cdot \per\, \, dx,
\end{split}
\label{eq:Bv}
\end{equation}
$B^v(\per,p)$ is linear in $\per$ and $p$, and we convert it to 
a matrix form by substituting basis functions for $\per$ and $p$.

 The contribution to the left-hand side of the adjoint equation is
 \begin{equation}
A^v(\psi,p) = \int_{\Omega_\prms} (\nabla_1 f(\nabla u ) : \nabla \psi) : \nabla p \, dx.
\label{eq:Av} 
\end{equation}
 Observe that the matrix is identical to the matrix used in the forward solve.

 \paragraph{Non-shape volumetric parameter derivatives} 
We assume that the force depends on $\fprm = \fprm(x)$, a function of the point in $\Omega_\prms$, defined by its values $\prm$ at the same nodes as the solution, and interpolated using the same basis $\phi$.

In this case, the  form $B$ is: 
\[
B^v(\per,p) = \int_{\Omega} (\pd_{\fprm} f \cdot \per) : \nabla p \,\, dx.
\]

 The contribution to the left-hand side of the adjoint equation is identical to the shape derivative case.

In our implementation we consider two versions of elastic forces, both 
defined by Lame parameters specified as functions on $\Omegar$: $q(x) = [\lambda(x), \mu(x)]$. The only quantities we need are derivatives of $f(\nabla u)$ with respect to 
$\nabla u$, and material parameters. 

\paragraph{Linear elasticity} 
For linear elasticity, we replace $f^v$ with 
\[f^{e}(\nabla u, \prm) =\stress(\nabla u,\prm)  = C(\prm) : \varepsilon(\nabla u) = \frac{1}{2} \, C(\prm) : (\nabla u^T + \nabla u),\] with $C_{ijkl}(\lambda, \mu) = \lambda \delta_{ij} \delta_{kl} + \mu (\delta_{ik} \delta_{jl} + \delta_{il} \delta_{jk})$.

For computing $A^e$ and $B^e$ we  use partial derivatives of 
$f^e$ with respect to material parameters:
\begin{align*}
    \nabla_1 f^e(\nabla u, \boldsymbol{\lambda}, \boldsymbol{\mu}) &= C, \\
    \partial_{\lambda} f^e(\nabla u, \boldsymbol{\lambda}, \boldsymbol{\mu})_{ij} &= \delta_{ij} \delta_{kl} \varepsilon_{kl},\\
    \partial_{\mu} f^e(\nabla u, \boldsymbol{\lambda}, \boldsymbol{\mu})_{ij} &= (\delta_{ik} \delta_{jl} + \delta_{il} \delta_{jk}) \varepsilon_{kl}.
\end{align*}

\paragraph{Neo-Hookean elasticity} 
For Neo-Hookean elasticity, the following formula is used for computing stress from the deformation gradient: 
\begin{align*}
    f^e(\nabla u, \prm) = \mu (F(\nabla u) - Q(\nabla u)) + \lambda \log(\det(F(\nabla u))) \,Q(\nabla u),
\end{align*}
where $F(\nabla u) = \nabla u + I$ and $Q(\nabla u) = F(\nabla u)^{-T}$.

We can then compute derivatives of $f(\nabla u)$:
\begin{align*}
    \nabla_1 f^e(\nabla u, \prm)_{ijkl} &=
    \mu (\delta_{ik} \delta_{jl} + Q_{il} Q_{kj}) \\
    &\quad+ \lambda(Q_{ij} Q_{kl} - \log(\det(F)) \, Q_{il}Q_{kj}), \\
    \partial_{\lambda} f^e(\nabla u, \prm) &= F(\nabla u) - Q(\nabla u), \\
    \partial_{\mu} f^e(\nabla u, \prm) &= \log(\det(F(\nabla u))) \,Q(\nabla u).
\end{align*}

\paragraph{Damping} %
For damping, we have material parameters controlling shear and bulk damping $\alpha,\beta$.
We use the strain-rate proportional damping described in ~\cite{damping2018}. Given deformation gradient $F=\nabla u + I$, the Green strain tensor $E=\frac{1}{2}(F^TF-I)$ is rotation-invariant. The viscous Piola-Kirchhoff stress is of the form
\[
P(\nabla u,\nabla \dot{u})=F(2\alpha \dot{E} +  \beta \Tr(\dot{E}) I),
\]
where $\Dot{E}$ denotes the time derivative, and the weak form of 
the corresponding force 
\[
\fh^d(u,\Dot{u},w) = \int_{\Omega_\prms} P(\nabla u,\nabla\Dot{u}) \nabla w dx.
\]

In our case, to fit this force into our differentiable formulation, we discretize $\Dot{F}$ using as $\Dot{F}^i=\frac{1}{\dt}(F^i-F^{i-1})$;
this yields a force expression of the form 

\[
\fh^d(u^i,u^{i-1},w) = \int_{\Omega_\prms} P(\nabla u^i,\nabla u^{i-1}) \nabla w dx,
\]
which is identical to \eqref{eq:vol-force}, except it depends on both 
$\nabla u^i$ and $\nabla u^{i-1}$. 
As a consequence, expressions for $A^d(\psi,p)$ and $B^d(\theta,p)$  are obtained in the same way as in \eqref{eq:Av} and \eqref{eq:Bv}, 
except two pairs of matrices are obtained, one for $\nabla u^i$ the 
other for $\nabla u^{i-1}$, using  $\nabla_1 P$ and $\nabla_2 P$ as $\nabla_1 f$ respectively.  

\subsection{Contact and Friction}
\label{sec:contact}
For the contact forces, we use a slightly modified version of the formulation of \cite{Li2020IPC}.  While the original formulation is introduced in a discrete form, it can be derived with minimal changes as a linear finite-element discretization of a continuum formulation \cite{Li:2022:ConvergentIPC}.
 The contact incremental potential uses log barrier function $b(y)$, where $b$ is a truncated log barrier function,  approaching infinity, if $y \rightarrow 0$, and vanishing for
 $y \geq \hat{d}$ for some small distance $\hat{d}$. 
 
 For any pair $k$ of primitives (vertices, edges, and faces) of the surface mesh $\partial\Omega_{\vx^d}$, defined by the vertex positions $\vx^d=M^*\prms+\vu$,  $d_k(\mathbf{x^d})$ denotes the distance between them;  $C$ is the set of primitive pairs in contact, i.e., pairs of primitives with $d_k < \hat{d}$.
 
 Recall that  the geometric map $\fprms$  always uses  piecewise-linear elements $\xi_\ell$, while the basis for the deformations $u$ can be of any order.  The matrix $M^*$ is an upsampling matrix to bring dimension of $\prms$ to the same as discrete solution $\vu$. The upsampling is performed by linear interpolation from $\hat{x}_\ell$ to nodes $\hat{z}_\ell$.

The contact forces are derived from the following potential:
\[
E(\vu, \prms) =\kappa \sum_{k\in C} b(d_k(\vx^d) ) A_k = \sum_{k \in C} W_k(\vu,\prms) A_k,
\]
where $\kappa > 0$ is a parameter controlling the barrier stiffness and $A_k$ corresponds to the sum of surface areas associated with each  primitive in $k$ (i.e., $\tfrac{1}{3}$ of the sum of areas of incident triangles for vertices and edges, and the area for triangles). See Section \ref{appsec:contact-area} in the Appendix.

We define $F^c_k(\vu, \prms) = \dpu W_k(\vu, \prms) = \kappa b'(d_k(\vx^d)) \partial_{\mathbf{x^d}} d_k$.

The contact force is given by 
\begin{align*}
    \pde^{c} =
    \sum_{k \in C} F^c_k(\vu, \prms)  A_k.\\
\end{align*}
The terms $B^c$ and $A^c$ have the form 
\begin{align*}
    B^c(p,\theta) &= \sum_k 
    \left( \dparms F^c_k \per \cdot p + F^c_k \cdot p \,\, \dparms A_k  \right)  A_k,\\
    A^c(p,\psi) &= \sum_k \dpu F^c_k \psi \cdot p \,\, A_k,
\end{align*}
where
\begin{align*}
    \dpu F^c_k &= \kappa \left( b'' (\partial_{\mathbf{x^d}} d_k) (\partial_{\mathbf{x^d}} d_k)^T + b' \partial_{\mathbf{x^d}}(\partial_{\mathbf{x^d}} d_k) \right),\\
    \dparms F^c_k &=  \dpu F^c_k  M^*
\end{align*}
and $\dparms A_k$ corresponds to the gradient of the area term, which varies depending on the type of primitive pairs corresponding to $k$. See Section \ref{appsec:contact-area} in the Appendix.

\paragraph{Friction}
In general, the friction coefficient $\friccoeff(x_1,x_2)$ 
is a function of pairs of surface material points in  $\partial\Omega_\prms$. 
As a simplification, in our implementation, we assume that each pair of objects $(m, n)$, in the simulation has a single coefficient $\friccoeff_{m,n}$, which can vary through the optimization.  To simplify notation, we use $\friccoeff_{k_1,k_2}$ for a pair of primitives $k_1$ and $k_2$ to indicate the friction coefficient between objects these primitives belong to. 

We follow the IPC definition of friction \cite{Li2020IPC}.  Its key feature  is that it is a differentiable function of displacements, which determine the contact forces,  and relative velocities, which, for dynamic problems, we discretize using  first-order approximation $u^{i}-u^{i-1}$, where $i$ is the time step.

The friction force for each active pair of primitives $k$ is
\begin{align}
  F^f_k(\vu^{i-1},\vu^i) = -\friccoeff_{k_1,k_2} N_k T_k f_\eta(\lVert \tau_k \rVert) \frac{\tau_k}{\lVert \tau_k \rVert},
\end{align}
where $N_k$ is the contact force magnitude, $T_k$ is a tangential frame matrix, constructed  as described in \cite{Li2020IPC}, and
$\tau_k$ and $f_\eta$ are defined as
\begin{align*}
    \tau_k &= T_k(\vx^{d,i-1} )^T (\vu^i - \vu^{i-1}), \\
    f_\eta(y) &= \begin{cases} 
      -\frac{y^2}{\eta^2} + \frac{2y}{\eta} & y \in [0, \eta) \\
      1  & y \geq \eta
     \end{cases}.
\end{align*}

The total friction force has the form
\begin{align*}
    \pde^f
  =\sum_{k \in C} F^f_k(\vu^i,\vu^{i-1}, \prms) \, A_k,
\end{align*}

with the form $B$ for shape derivatives given by 
\[
B^f(\adj,\per) = \sum_k \dparms F^f_k \per \cdot p \, A_k + F^f_k \cdot p \,\,  \dparms A_k  \,\, A_k.\]
Additional details on the computation of $\dparms F^f_k$ are in the Appendix (Section \ref{appsec:friction-derivative}).
The derivative with respect to friction coefficient values is easily obtained as the force is linear in friction coefficients.
If $\prm$  is a vector of friction coefficients,
\begin{align*}
    \partial_{q_\ell} F^f_k 
    = \begin{cases}
      -N_k T_k f_\eta(\lVert \tau_k \rVert) \frac{\tau_k}{\lVert \tau_k \rVert} & \text{if $q_\ell$ corresponds to $\friccoeff_{k_1,k_2}$}\\ 
      0  & \text{otherwise}.\\ 
  \end{cases}
\end{align*}

Two forms  $A^f$, for $\partial_{\vu_i}$ and 
$\partial_{\vu_{i-1}}$ are needed for the adjoint equation. Both have the general form 
\[
A^f(\psi,p)= \sum_k \dpu F_k \psi \cdot p \, \, A_k,
\]
which reduces to computing the derivative of each $F_k$ term with respect to $u^i$ and $u^{i-1}$, which can be be found in Appendix \ref{appsec:friction-derivative}.

\section{Objective derivatives} 
\label{sec:all-objectives}

In this section, we define the $\dpar\obj$ and $\dpu\obj$ terms needed for the gradient computation \eqref{eq:static-deriv-final}:
For each objective-optimization parameter pair, 
$\dpar \obj^\ell \, \vper$ and  $\dpu \obj^\ell \vpsi$, i.e., two  vectors of size $D_s n_N^z$.

Similar to Section~\ref{sec:general-forces}, we present all objective derivatives with respect to all types of optimization parameters, including shape in a unified way. We consider a comprehensive set of objectives used in many previous works, that can be easily extended with additional ones. In Section~\ref{sec:objectives}
we present general forms that all objectives can be reduced to.

\subsection{General forms of objectives}
\label{sec:objectives}

Typically, objectives do not depend directly on the optimization parameters other than shape, so we focus primarily on derivatives of objectives with respect to shape parameters $\prms$ and solution $\vu$.

We consider objectives of the form 
\begin{equation}
    J(\vu,\prms) = J(J_1(\vu,\prms),\ldots J_{n_J}(\vu,\prms)),
    \label{sec:general-objective}
\end{equation}
where $J$ is a differentiable function, and $J_i$, $i=1\ldots n_J$
are \emph{objective terms} each of which typically has one of the integral forms described below.  $J$ can be as simple $J(J_1) = J_1$, 
or can depend on several terms, as e.g., the center of mass optimization. 
The derivatives of objective are reduced to the the derivatives of the objective terms by a direct application of a chain rule, so we focus on these. 

We first consider two general forms of objective terms which will be used for a number of specific objectives in Section~\ref{sec:specific-objectives}.
This includes inequality constraints in penalty form.

For each  objective term $J^o$, 
we obtain vectors $R^o(\psi)$ and $S^o(\per)$ corresponding to the partial derivatives $\dpu\obj^o$ and $\dpar\obj^o$, which are necessary to compute the adjoint solution and the full shape derivative.
As for the derivatives of the objective vectors $\dpu\obj^o$ and $\dpar\obj^o$ are obtained by 
plugging in the basis functions $\phi_\ell$ int $R^o$ and
$S^o$.

\paragraph{Objectives depending on gradient of solution and shape}

Consider an objective term that depends on both the solution of the PDE and 
the domain:

\begin{align}
    J^o(\nabla u, \fprms) = \int_{\Omega_\prms} j(\nabla u, x) dx.
    \label{eq:objective-gradient}
\end{align}

In this case, as derived in the supplementary document, 
\begin{align}
    S^o(\per) = \int_{\Omega_\prms} - \nabla_1 j : \nabla u\, \nabla \per + \nabla_2 j\cdot \per + j \nabla \cdot \per \,\, dx
\end{align}
and
\begin{align}
    R^o(\psi) = \int_{\Omega_\prms} \nabla_1 j : \nabla \psi \,\, dx.
\end{align}

\paragraph{Objective terms depending on solution and shape}
We also use objective terms depending on both the solution of the PDE and the domain:
\begin{align}
    J^o(u, \fprms) = \int_{\Omega_\prms} j(u, x) dx.
    \label{eq:objective-solution}
\end{align}
In this case, 
\begin{align}
    S^o(\per) = \int_{\Omega_\prms} \nabla_2 j \cdot \per + j \nabla \cdot \per \,\, dx
\end{align}
and
\begin{align}
    R^o(\psi) = \int_{\Omega_\prms} \nabla_1 j \cdot \psi \,\, dx.
\end{align}

\subsection{Specific objectives} 
\label{sec:specific-objectives}

\newcommand{\xd}{x^d}
\paragraph{$L_p$ norm of stress} 
For $p=2$ this objective measures the overall average stress, and for high $p$,  $L_p$-norm of stress approximates maximal stress:
\begin{align}
J^\sigma = \left(\int_{\Omega_\prms} \|\sigma(\nabla u)\|_F^p dx\right)^{1/p},
\label{eq:obj:stress}
\end{align}
where $\sigma(\nabla u)=f(\nabla u)$ represents stress, which depends on $\nabla u$.
Following the chain rule,
this objective is a function of a single objective term
 $J^{\sigma} = (J_1^\sigma)^p$ which is of the form~\eqref{eq:objective-gradient}. with $j =\|\sigma(\nabla u)\|_F^p$ for which $\nabla_2 j = 0$, and
\begin{align*}
    \nabla_1 j 
    &= p \, \|\sigma\|^{p-2} \,\, \sigma \, :\, \nabla f(\nabla u).
\end{align*}

\paragraph{Weighted difference from target deformations} 
\begin{align}
J^{trj}(x,u) = \int_{\Omega_\prms} w(\fprms^{-1}(x)) \,\, \|\xd - x^{trg}(\fprms^{-1}(x))\|^2 \,\, dx
\label{eq:obj:targetdeformation}
\end{align}
where $\xd = x + u$, the deformed state of the object, weight $w$ determines relative importance of points,  and $x^{trg}$ is the target configuration, defined as function on $\Omegar$. 

The formulas for the general objective~\eqref{eq:objective-solution}, apply, with
\begin{align*}
    \nabla_1 j = \nabla_2 j = 2 w(\fprms^{-1}(x)) (\xd - x^{trg}(\fprms^{-1}(x)).
\end{align*}

If we define only the shape  on the boundary as the target, then we have:
\[J^{btrj} = \int_{\partial \Omega_\prms} w(\fprms^{-1}(x)) \,\, \|\xd - x^{trg}(\fprms^{-1}(x))\|^2 \,\, dx\].
Formulas for the derivatives are similar:

\begin{align*}
    S^{btrj} = \int_{\partial \Omega_\prms} \nabla_2 j \cdot \per + j(u,x) \, \nabla_s \cdot \per \,\, dx,\\
    R^{btrj} = \int_{\partial \Omega_\prms} \nabla_1 j \cdot \psi \,\, dx,
\end{align*}
where $\nabla_s$ denotes the surface derivative. 

\paragraph{Target center of mass trajectory} 
A related objective is  the deviation of  the center of mass of the object from a target trajectory.

\begin{equation}
\begin{aligned}
J^{ctr}(J^P,J^D) &= \bigg\lVert \frac{J^P}{J^D} - x^{ctr}\bigg\rVert ^2 = \bigg\lVert \frac{\int_{\Omega_\prms} \rho(x) \, \xd \, dx}{\int_{\Omega_\prms} \rho(x) \, \, dx} - x^{ctr}\bigg\rVert ^2 \\
&= \sum_{i}^{D_d} \left( \frac{J^P_i}{J^D}  - x^{ctr}_i\right)^2.
\end{aligned}
\label{eq:obj:centermasstrajectory}
\end{equation}

Using the chain rule, we can reach a formulation where $S^{ctr}$ and $R^{ctr}$ depend on respective derivatives from each $J^P_i$ and $J^D$:
\[S^{ctr} = \sum_i (\partial_1 J^{ctr})_i S^P_i + \partial_2 J^{ctr} S^D,\]
\[R^{ctr} = \sum_i (\partial_1 J^{ctr})_i R^P_i + \partial_2 J^{ctr} R^D.\]

We then need to compute shape derivative and adjoint terms for both of our scalar integrals $J^P_i$ and $J^D$, following general formulas for \ref{eq:objective-solution}. For each $J^P_i$, we have:
\begin{align*}
    \nabla_1 j = \nabla_2 j = \rho(x) e_i,
\end{align*}
where $e_i \in \R^{D_d}$ is a vector with $0$s everywhere except at index $i$, where the value is $1$.

Finally, assuming that densities are constant per point, for $J_D$,
\begin{align*}
    \nabla_1 j = \nabla_2 j = 0.
\end{align*}
\normalsize

\paragraph{Height}  This functional aims to maximize the height of the center of mass: 
\begin{align}
    J^{z_{\text{max}}} = -\frac{\int_{\Omega_\prms} \rho(x) \xd_z  dx}{\int_{\Omega_\prms} \rho(x) dx},
\label{eq:obj:centermassheight}
\end{align}
where $u_z$ is the z (vertical) component of the solution (displacement) $u$, $x_z$ is the z component of the original position x. 
We can rewrite this formula using $J^P_z$ and $J^D$ from previous subsection:
\begin{align}
    J^{z_{\text{max}}}(J^P_z, J^D) = - \frac{J^P_z}{J^D}.
\end{align}

This way, similar to $J^{ctr}$, we have:
\[S^{z_{\text{max}}} = \partial_1 J^{z_{\text{max}}} S^P_z + \partial_2 J^{z_{\text{max}}} S^D,\]
\[R^{z_{\text{max}}} = \partial_1 J^{z_{\text{max}}} R^P_z + \partial_2 J^{z_{\text{max}}} R^D.\]

Then, as for previous case, we can compute $S^P_z$, $R^P_z$, $S^D$ and $R^D$ through general formula \ref{eq:objective-solution}, using $\nabla_1 j = \nabla_2 j = \rho(x) e_z$ for $J^P_z$ and $\nabla_1 j = \nabla_2 j = 0$ for $J^D$.

\paragraph{Upper bound for volume} A constraint on the volume of the optimized object in penalty form is 
\label{sec:volume_constraint}
\begin{align}
    J^V &= \varphi(V(\Omega_\prms) - V_t), %
\label{eq:obj:volumeconstraint}
\end{align}
where $V$ corresponds to  ($\int_{\Omega_\prms} \, dx$), the volume of shape $\Omega_\prms$, $V_t$ to the target volume, and $\varphi(z)$ is a quadratic penalty function equal to $z^2$ for positive $z$ and zero for negative $z$.
This functional reduces to the general objective~\eqref{eq:objective-solution},
with $\nabla_1 j = \nabla_2 j = 0$, since $j(u,x) = 1$.

\paragraph{Upper bound for stress}
Similarly, we can impose an approximate upper bound on stress via a penalty:
\begin{align}
    J^{\sigma_t} &= \int_{\Omega_\prms} \varphi(\|\sigma\| - s_t) dx,
\label{eq:obj:upperbound}
\end{align}
where $s_t$ is the stress magnitude target. 
As for $L_p$ stress energy, our integrand $\varphi(\|\sigma\| - s_t)$ depends only on $\nabla u$ and 
\eqref{eq:objective-gradient} applies with
\begin{align*}
    \nabla_1 j &= \varphi' \, \frac{f(\nabla u)}{\|\sigma\|} \, : \, \nabla f(\nabla u),\\
    \nabla_2 j &= 0.
\end{align*}

\subsection{Regularization terms}
\label{sec:regularization}
In addition to the physical objectives described in the previous sections, we use two discrete regularization terms essential for numerical stability for a number of problems. 

\paragraph{Scale-invariant smoothing}
\label{sec:scale_invariant_smoothing}
\begin{align}
J^{\text{smooth}}=\sum_{i\in B} \|s_i\|^p, \qquad s_i=\frac{\sum_{j\in N(i)\cap B}(v_i - v_j)}{\sum_{j\in N(i)\cap B}\|v_i - v_j\|},
\label{eq:obj:smoothing}
\end{align}

where $B$ contains the indices of all boundary vertices, $N(i)$ contains the indices of all neighbor vertices of vertex $i$, and $v_i$ is the position of vertex $i$. The value of $p$ can be adjusted to obtain smoother surfaces at the cost of less optimal shapes, normally we use $p=2$. This term is scale-invariant and pushes the triangles/tetrahedra of the mesh toward equilateral. The derivative of this smoothing term with respect to optimization parameters $v_i$ can be seen in the first paragraph of Appendix \ref{appsec:regularization-derivatives}.

\paragraph{Material parameter spatial smoothing}
\begin{align}
    J^{\lambda, \mu \ \text{smooth}} = \sum_{t \in T} \sum_{t' \in Adj(t)} \left( 1 - \frac{\lambda_{t'}}{\lambda_{t}} \right)^2 + \left( 1 - \frac{\mu_{t'}}{\mu_{t}} \right)^2,
\label{eq:obj:matsmoothing}
\end{align}
where $T$ is the set of all triangles/tetrahedra, $Adj(t)$ is the set of triangles/tetrahedra adjacent to $t$. $\lambda, \mu$ are the material parameters defined per triangle. The derivative of this term can be seen in the last part of Section \ref{appsec:regularization-derivatives} (Appendix).

\section{Results}\label{sec:results}

We partition our results into three groups depending on the type of the dofs used in the objective function: shape (Section~\ref{sec:res:shapenocontact}), initial conditions (Section~\ref{sec:res:initial_c}), or material (Section~\ref{sec:res:mat_opt}). For each group, we provide a set of examples of static and dynamic scenes of increasing complexity.
In Section \ref{sec:res:comparisons}, we compare our solver, \cite{du2021diffpd}, and \cite{gradsim}  to evaluate the effect of different material and contact models. We also compare against a baseline implementation using finite differences.
We run our experiments on a workstation with a Threadripper Pro 3995WX with 64 cores and 512 Gb of memory. For a selection of problems, we validate our results with physical experiments using items fabricated in silicon rubber (we use 1:1 SMOOTH-ON OOMOO 30 poured into a 3D printed PVA mold) or 3D printed PLA plastic.

We additionally provide a video showing the intermediate optimization step for all the results in the paper as part of our additional material.

\paragraph{Statistics}
We provide statistics for our experiments in Table \ref{Table:stats}, including the size of the meshes, material model, running time, and memory used.

We observe that the time to compute the gradients of the objective function is negligible compared to the forward solve time (usually less than 10\%). This implies that as long as a physical system can be simulated in PolyFEM, our approach enables optimizing functionals depending on it with a comparable running time per optimization iteration. %

We recall that the gradient computation requires solving one linear system for each time step of the forward simulation. For linear problems, the system to solve has the same stiffness matrix and we can thus reuse the factorization. For non-linear problems requiring Newton iterations, the forward step requires multiple Netwon steps, while the solve for the gradient is always a single linear system solve.

An additional acceleration strategy that we employ is noting that the optimization algorithm needs to solve many, often similar, forward simulations. We thus initialize, for non-linear problems, the forward solver with the solution at the previous step, which is often a good initialization.

\begin{table*}[htb]
\caption{Columns from left to right are: example names, number of vertices, degree of freedom of the simulation, physical formulation, objective functional of the optimization, total running time (sec), peak memory (Mb), number of iterations of the optimization, average running time of the simulation (sec), average running time of computing gradients (sec), total number of Newton iterations (linear solves) in the simulation, number of Newton solves in the simulation.}
\resizebox{.96\textwidth}{!}{
\begin{tabular}{lrrllllrrrrr}
\toprule
                    Example &  Vertices &   Dofs &      Model &  Objective & Total time & Memory &  Iter &  Solve time &  Grad time &  Newton Iter. &  Newton Solve \\
\midrule
          Bridge (Figure \ref{fig:bridge-shape}) &      4641 &   9282 &     Linear &     Target &             16.1 &       162.3 &    55 &          0.0343 &         0.0296 &            0 &             0 \\
          Bridge (Figure \ref{fig:stress-bridge}) &     18598 & 143378 &     Linear &     Stress &           1665.8 &      1690.8 &   402 &          1.3859 &         0.1861 &            0 &             0 \\
         3D Beam (Figure \ref{fig:stress-beam}) &      9939 & 209409 & NeoHookean &     Stress &          95738.6 &    101786.3 &   171 &        192.1587 &        37.2651 &         1083 &           361 \\
    Interlocking (Figure \ref{fig:interlocking}) &      1290 &   9946 &        IPC &     Stress &            303.3 &       188.5 &   101 &          0.8394 &         0.0590 &         4900 &           335 \\
         2D Hook (Figure \ref{fig:hook2d}) &      1760 &  13348 &        IPC &     Stress &            220.5 &       726.5 &    60 &          1.7224 &         0.0802 &         2548 &           126 \\
       3D Hanger (Figure \ref{fig:hanger3D}) &      4190 &  80412 &        IPC &     Stress &          14129.4 &      6554.0 &    29 &        204.2374 &         3.2117 &         4708 &            64 \\
   Bouncing Ball (Figure \ref{fig:bouncing-ball}) &        73 &    146 &        IPC &     Target &            961.7 &        29.2 &   202 &          1.1061 &         0.0898 &       211611 &         41200 \\
   Sliding Ball (Figure \ref{fig:sliding-ball}) &       526 &   6849 &        IPC &     Stress &           1610.1 &      2184.6 &    29 &         24.3086 &         1.2580 &            0 &             0 \\
   Shock Protection (Figure \ref{fig:shock-protect}) &       53879 &   107758 &        IPC &     Stress &     33301 & 10100 &    9 &         1264.395 &   129.96 & 35553  &   4800 \\
   Puzzle Piece (Figure \ref{fig:bouncing-puzzle}) &       370 &    740 &        IPC & Trajectory &             47.4 &       107.6 &    19 &          1.5877 &         0.2129 &         2917 &           630 \\
    Throw Bunny (Figure \ref{fig:bunny-pool}) &      2174 &   6522 &        IPC &     Target &            602.0 &      3344.1 &     9 &        209.2998 &         4.3731 &        15324 &          1000 \\
    Colliding Tentacles (Figure \ref{fig:tentacles}) & 6896 & 20688 & IPC & Trajectory & 13070 & 8325 & 5 & 2043 & 41.25 & 14447 & 720 \\
           Sine (Figure \ref{fig:sine}) &       651 &   1302 &     Linear &     Target &              0.3 &        34.4 &    12 &          0.0042 &         0.0022 &            0 &             0 \\
         Bridge (Figure \ref{fig:bridge-mat}) &     18598 &  37196 &     Linear &     Target &             32.7 &       655.2 &    39 &          0.1416 &         0.0398 &            0 &             0 \\
           Cube (Figure \ref{fig:physical-cube}) &      4631 & 103383 & NeoHookean &     Target &            455.8 &      6316.6 &     8 &         37.9947 &         2.6101 &           33 &            11 \\
Micro-Structure (Figure \ref{fig:physical-microstructure}) &      3268 &   9804 &        IPC &     Target &            602.3 &     33502.3 &    11 &         42.0954 &         0.0746 &          249 &            14 \\
       Kangaroo (Figure \ref{fig:kangaroo}) &       231 &    462 &        IPC & Trajectory &             21.6 &       224.6 &     6 &          1.6704 &         0.1624 &         2987 &           660 \\
  Sliding Bunny (Figure \ref{fig:bunny}) &      5682 &  17046 &        IPC &     Target &          11734.0 &      2304.3 &     8 &        547.3644 &         1.6478 &        61517 &           880 \\
  Bouncing Ball (Figure \ref{fig:ball-h}) &       720 &   1440 &        IPC &     Height &            612.3 &        86.4 &    79 &          3.3482 &         0.2003 &        33609 &          5160 \\
  Bouncing Ball (Figure \ref{fig:bounce-ball}) &       646 &   1938 & NeoHookean & Trajectory &            206.3 &       152.8 &    24 &          3.0548 &         0.7396 &         3264 &          1632 \\
  Bouncing Ball (Figure \ref{fig:bounce-ball}) &      1251 &   3753 &        IPC & Trajectory &          10546.6 &      1547.0 &    49 &        113.4214 &         8.8056 &       105401 &         20160 \\
\bottomrule
\end{tabular}
}
\label{Table:stats}
\end{table*}

\paragraph{Color Legend} We use
green arrows to indicate Neumann boundary conditions, and black squares to indicate nodes that have a Dirichlet boundary condition. To reduce clutter, we use a uniform gray to indicate objects with a uniform Dirichlet boundary condition on all nodes.

To avoid singularities in the optimization we add, to the objective function, a boundary smoothing term \eqref{eq:obj:smoothing} in all our shape optimization experiments, and a material regularization term \eqref{eq:obj:matsmoothing} to all our material optimization experiments.

\subsection{Implementation}\label{sec:implementation}

\paragraph{FE Solver}
We implemented our solver in C++ using the PolyFEM library \cite{polyfem} for the forward solve, the IPC~Toolkit~\cite{ipc_toolkit} for computing contact and friction potentials, and Pardiso~\cite{pardiso-7.2a,pardiso-7.2b,pardiso-7.2c} for solving linear systems.

\paragraph{Optimization} Our optimization algorithm (\cref{alg:the_alg}) uses the L-BFGS implementation in \cite{cppoptlib}, with backtracking line search.

\paragraph{Remeshing} Shape optimization might negatively affect the element shape, and for large deformation introduce close to singular elements that force the optimization to take tiny steps. After every optimization iteration, we evaluate the element quality using the scaled Jacobian quality measure \cite{Knupp:2001}, and optimize the mesh if it is below a threshold experimentally set to $10^{-3}$.

\begin{figure}\centering
    \centering
    \includegraphics[width=\linewidth]{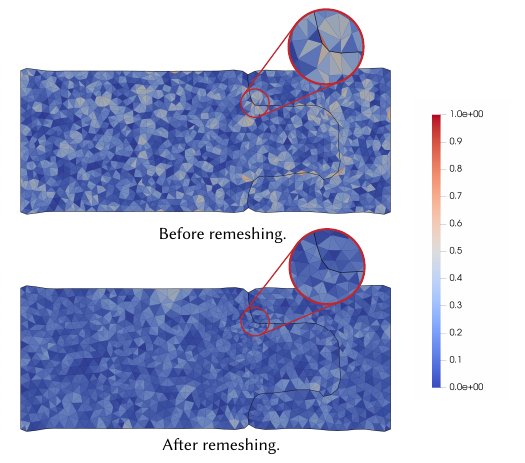}
    \caption{An example of remeshing in the shape optimization. The quality is shown for each triangle. Triangles with bad quality have higher values.}
    \label{fig:remesh}
    \Description{}
\end{figure}

For 2D examples, we keep the mesh boundary fixed and we regenerate the interior using GMSH \cite{GMSH} (\cref{fig:remesh}). For 3D examples, we similarly fix the boundary and then use the mesh optimization procedure of fTetWild \cite{fTetWild} to improve the quality of the interior until its quality is above the threshold.

The reason why we can remesh without damaging the optimization convergence is that our optimization objectives have little dependence on interior node positions. The objectives are in the form of an integral over the domain or boundary, so remeshing only leads to small errors due to projections between the meshes.

The reason for fixing the boundary in the remeshing is that our optimization objectives (Section \ref{sec:objectives}) often depend on quantities on the boundary vertices: if the boundary is remeshed, we will need a bijective map between the two boundaries. Meshing methods providing this map exist \cite{Jiang:2020}, but their integration in our framework, while trivial from a formulation point of view, is an engineering challenge that we leave as future work.

\paragraph{Reproducibility} The reference implementation of our solver and applications will be released as an open-source project.

\subsection{Shape Optimization} \label{sec:res:shapenocontact}

We start our analysis with shape optimization problems both without and with contact or friction forces.

\begin{figure}\centering\footnotesize
\includegraphics[trim={0 350 0 400},clip=True,width=.33\linewidth]{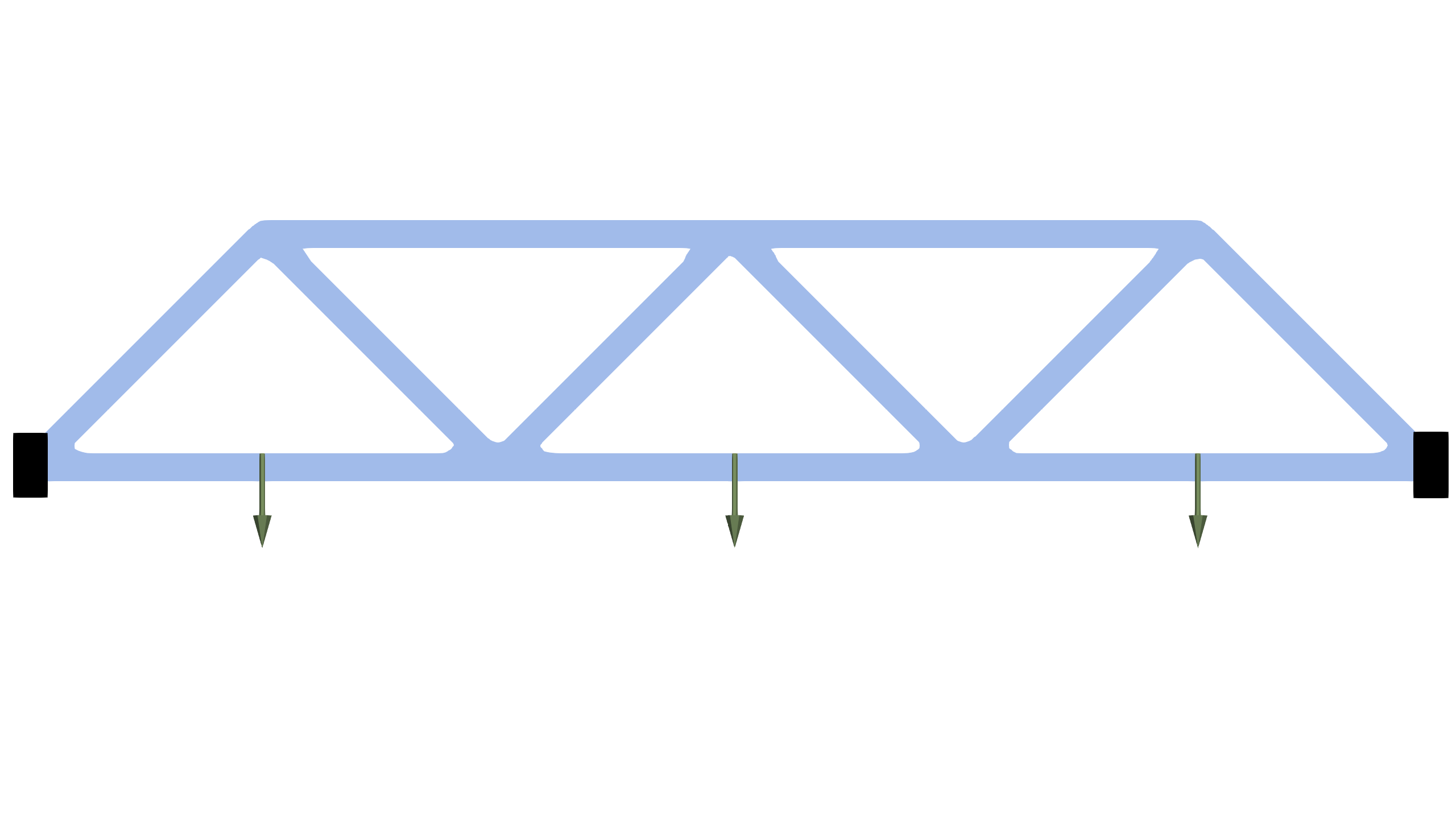}\hfill
\includegraphics[width=.33\linewidth]{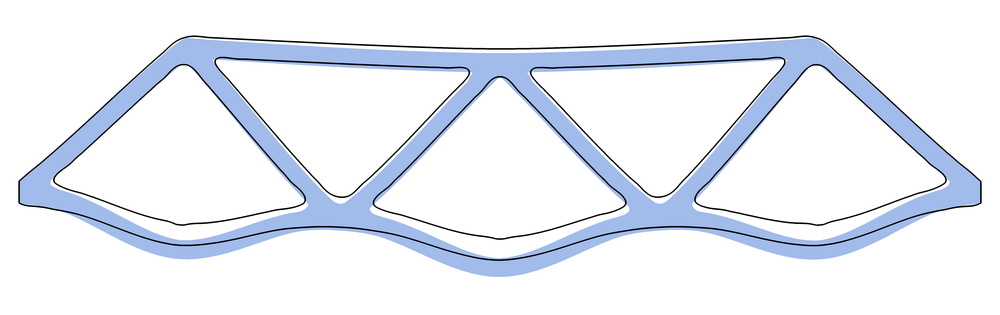}\hfill
\includegraphics[width=.33\linewidth]{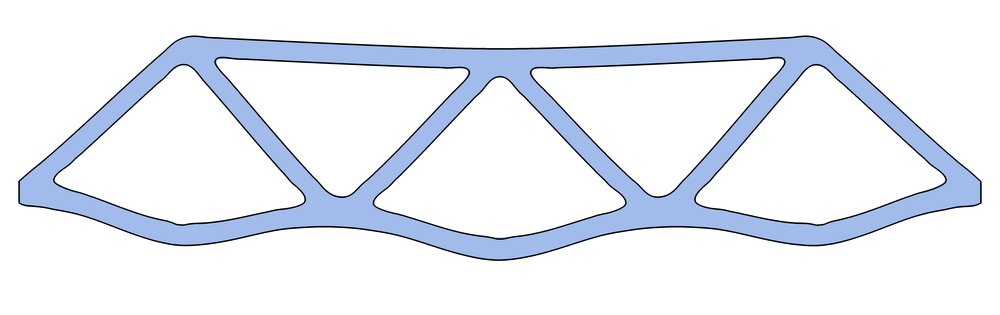}
\parbox{.33\linewidth}{\centering Problem setup.}\hfill
\parbox{.33\linewidth}{\centering Initial shape.}\hfill
\parbox{.33\linewidth}{\centering Optimized shape.}
\caption{\textbf{Static: Bridge With Fabricated Solution.} The result of the shape optimization (blue surface) matches the target shape (wire-frame).}
\label{fig:bridge-shape}
\Description{}
\end{figure}

\paragraph{Static: Bridge With Fabricated Solution} We fabricate a 2D solution to verify the correctness of our formulation and implementation. Starting from the shape of a bridge (Figure~\ref{fig:bridge-shape}) we run a forward linear elasticity simulation with the two sides fixed and gravity forces. We now perturb the geometry of the rest pose and solve a shape optimization problem to recover the original rest pose, i.e. we remove the perturbation we introduced by minimizing the objective in \eqref{eq:obj:targetdeformation}.

\begin{figure}\centering\footnotesize
\parbox{.88\linewidth}{\centering
\includegraphics[width=\linewidth]{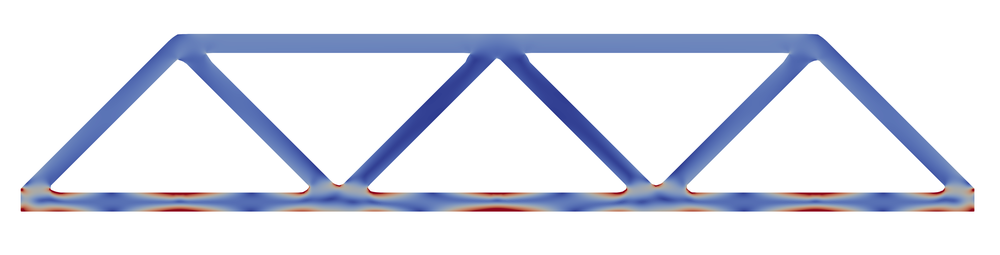}
Initial stress distribution.\\
\includegraphics[width=\linewidth]{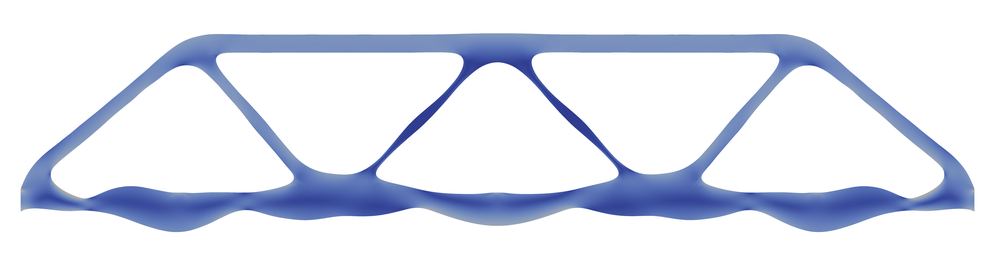}
Optimized stress distribution.
}\hfill
\parbox{.1\linewidth}{\includegraphics[width=\linewidth]{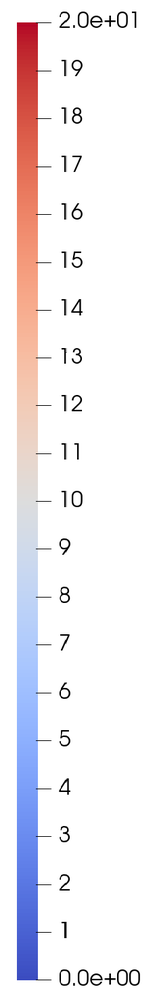}}
\caption{\textbf{Static: Bridge.} Result of shape optimization to minimize the average stress.}
\label{fig:stress-bridge}
\Description{}
\end{figure}

\paragraph{Static: Bridge}
We use the same model for a more challenging problem (\cref{fig:stress-bridge}): we use the same Dirichlet conditions and material model, replace the gravity forces by 3 Neumann conditions on the lower beams, and minimize
the $L^8$ norm of stress \eqref{eq:obj:stress}. To avoid trivial solutions we add a constant volume constraint (Section \ref{sec:volume_constraint}). The maximum stress is reduced from $68.789$ to $22.232$.

\paragraph{Static: 3D Beam}
\begin{figure}\centering\footnotesize
\parbox{0.6\linewidth}{\centering
\includegraphics[width=0.9\linewidth]{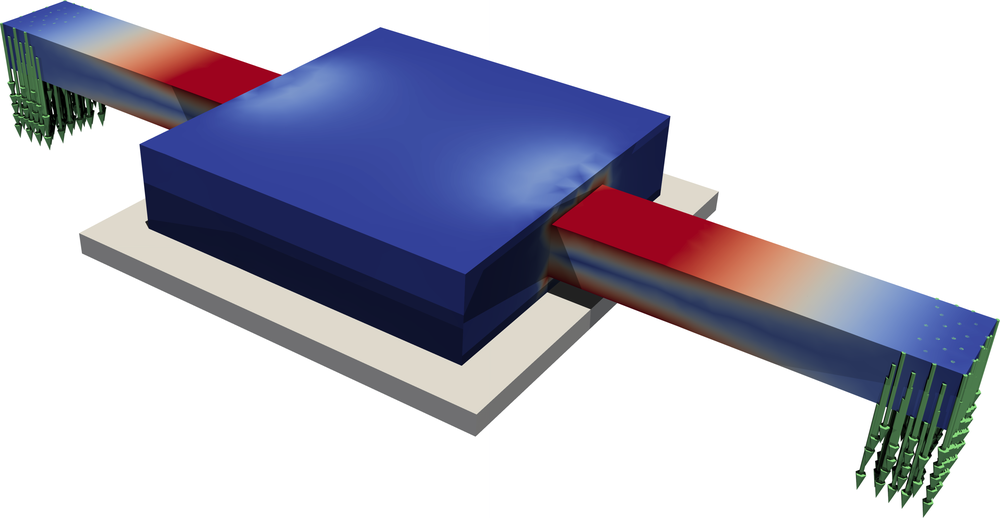}\\
\parbox{\linewidth}{\centering Initial stress distribution.}\\[1em]
\includegraphics[width=0.9\linewidth]{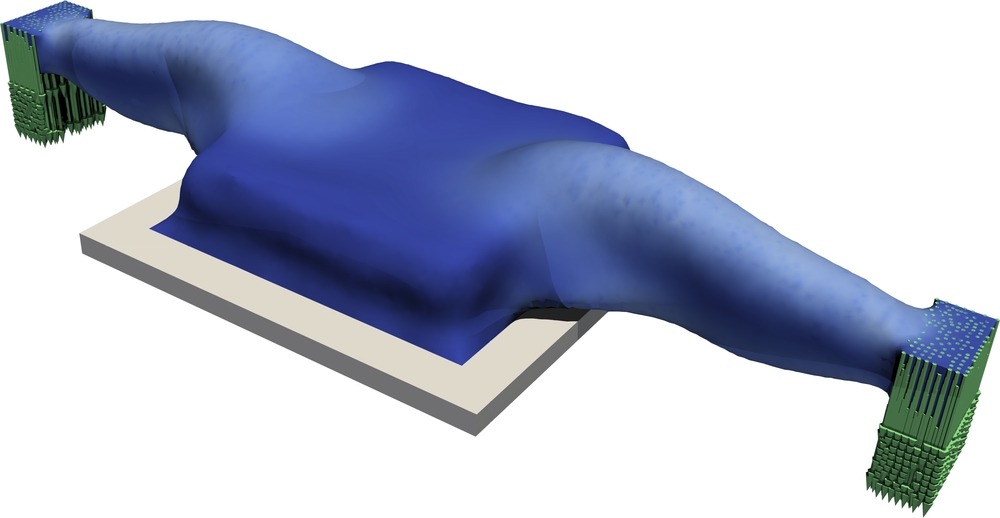}\\
\parbox{\linewidth}{\centering Optimized stress distribution.}}
\parbox{.1\linewidth}{\includegraphics[width=\linewidth]{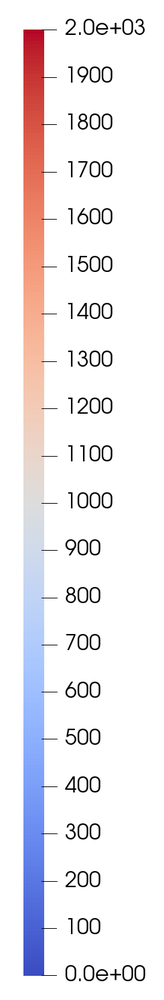}}
\caption{\textbf{Static: 3D Beam.} Result of stress minimization on a beam standing on a platform, with two loads on its sides.}
\label{fig:stress-beam}
\Description{}
\end{figure}
Moving to 3D (Figure \ref{fig:stress-beam}), we perform static optimization of
the $L^8$ norm of stress using Neo-Hookean materials on a beam standing on a fixed support at the center (nodes on the bottom surface of the beam have zero Dirichlet boundary conditions), and with two side loads applied as Neumann boundary conditions. We use \eqref{eq:obj:volumeconstraint} to bound the volume of the beam during optimization in order to avoid trivial solutions. The maximum stress is reduced from $3,377$ to $920$. Note that this scene is not using contact, the lower region of the central part of the beam is fixed with Dirichlet boundary conditions.

\paragraph{Static: Interlocking}
\begin{figure}\centering\footnotesize
\parbox{.33\linewidth}{\centering
\includegraphics[width=\linewidth]{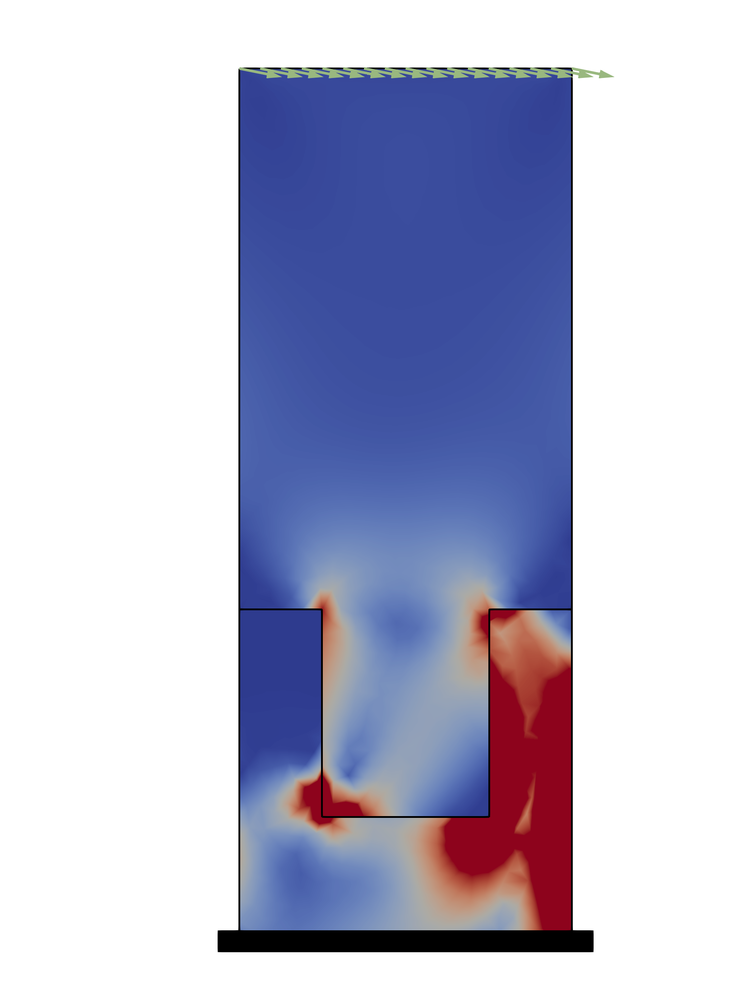}}\hfill
\parbox{.33\linewidth}{\centering
\includegraphics[width=\linewidth]{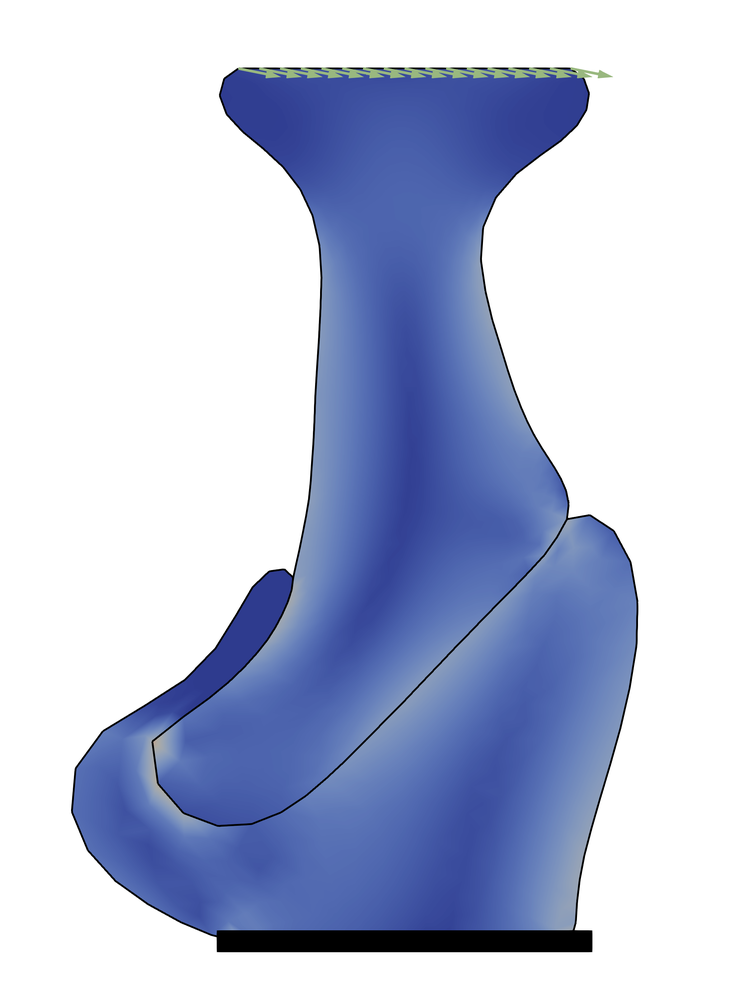}}\hfill
\parbox{.1\linewidth}{\includegraphics[width=\linewidth]{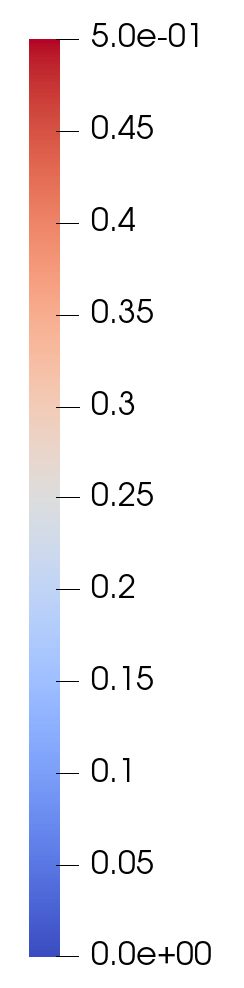}}\hfill
\parbox{.2\linewidth}{ \centering
\includegraphics[width=\linewidth]{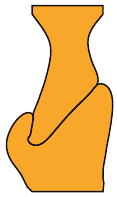}}
\parbox{.33\linewidth}{\centering Initial stress distribution.}\hfill
\parbox{.33\linewidth}{\centering Optimized stress distribution.} \hfill
\parbox{.1\linewidth}{~} \hfill
\parbox{0.2\linewidth}{\centering \protect\cite{Tozoni2021}.}
\caption{\textbf{Static: Interlocking.} Result of shape optimization to minimize the $L^8$ norm of stress.}
\label{fig:interlocking}
\Description{}
\end{figure}
Our framework supports contact and transient friction forces between objects without requiring explicit definition of contact pairs.
We borrow the experimental setup used in \cite{Tozoni2021}: we optimize the shape of two interlocking 2D parts (Figure \ref{fig:interlocking}) to minimize the $L^8$ norm of the stress \eqref{eq:obj:stress}. The bottom part is fixed and a force pointing down-right is applied to the top. Figure~\ref{fig:interlocking} shows how the shape changes to reduce the maximum stress from $\qty{3.2}{\Pa}$ to $\qty{0.29}{\Pa}$. %

Note that unlike \cite{Tozoni2021}, our contact model does not support overlapping boundary nodes, which are used in \cite{Tozoni2021} to keep the contact over the optimization. To mimic this behaviour in our setting, we create small displacements on the overlapped boundary nodes along the normal directions as the initial guess for the forward simulation, so that each object is shrinked by a tiny amount and there is no overlap in the initial guess.

We note that our result is expected to be different from \cite{Tozoni2021}, as the contact models are different and the solutions of these problems are in general not unique. Despite their differences, we observe in both cases a reduction in maximal stress of similar magnitude (around 10 times reduction).
\begin{figure}\centering\footnotesize
\parbox{.46\linewidth}{
\parbox{.7\linewidth}{\centering
\includegraphics[width=0.8\linewidth]{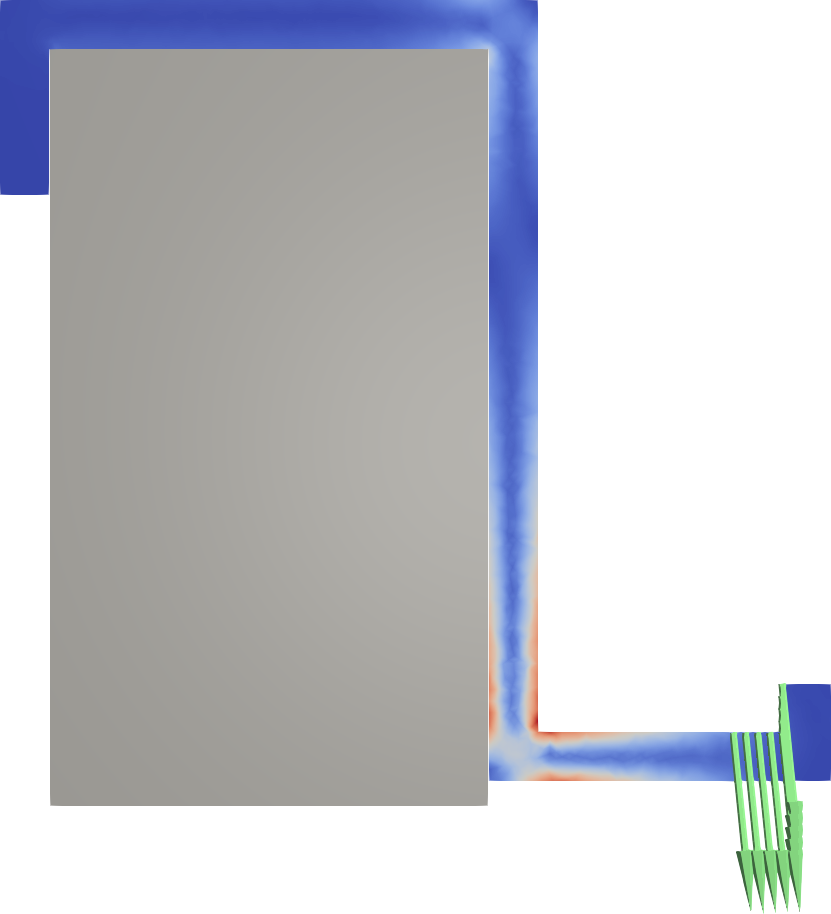}
Initial stress distribution.\\[1em]
\includegraphics[width=0.8\linewidth]{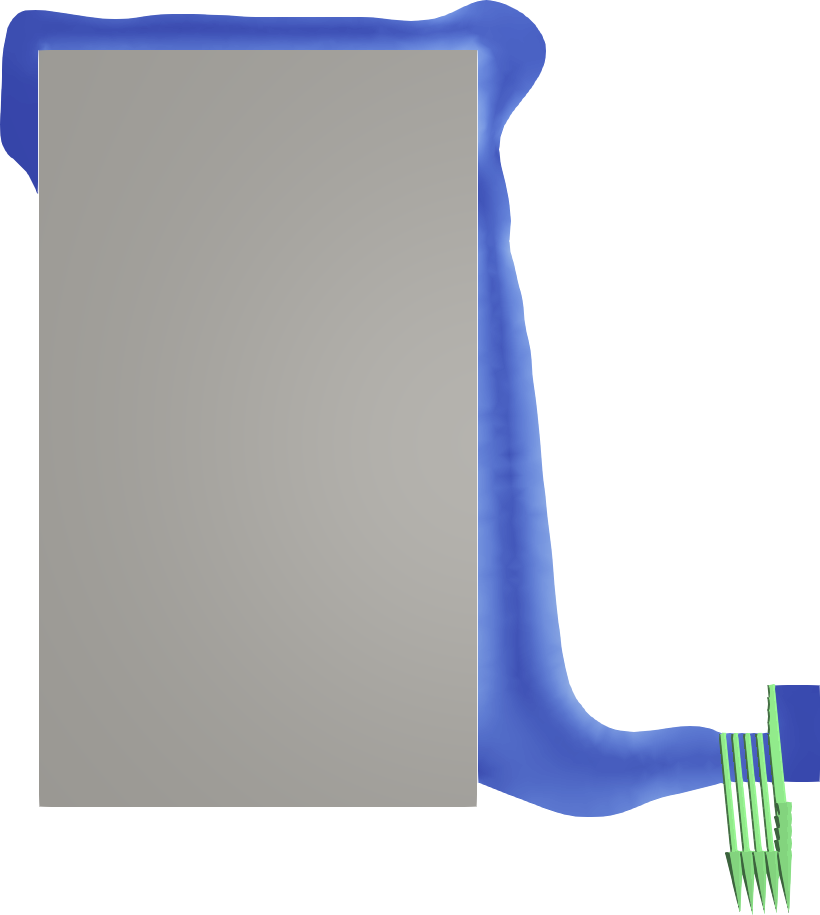}
Optimized stress distribution.\\[1em]
\includegraphics[width=0.8\linewidth]{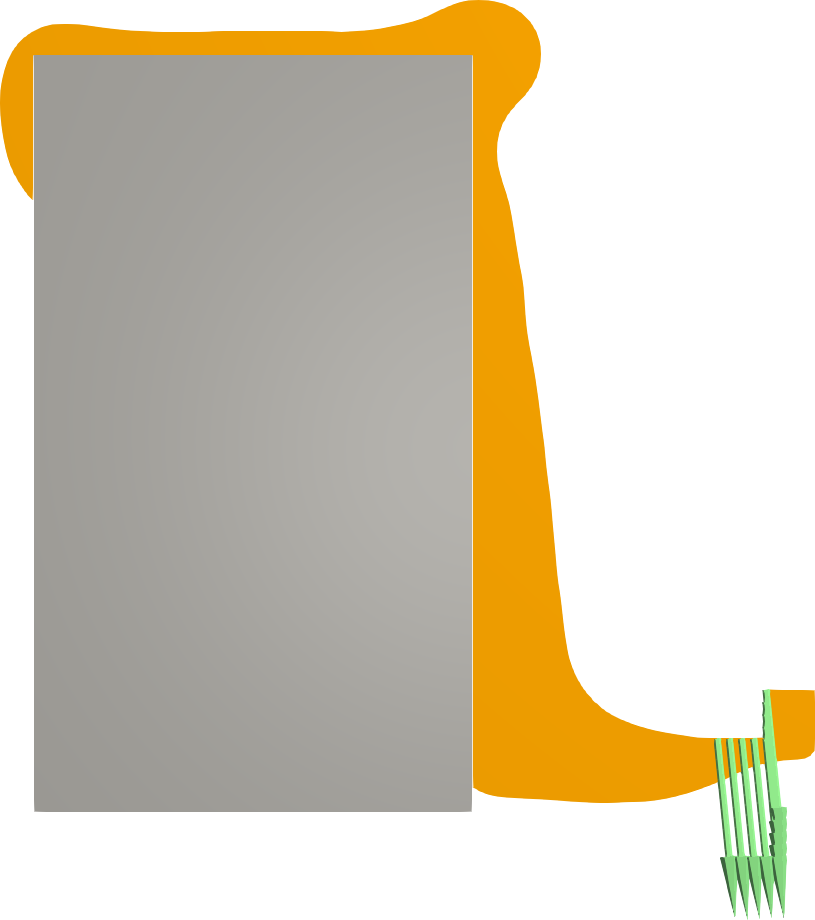}
Optimization result from \protect\cite{Tozoni2021} }\hfill
\parbox{.28\linewidth}{\includegraphics[width=\linewidth]{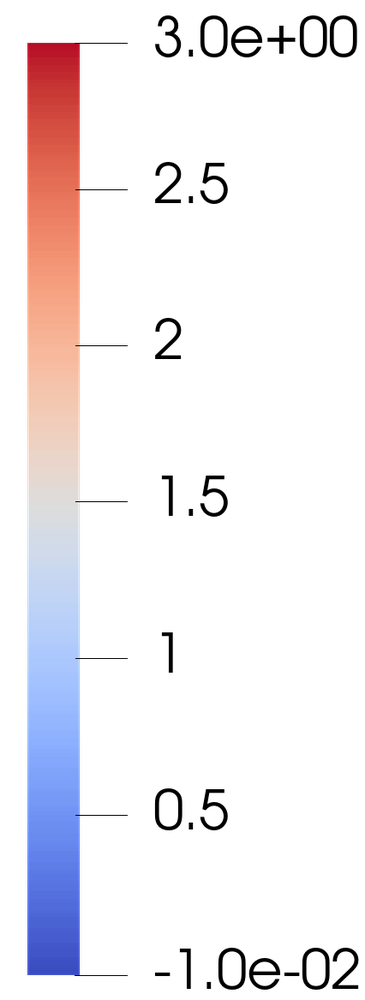}}
}\hfill
\parbox{.22\linewidth}{\includegraphics[width=\linewidth]{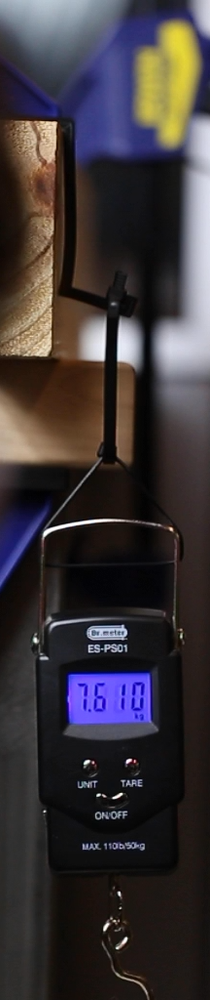}}\hfill
\parbox{.22\linewidth}{\includegraphics[width=\linewidth]{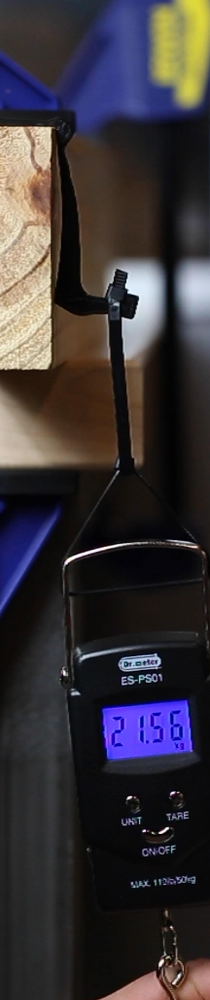}}
\caption{\textbf{Static: 2D Hook.} Shape optimization of a hook to reduce stress concentration (left). Fabricated results with maximum load before failure (right).}
\label{fig:hook2d}
\Description{}
\end{figure}

\paragraph{Static: 2D Hook}
To physically validate our shape optimization results we reproduce the experiment in \cite[Figure 21]{Tozoni2021}, where a hook is optimized to minimize
the maximum stress \eqref{eq:obj:stress} when a load is applied to one of its ends (Figure \ref{fig:hook2d}). The grey block is fixed with zero Dirichlet conditions on all nodes. We physically validate that the optimized shape is able to withstand a load of over 3$\times$ the unoptimized shape before breaking (Figure \ref{fig:hook2d}). The hook has been fabricated using an Ultimaker 3 3D printer, using black PLA plastic. Despite the different contact model, the result is quite similar to the one presented in \cite{Tozoni2021}: our approach has the advantage of not requiring manual specification of the contact surfaces.

\begin{figure}\centering\footnotesize
\parbox{.33\linewidth}{\centering
  \includegraphics[width=0.8\linewidth]{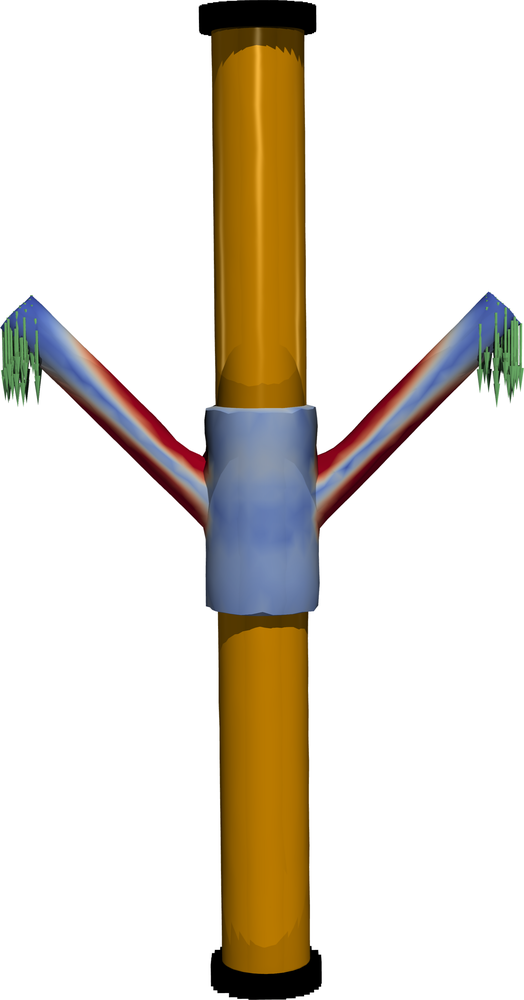}}\hfill
\parbox{.33\linewidth}{\centering
  \includegraphics[width=0.8\linewidth]{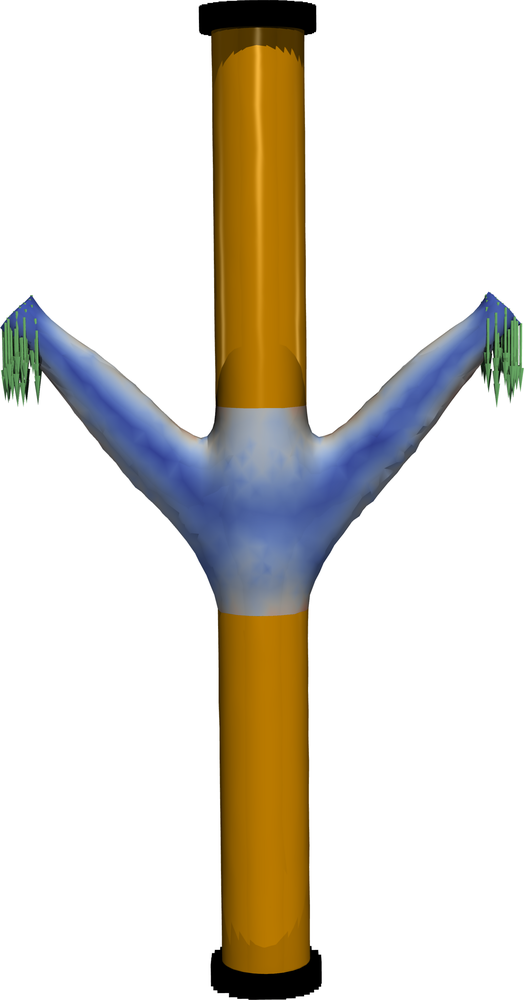}}\hfill
\parbox{.1\linewidth}{\includegraphics[width=\linewidth]{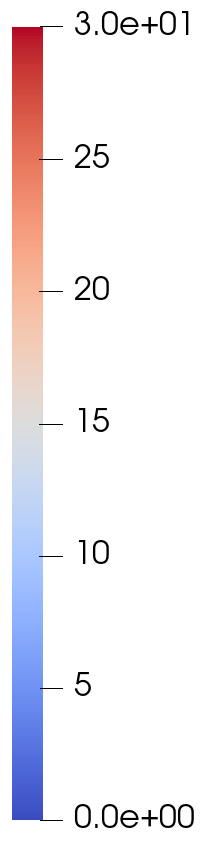}}
\parbox{.23\linewidth}{\includegraphics[width=\linewidth]{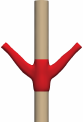}}\\[1em]
\parbox{.33\linewidth}{\centering Initial configuration.}\hfill
\parbox{.33\linewidth}{\centering Optimized configuration.}\hfill
\parbox{.1\linewidth}{~}\hfill
\parbox{.23\linewidth}{\centering \protect\cite{Tozoni2021}}
\caption{\textbf{Static: 3D Hanger.} Result of shape optimization of a hanger to reduce stress concentration.}
\label{fig:hanger3D}
\Description{}
\end{figure}

\paragraph{Static: 3D Hanger}
We also reproduce the experiment \cite[Figure 29]{Tozoni2021}: a coat hanger is composed of two cylinders and a hanger keeping the together. The shape of the hanger is optimized to minimize
the maximum internal stress \eqref{eq:obj:stress} when two loads are applied on its arms (Figure \ref{fig:hanger3D}). The maximal stress is reduced from $\qty{89.93}{\Pa}$ to $\qty{25.74}{\Pa}$. When comparing with \cite{Tozoni2021}, we observe a similar optimized shape and an equivalent stress reduction rate (around 3 times).

\paragraph{Transient: Bouncing ball}
\begin{figure}\centering\footnotesize
\includegraphics[width=.48\linewidth]{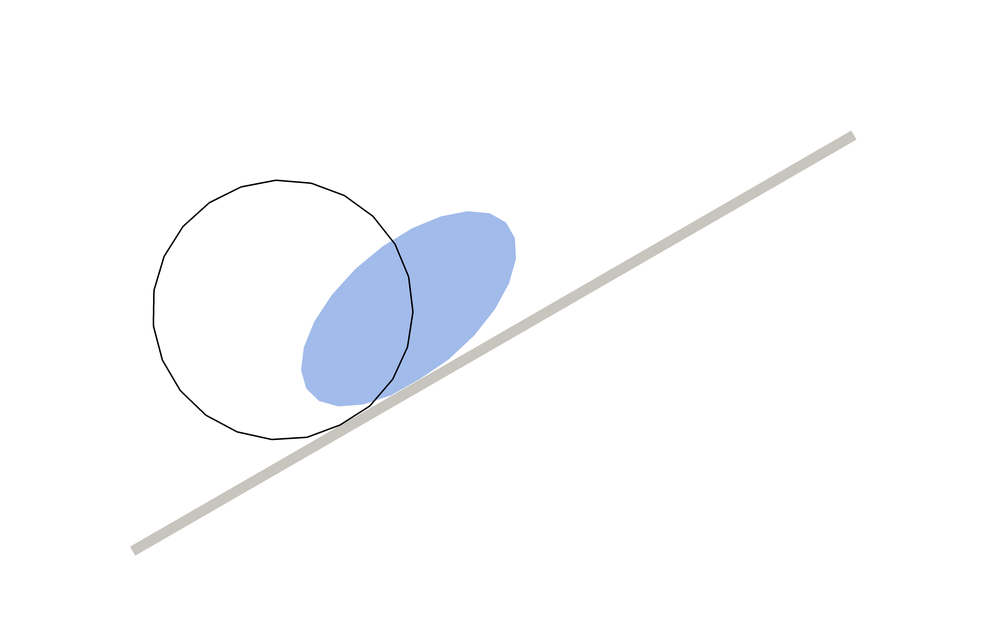}\hfill
\includegraphics[width=.48\linewidth]{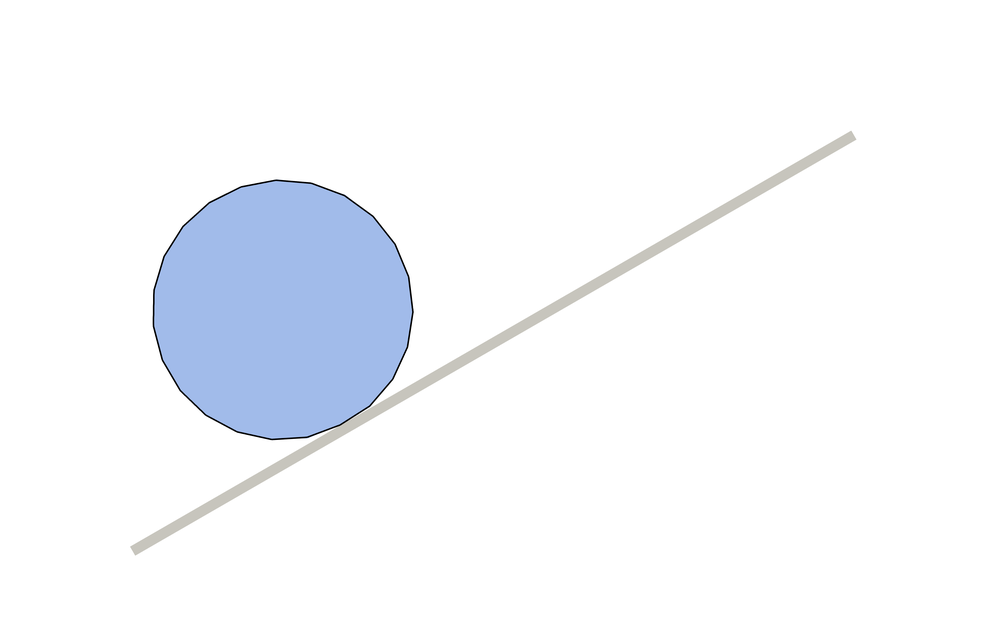}\par
\parbox{.48\linewidth}{\centering Initial shape.}\hfill
\parbox{.48\linewidth}{\centering Optimized shape.}\par
\includegraphics[width=.32\linewidth]{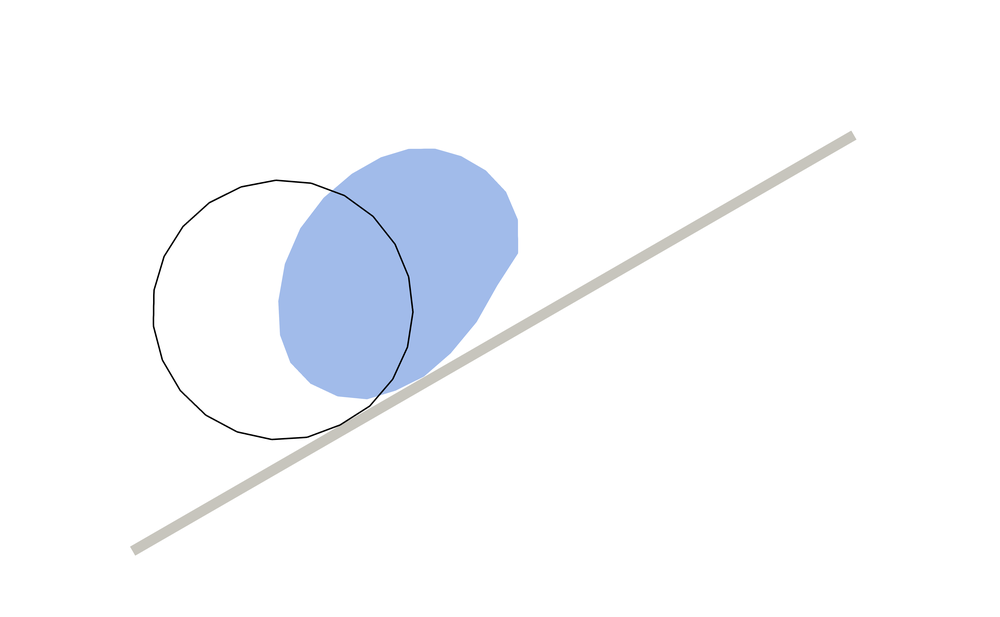}\hfill
\includegraphics[width=.32\linewidth]{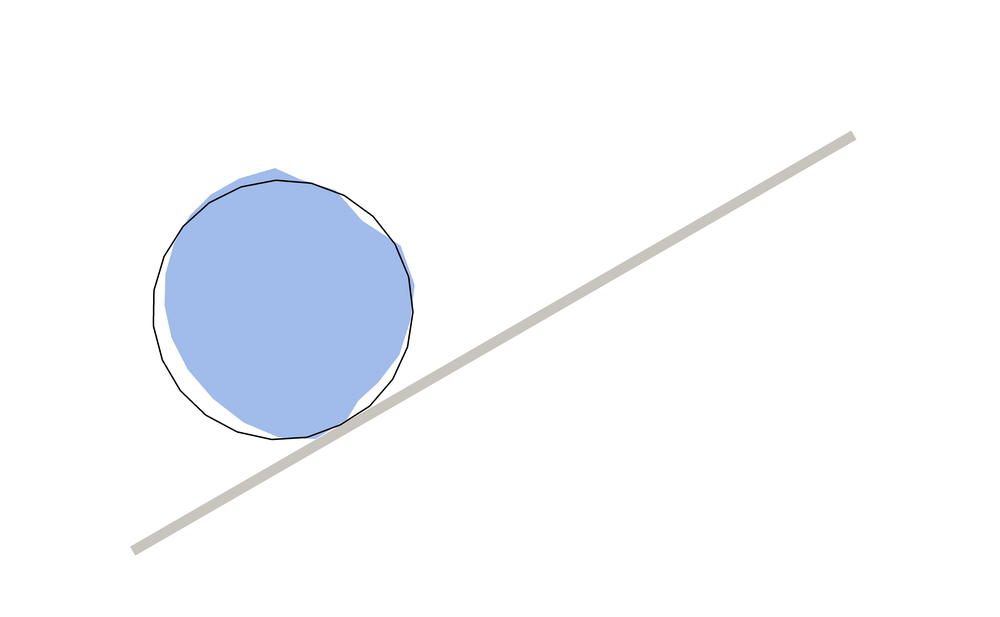}\hfill
\includegraphics[width=.32\linewidth]{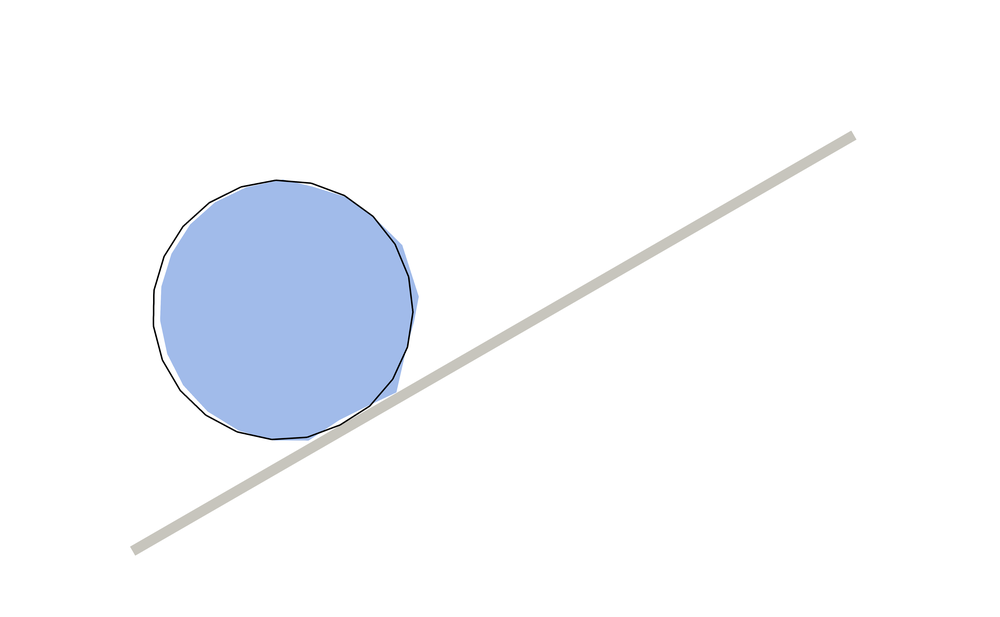}\par
\parbox{.32\linewidth}{\centering Iteration 20.}\hfill
\parbox{.32\linewidth}{\centering Iteration 50.}\hfill
\parbox{.32\linewidth}{\centering Iteration 70.}
\caption{\textbf{Transient: Bouncing ball.} The result of the shape optimization (blue surface) matches the desired trajectory (wire-frame).}
\label{fig:bouncing-ball}
\Description{}
\end{figure}
As a demonstration of shape optimization in a transient setting, we run a forward non-linear simulation of a ball bouncing on a plane and use its trajectory as the optimization goal \eqref{eq:obj:centermasstrajectory}. We then deform the initial shape into an ellipse and try to recover the original shape (\cref{fig:bouncing-ball}).

\begin{figure}\centering\footnotesize
\parbox{.29\linewidth}{\includegraphics[width=\linewidth]{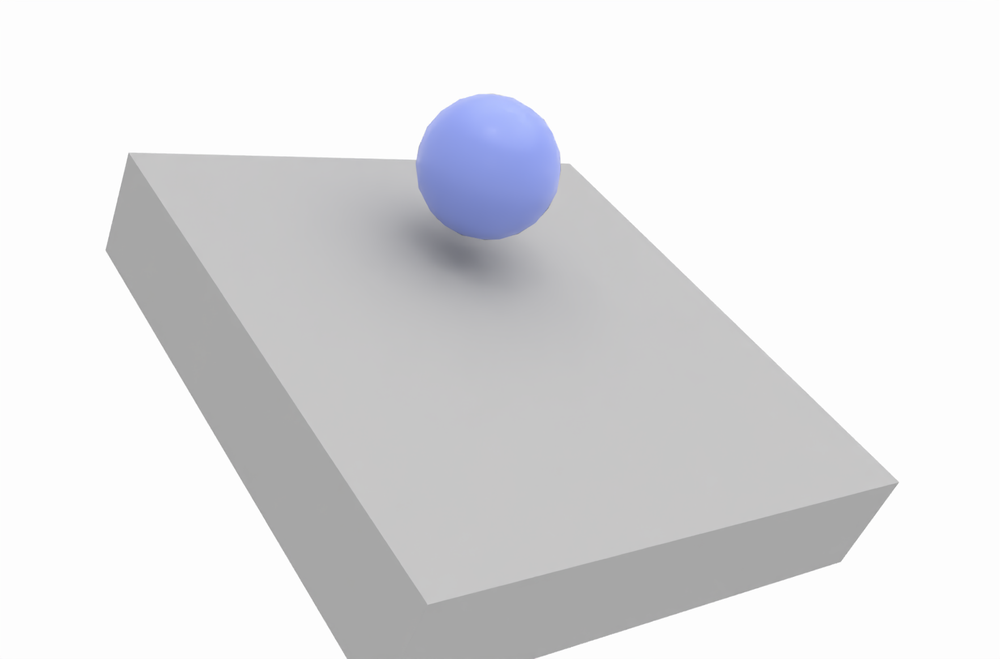}}\hfill
\parbox{.29\linewidth}{\includegraphics[width=\linewidth]{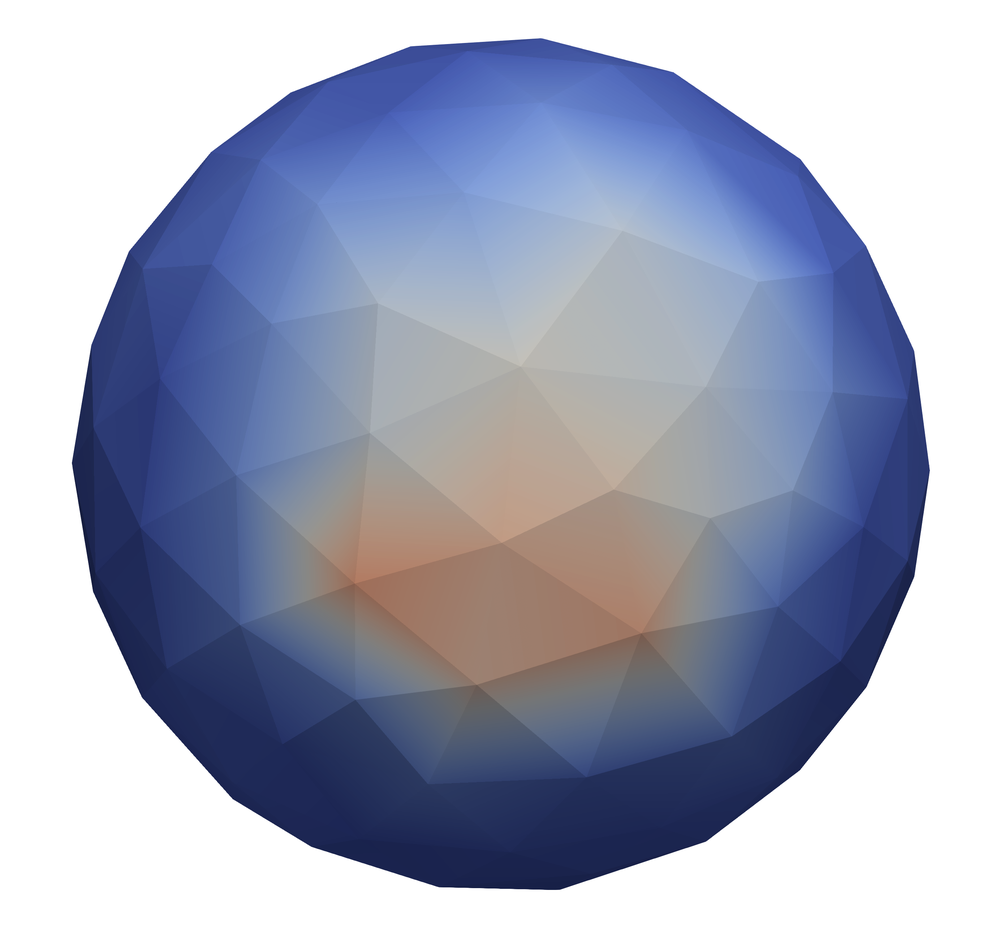}}\hfill
\parbox{.29\linewidth}{\includegraphics[width=\linewidth]{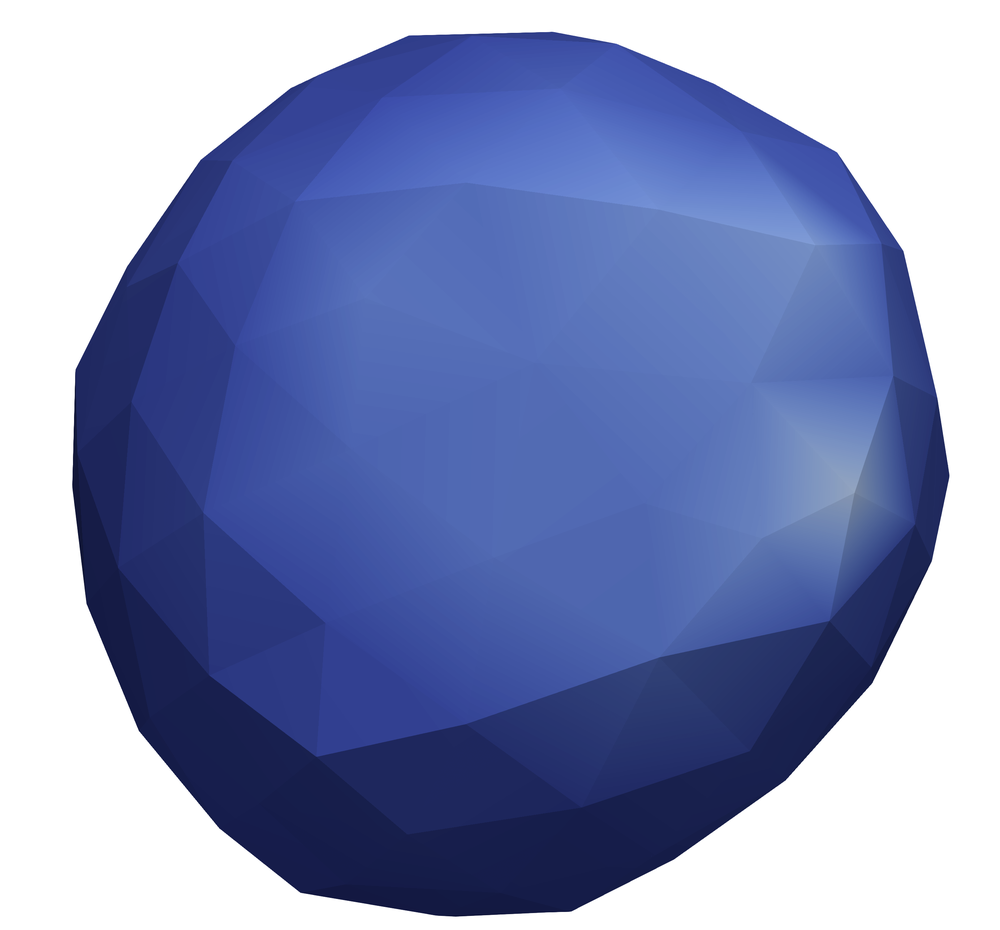}}\hfill
\parbox{.07\linewidth}{\includegraphics[width=\linewidth]{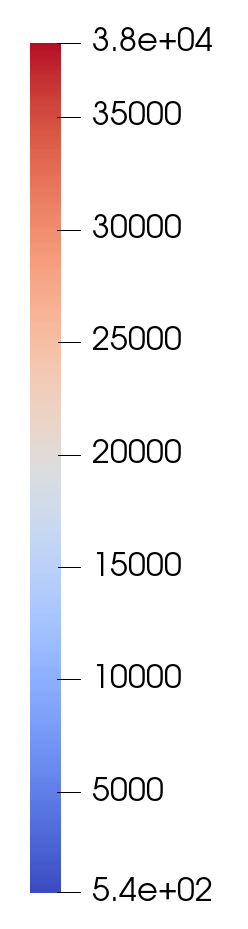}}
\parbox{.29\linewidth}{\centering Problem setup.}\hfill
\parbox{.29\linewidth}{\centering Initial stress distribution at the frame of contact.}\hfill
\parbox{.29\linewidth}{\centering Optimized stress distribution at the frame of contact.}\hfill
\parbox{.07\linewidth}{~}
\caption{\textbf{Transient: Sliding Ball.} Result of shape optimization to reduce stress.}
\label{fig:sliding-ball}
\Description{}
\end{figure}

\begin{figure}\centering\footnotesize
\includegraphics[width=\linewidth]{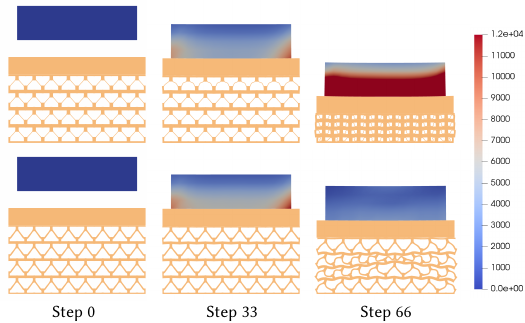}
\caption{\textbf{Transient: Shock Protection.} Shape optimization of the shock-protecting microstructure to reduce the stress on the falling load. The stress distribution at different time steps is shown for the initial shape (top) and optimized shape (bottom). }
\label{fig:shock-protect}
\Description{}
\end{figure}

\paragraph{Transient: Shock Protection}
We optimize the shape of a shock-protecting microstructure from \cite{bistable2015} so that the stress \eqref{eq:obj:stress} of the load being dropped onto the microstructure is minimized. To accelerate convergence, we adopt a low-parametric shape representation from \cite{Panetta2015}. In Figure \ref{fig:shock-protect}, the maximal stress is reduced from $\qty{32}{\kPa}$ to $\qty{12}{\kPa}$. This example involves complex self-contact inside the microstructure. Unlike penalty-based contact, our method is intersection-free regardless of the contact parameters, so able to produce plausible results with the same configuration even though the thickness of beams inside the microstructure changes drastically in the optimization.

\paragraph{Transient: Sliding Ball}
We optimize the shape of a ball sliding down a ramp to minimize the internal stress \eqref{eq:obj:stress}. To avoid trivial solutions, we add a volume constraint to not allow its volume to decrease. Perhaps unsurprisingly, the ball gets flattened on the side it contacts with the ramp as this leads to a major reduction of max stress, from $\qty{38}{\kPa}$ to $\qty{14}{\kPa}$.

\subsection{Initial Conditions}\label{sec:res:initial_c}

Our formulation supports the optimization of objectives depending on the initial conditions. We show three examples: the first involves an object sliding on a ramp with a complex geometry, the second simulates a game of pool, using bunnies instead of spheres, and the third demonstrates complex contact between tentacles.

\begin{figure}\centering\footnotesize
\includegraphics[width=.48\linewidth]{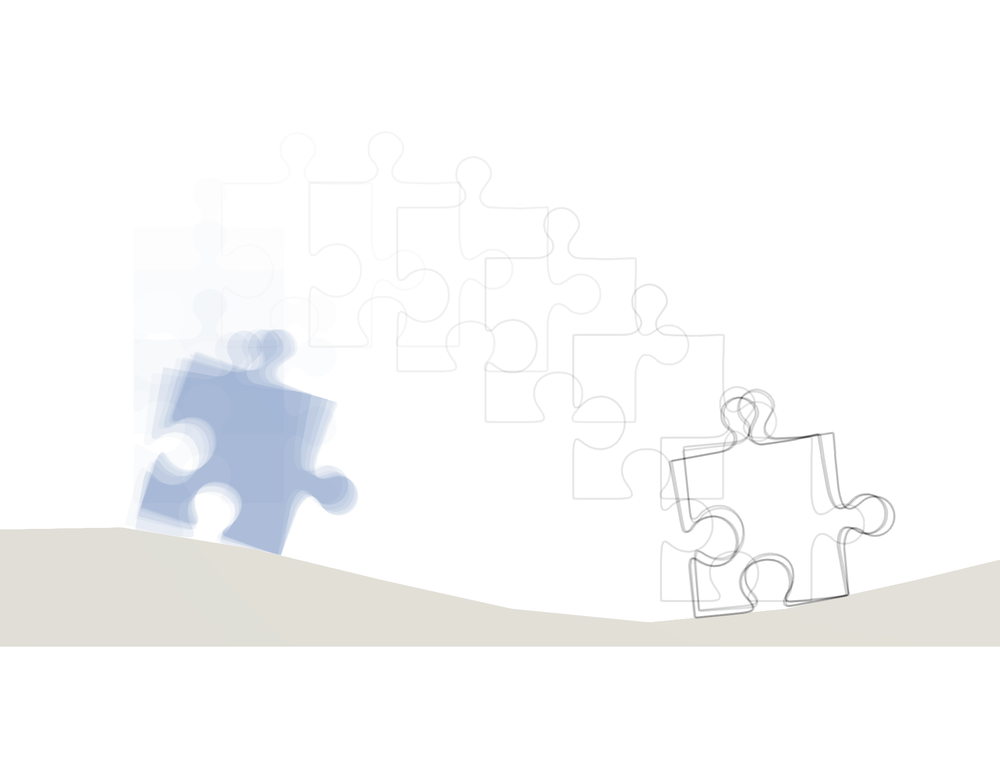}\hfill
\includegraphics[width=.48\linewidth]{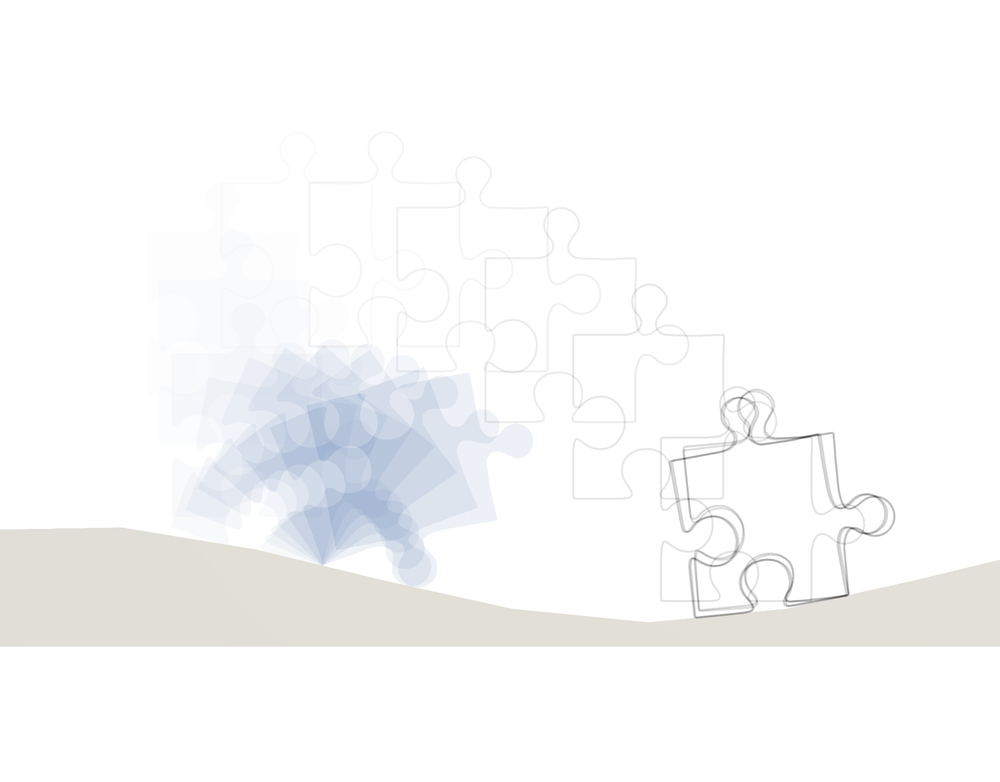}\hfill
\parbox{.48\linewidth}{\centering Initial trajectory.}\hfill
\parbox{.48\linewidth}{\centering Iteration 3.}\hfill
\includegraphics[width=.48\linewidth]{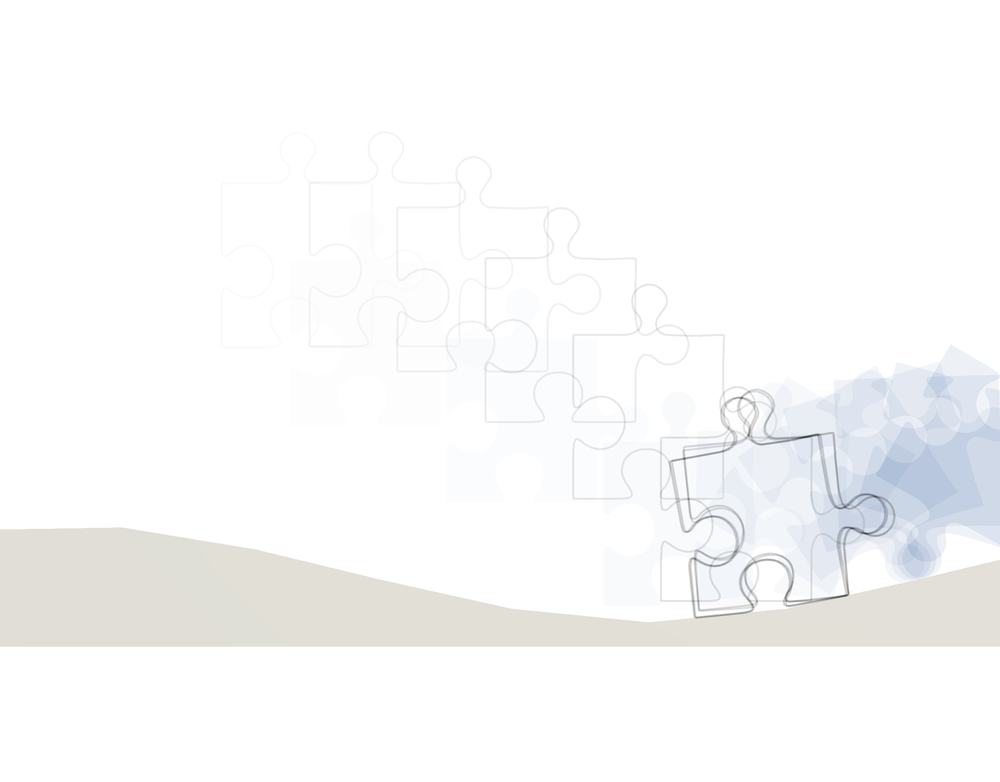}\hfill
\includegraphics[width=.48\linewidth]{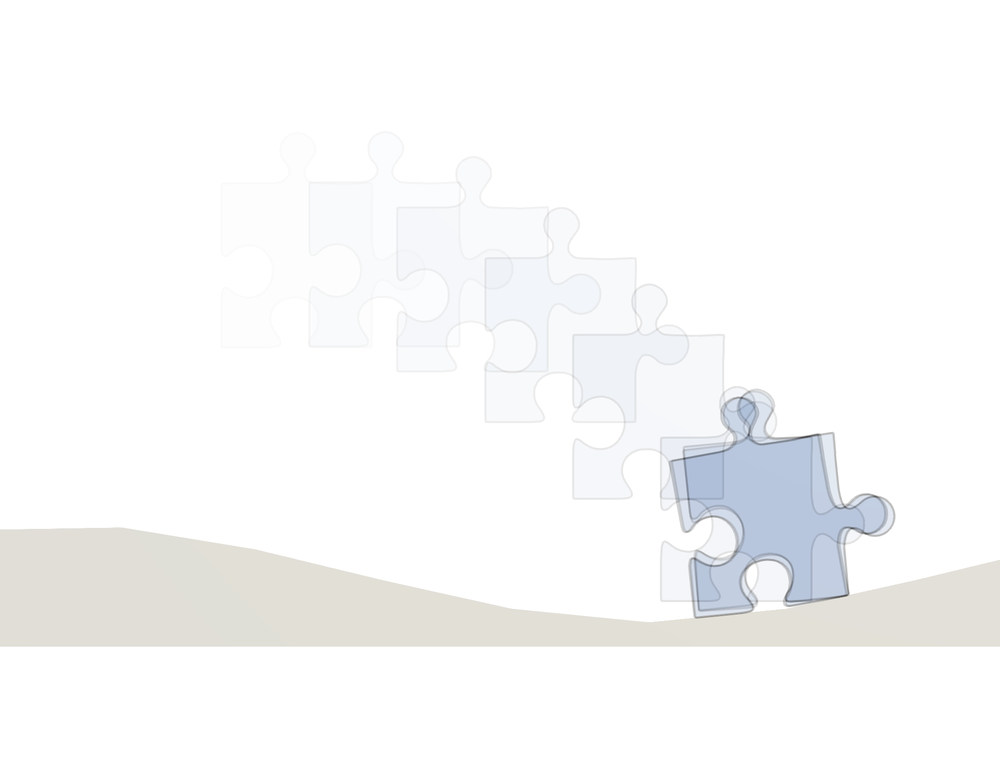}\hfill
\parbox{.48\linewidth}{\centering Iteration 5}.\hfill
\parbox{.48\linewidth}{\centering Optimized result}\hfill
\caption{\textbf{Transient: Puzzle Piece.} Optimizing the initial velocity of a bouncing puzzle. Target is shown as a black outline while the trajectory being optimized is blue.}
\label{fig:bouncing-puzzle}
\Description{}
\end{figure}
\paragraph{Transient: Puzzle Piece}
We synthesise a trajectory using a forward simulation, and we then perturb the initial conditions and try to reconstruct them minimizing \eqref{eq:obj:targetdeformation}, with an additional integration over time (Figure \ref{fig:bouncing-puzzle}). %
The puzzle piece uses a Neo-Hookean material.

\paragraph{Transient: Throw Bunny}
We use our solver to optimize the throw (initial velocity) of a bunny to hit and displace a second bunny into the prescribed circle (Figure \ref{fig:bunny-pool}), minimizing \eqref{eq:obj:centermasstrajectory}. This example involves complex contact between the bunnies and the pool table, and also friction forces slowing down the sliding after contact.

\paragraph{Transient: Colliding Tentacles}
\begin{figure}\centering\footnotesize
\includegraphics[width=\linewidth]{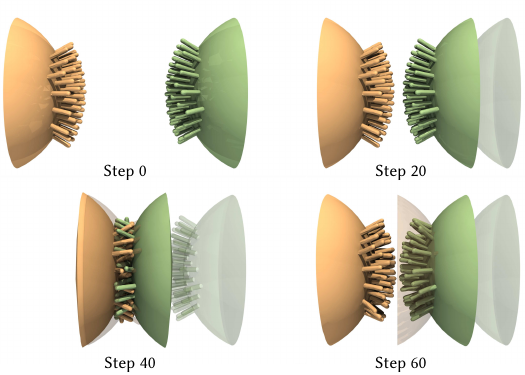}\par
\caption{\textbf{Transient: Colliding Tentacles.} We optimize the initial velocity so that the mass trajectory matches the reference simulation. The faded view represents the initial configuration. The optimized simulation matches exactly with the reference simulation.}
\label{fig:tentacles}
\Description{}
\end{figure}
We optimize the initial velocity of the green object in the scene of two colliding half spheres with tentacles (Figure \ref{fig:tentacles}), minimizing the difference of the mass trajectory with respect to a trajectory obtained from a reference simulation \eqref{eq:obj:centermasstrajectory}. Our method manages to resolve the complex contact between the soft tentacles.

\subsection{Material Optimization}\label{sec:res:mat_opt}

Next, we look at material optimization problems, where our differentiable simulator is used to estimate the material properties of an object from observations of its displacement.

\begin{figure}\centering\footnotesize
\parbox{.48\linewidth}{\includegraphics[width=\linewidth]{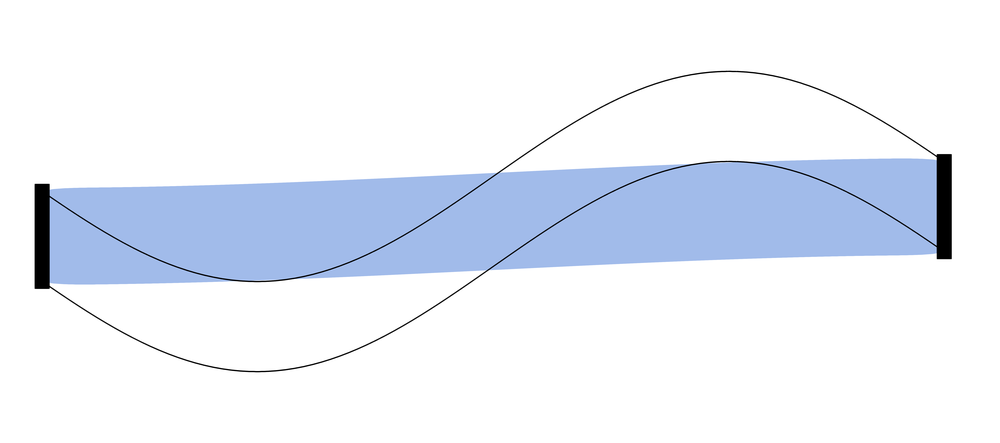}}\hfill
\parbox{.48\linewidth}{\includegraphics[width=\linewidth]{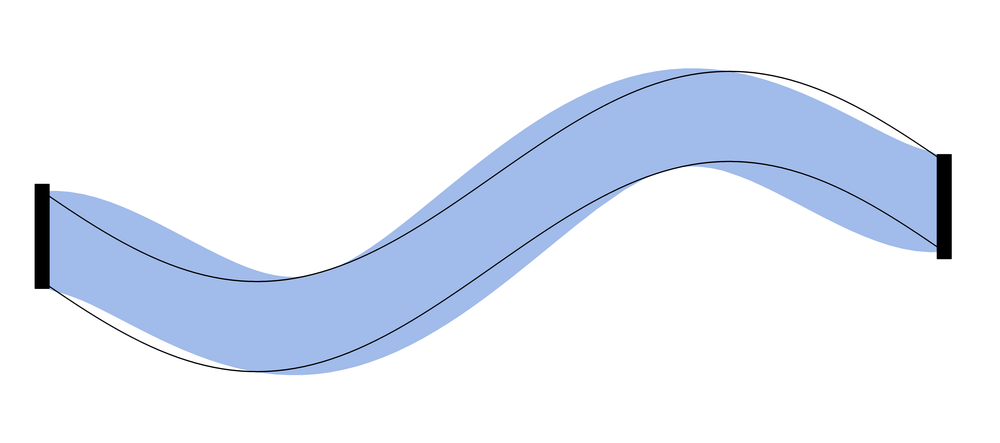}}\par
\parbox{.48\linewidth}{\centering Initial displacement.}\hfill
\parbox{.48\linewidth}{\centering Optimized displacement.}\\[2ex]
\parbox{.48\linewidth}{\centering
\includegraphics[width=.88\linewidth]{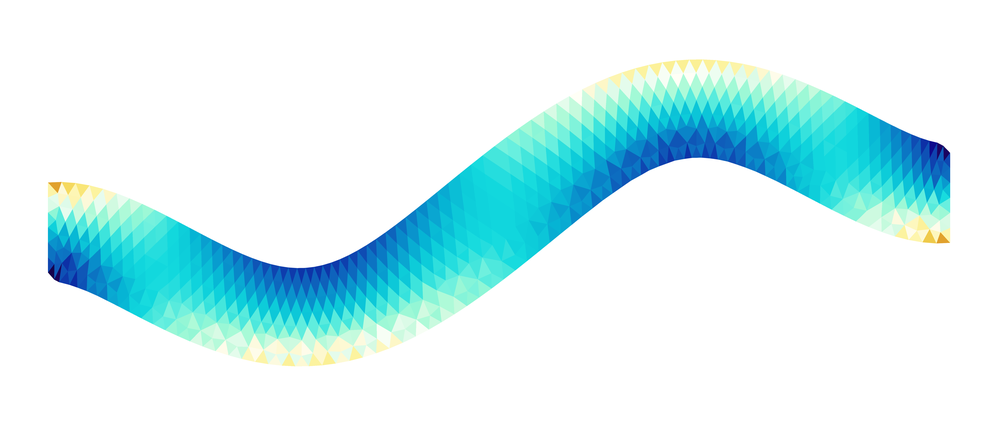}
\includegraphics[width=.1\linewidth]{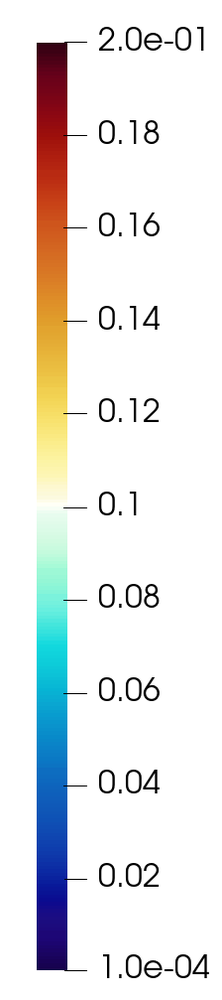}
}\hfill
\parbox{.48\linewidth}{\centering
\includegraphics[width=.88\linewidth]{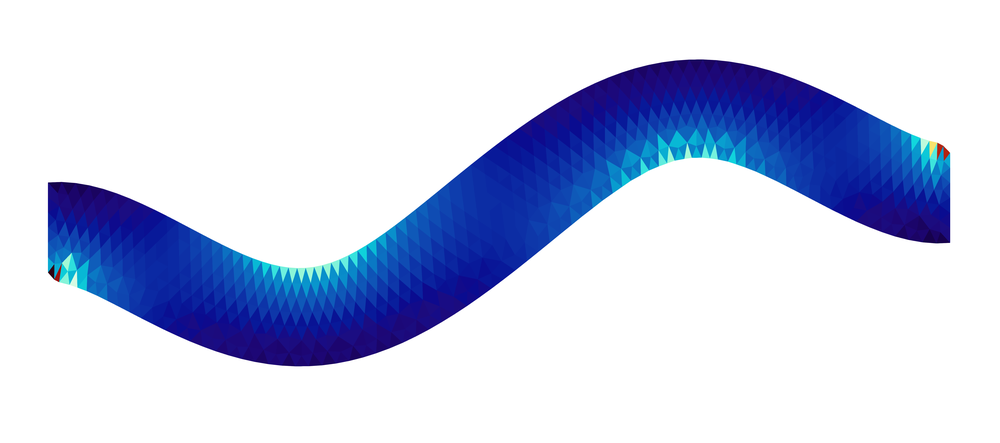}
\includegraphics[width=.1\linewidth]{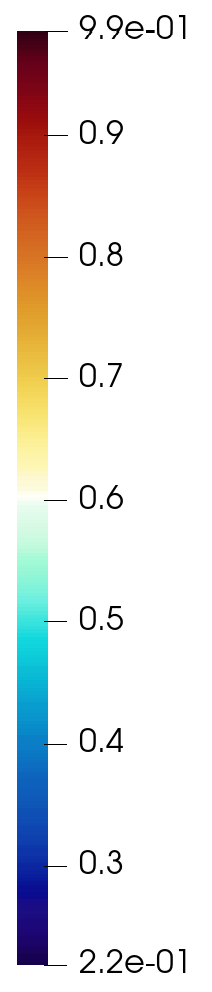}
}
\parbox{.48\linewidth}{\centering Optimized $E$ pattern.}\hfill
\parbox{.48\linewidth}{\centering Optimized $\nu$ pattern.}\hfill
\caption{\textbf{Static: Sine.} Optimized material parameters to obtain a displacement  (blue surface) in $y$-direction similar to a sine function for a linear material model (wire-frame).}
\label{fig:sine}
\Description{}
\end{figure}
\paragraph{Static: Sine}
We optimize the material of a bar to match the shape of a sine function (wire-frame) when Dirichlet boundary conditions are applied at its ends \eqref{eq:obj:targetdeformation}. The rest shape of this bar is a rectangle $[-4,4]\times[-0.3,0.3]$, the left and right surfaces are fixed by Dirichlet boundary condition of $u_y=0.7\sin(x + u_x)$ and $u_x=-sign(x)$, and no body force is applied. Figure~\ref{fig:sine} shows that deformed bar is aligned with a sine function.

\begin{figure}\centering\footnotesize
\parbox{.32\linewidth}{\includegraphics[width=\linewidth]{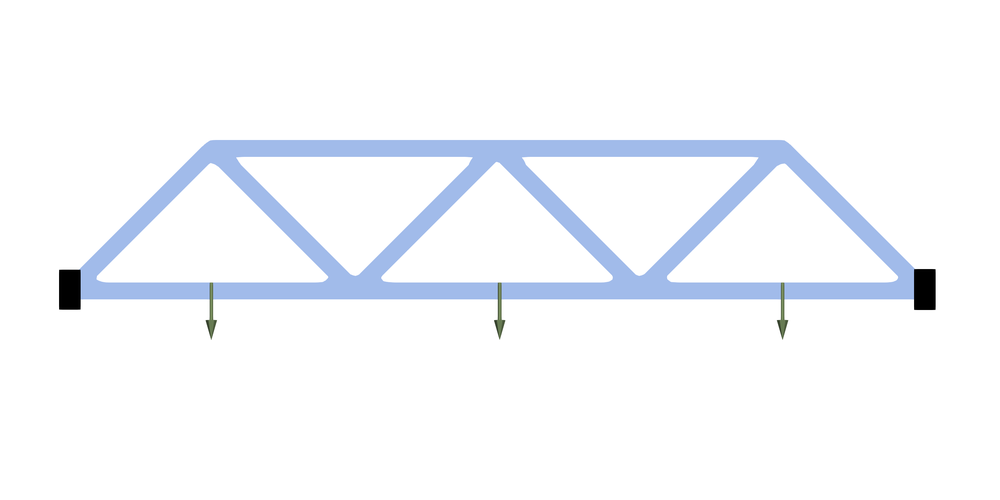}}\hfill
\parbox{.32\linewidth}{\includegraphics[width=\linewidth]{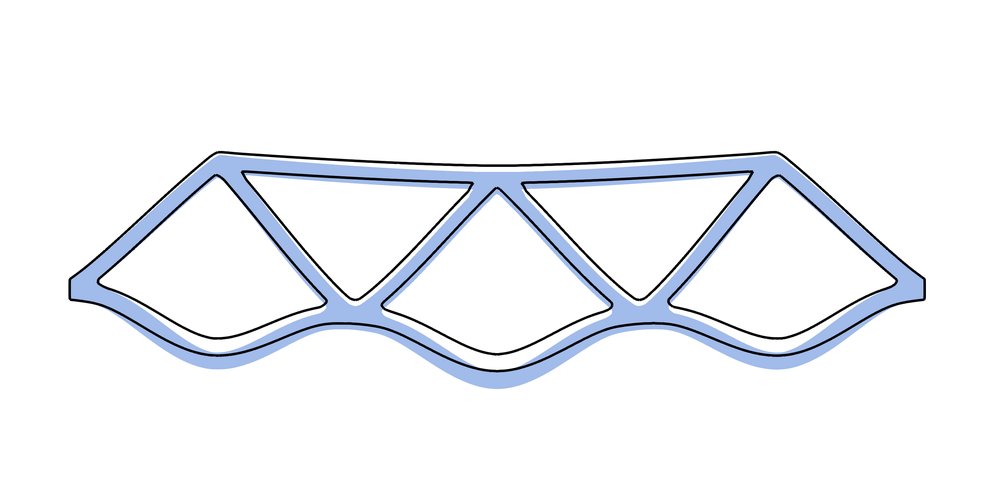}}\hfill
\parbox{.32\linewidth}{\includegraphics[width=\linewidth]{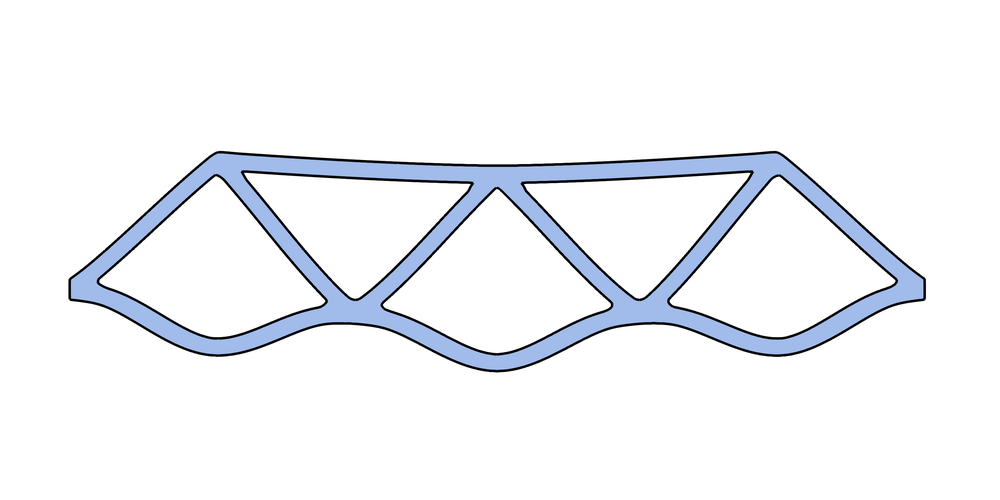}}\par
\parbox{.32\linewidth}{\centering Problem setup.}\hfill
\parbox{.32\linewidth}{\centering Initial displacement.}\hfill
\parbox{.32\linewidth}{\centering Optimized displacement.}\par
\parbox{.45\linewidth}{\centering
\parbox{.88\linewidth}{\includegraphics[width=\linewidth]{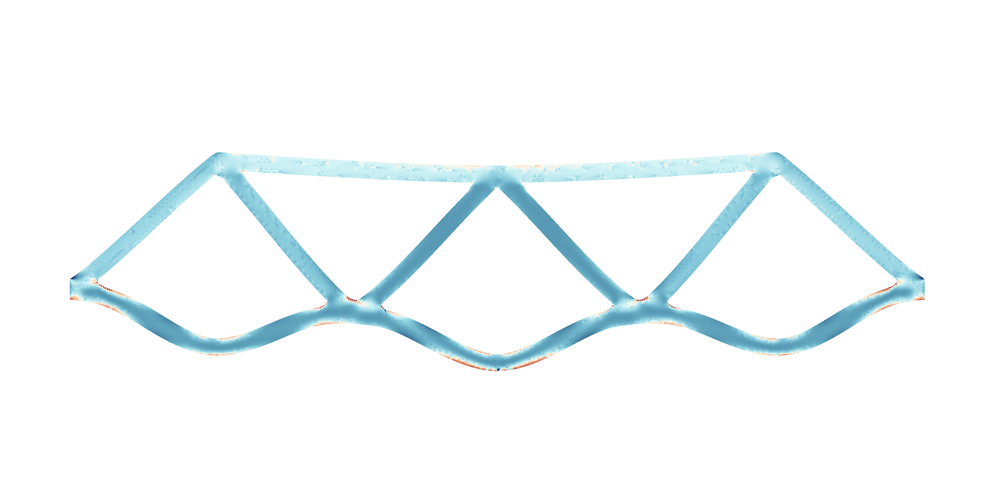}}\hfill
\parbox{.1\linewidth}{\includegraphics[width=\linewidth]{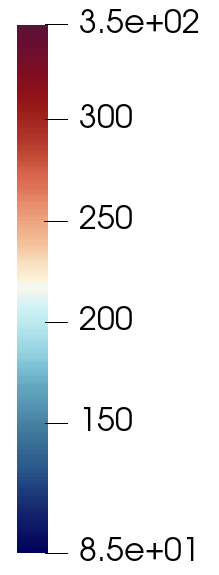}}\par
Optimized $E$ pattern.
}\hfill
\parbox{.45\linewidth}{\centering
\parbox{.88\linewidth}{\includegraphics[width=\linewidth]{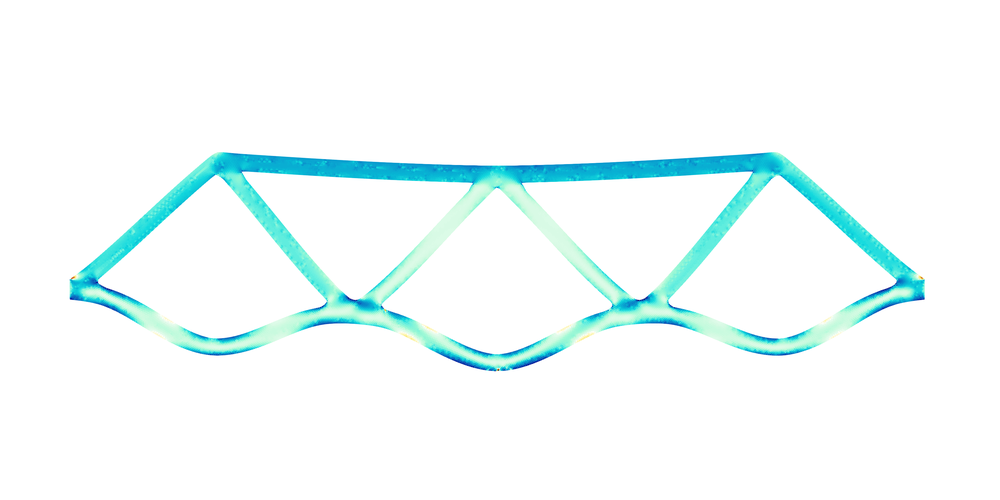}}\hfill
\parbox{.1\linewidth}{\includegraphics[width=\linewidth]{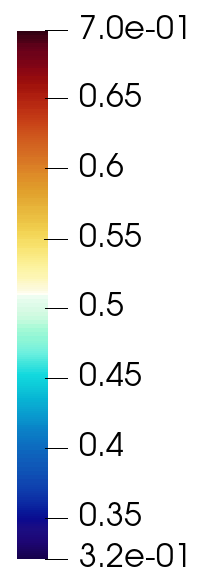}}
Optimized $\nu$ pattern.
}
\caption{\textbf{Static: Bridge.} Optimization of the materials of a bridge (blue surface) to match a forward simulation (wire-frame).}
\label{fig:bridge-mat}
\Description{}
\end{figure}
\paragraph{Static: Bridge}
We assign material parameters $\lambda=160, \mu=80$ to a bridge shape and run a linear forward simulation to obtain the target displacement $u^\star$ (Figure~\ref{fig:bridge-mat} in gray), using the same set of boundary conditions as Figure \ref{fig:bridge-shape}. We initialize the optimization using uniform material $\lambda=100, \mu=50$ and minimize \eqref{eq:obj:targetdeformation}, successfully recovering $\lambda, \mu$ from $u^\star$.

\begin{figure}\centering\footnotesize
\includegraphics[width=\linewidth]{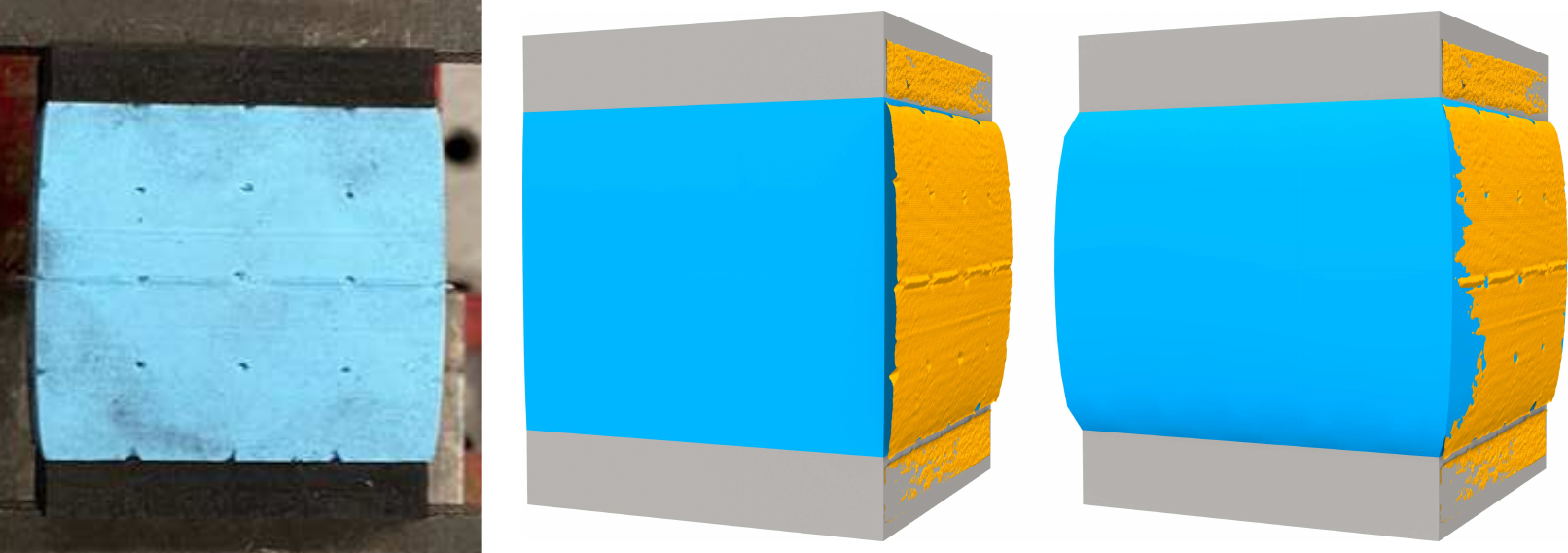}\\
\parbox{0.31\linewidth}{\centering Physical Experiment Setup.}\hfill
\parbox{0.34\linewidth}{\centering Initial shape.}\hfill
\parbox{0.34\linewidth}{\centering Optimized.}
\caption{\textbf{Static: Cube.} Material optimization (blue) to match real data (orange).}
\label{fig:physical-cube}
\Description{}
\end{figure}
\paragraph{Static: Cube}
We set up a physical experiment with a silicon rubber cube compressed by a vise. The deformation is acquired using an HP 3D scanner, and a set of marker points is manually extracted from the scan. We minimize  \eqref{eq:obj:targetdeformation} to find the material parameters which produce the observed displacements. We found that the material parameter that leads to the smallest error is $\nu=0.4817$ (Young's modulus does not affect its deformation in this setting) and the L2 error in markers position is  $\qty{3.85e-3}{\m}$.

\begin{figure}\centering\footnotesize
\includegraphics[width=0.32\linewidth]{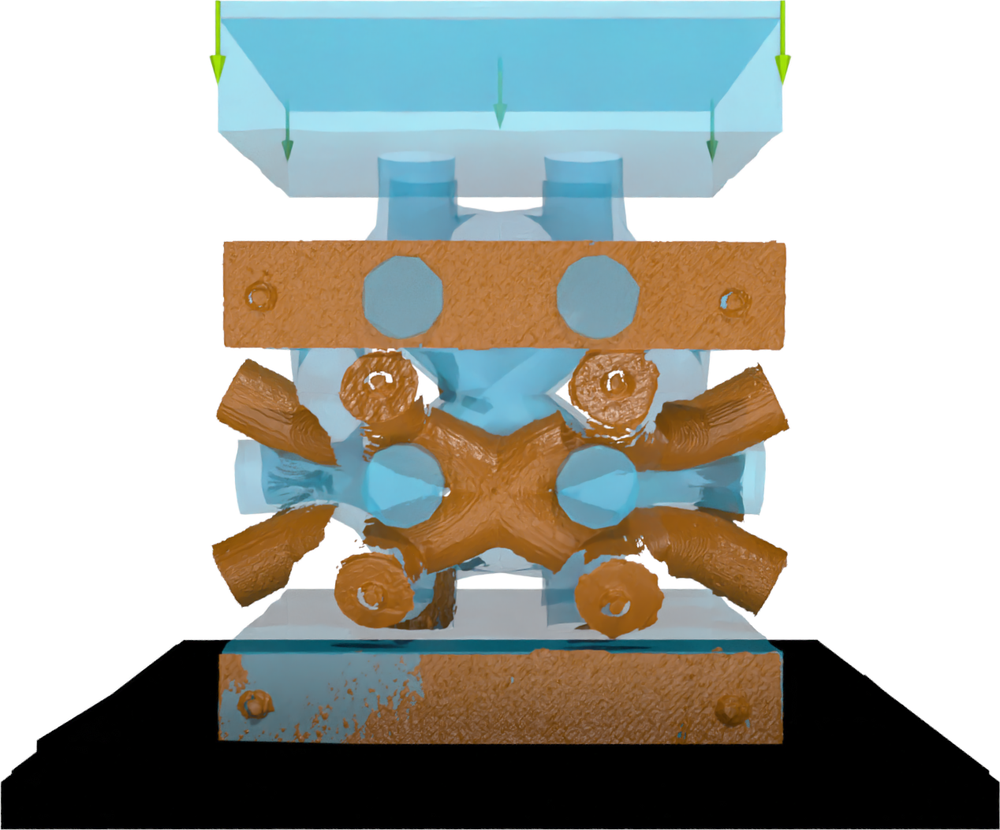}\hfill
\includegraphics[width=0.32\linewidth]{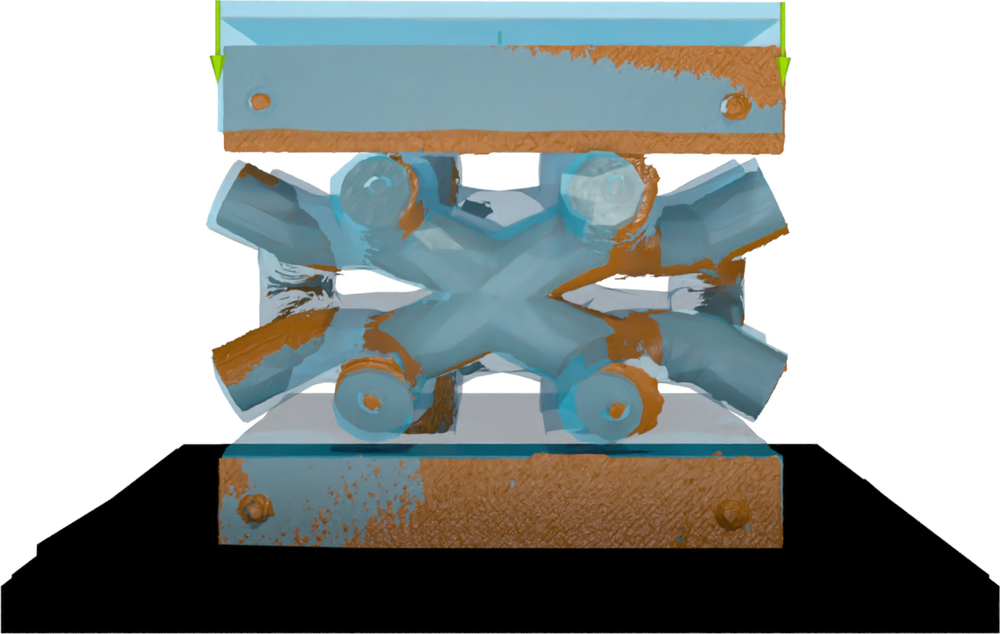}\hfill
\includegraphics[trim={700 0 700 0},clip=True,width=0.32\linewidth]{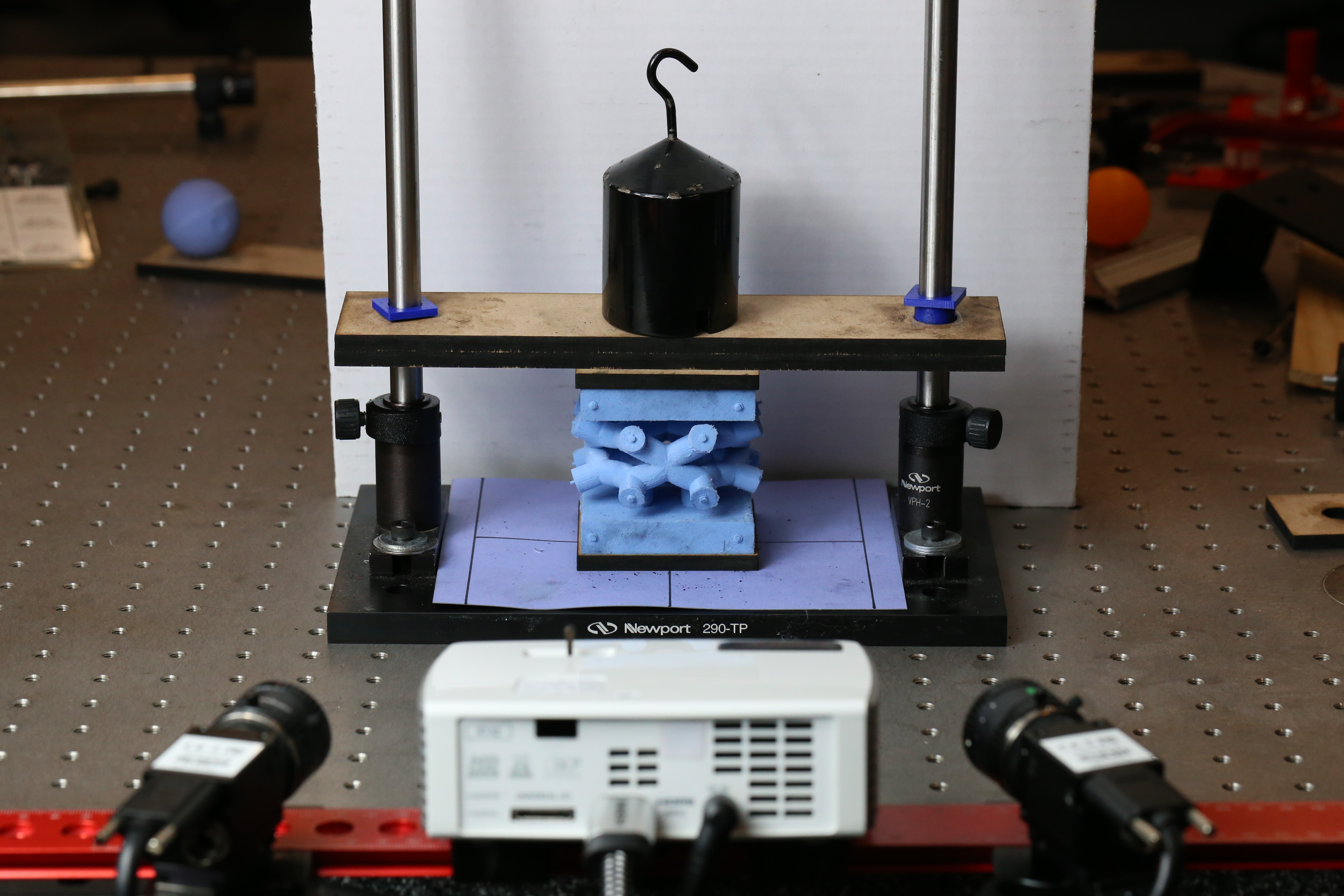}\\
\parbox{.32\linewidth}{\centering Initial.}\hfill
\parbox{.32\linewidth}{\centering Optimized.}\hfill
\parbox{.32\linewidth}{\centering Physical Experiment Setup.}
\caption{\textbf{Static: Micro-Structure}
Material optimization of a complex microstructure in a deformed state with contact (blue) to match real data (orange).
}
\label{fig:physical-microstructure}
\Description{}
\end{figure}
\paragraph{Static: Micro-Structure}
We repeat the same experiments with the complex geometry of a micro-structure tile from \cite{panetta2017}. This is a challenging example, as the micro-structure beams come in contact after compression, and physical models without self-contact handling may lead to penetration. The optimization is initialized with $E=\qty{e6}{\Pa}$ and $\nu=0.3$, and our solver converges to a material properties of $E=\qty{2.27e5}{\Pa}$ and $\nu=0.348$ with a L2 error on the markers of $\qty{8.8e-3}{\m}$. %

\begin{figure}\centering\footnotesize
\includegraphics[width=.48\linewidth]{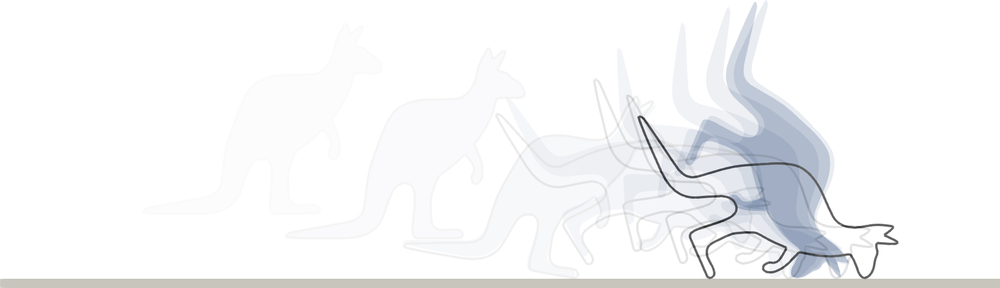}\hfill
\includegraphics[width=.48\linewidth]{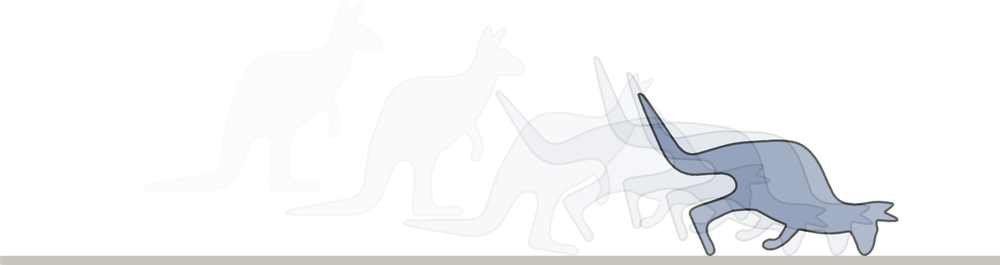}\par
\parbox{.48\linewidth}{\centering Initial displacement.}\hfill
\parbox{.48\linewidth}{\centering Optimized displacement.}
\caption{\textbf{Transient: Kangaroo.} Non-linear transient simulation of a kangaroo (blue surface) bouncing on a plane to match a target shape (wire-frame).}
\label{fig:kangaroo}
\Description{}
\end{figure}
\paragraph{Transient: Kangaroo}
As an example of reconstruction of material parameters from a transient simulation, we run a forward simulation to obtain a transient non-linear target displacement. Then we minimize \eqref{eq:obj:targetdeformation} to reconstruct the material parameters (Figure~\ref{fig:kangaroo}). The initial material parameters are $E=\qty{3e6}{\Pa}$ and $\nu=0.5$, and the target material parameters are $E=\qty{e7}{\Pa}$ and $\nu=0.3$.

\begin{figure}\centering\footnotesize
\parbox{0.495\linewidth}{
\centering
\includegraphics[width=\linewidth]{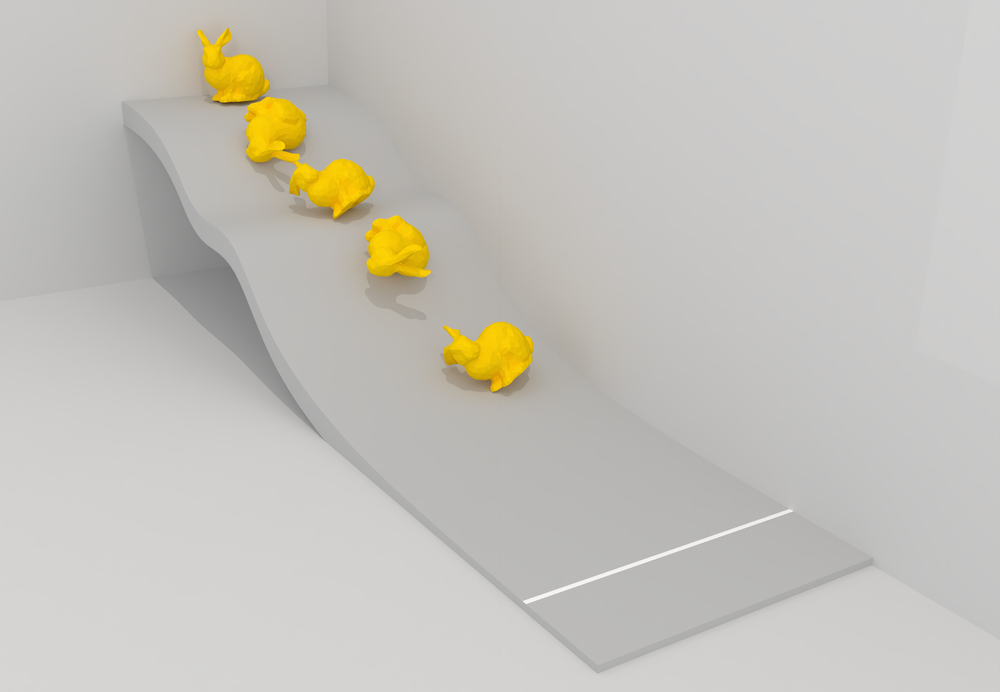}\par
Initial guess.}
\parbox{0.495\linewidth}{
\centering
\includegraphics[width=\linewidth]{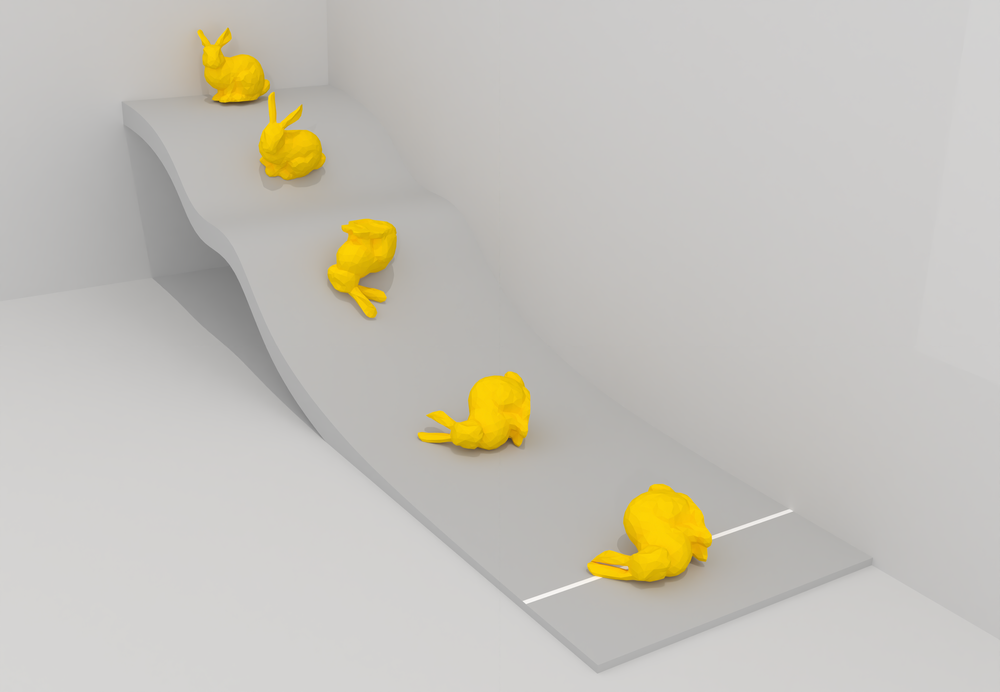}\par
Optimized result.}
\caption{\textbf{Transient: Sliding Bunny.} Optimize the friction coefficient so that the bunny can reach the white line at $t=2$.}
\label{fig:bunny}
\Description{}
\end{figure}
\paragraph{Transient: Sliding Bunny}
We use our solver to optimize the friction coefficient to ensure that the bunny is on the white line at time $t=2$. The initial friction coefficient is $\friccoeff=0.5$, and the optimized friction coefficient is $\friccoeff=0.0974$ (Figure ~\ref{fig:bunny}). This example involves complex self-contact and friction of the bunny with the floor.

\begin{figure}\centering\footnotesize
\includegraphics[width=\linewidth]{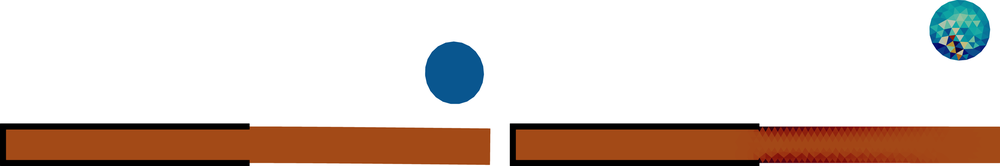}\\
\parbox{.495\linewidth}{\centering Initial $E$.}\hfill
\parbox{.495\linewidth}{\centering Optimized $E$.}\\[1em]
\parbox{\linewidth}{\centering $E = $ \includegraphics[width=0.7\linewidth]{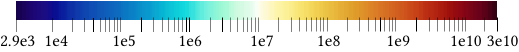}}
\includegraphics[width=\linewidth]{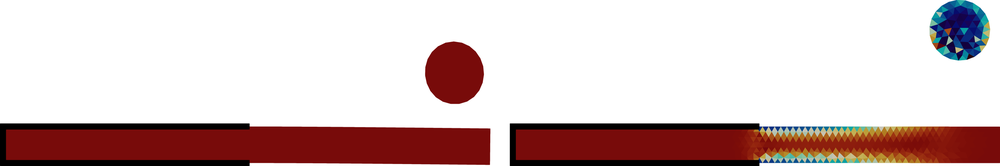}
\parbox{.495\linewidth}{\centering Initial $\nu$.}\hfill
\parbox{.495\linewidth}{\centering Optimized $\nu$.}\\[1em]
\parbox{\linewidth}{\centering $\nu = $ \includegraphics[width=0.7\linewidth]{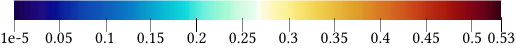}}
\caption{\textbf{Transient: Bouncing Ball.} Material optimization to increase the bouncing height.}
\label{fig:ball-h}
\Description{}
\end{figure}
\paragraph{Transient: Bouncing Ball}
We show that the height of the bounce of a ball can be optimized by changing the material parameters (Figure \ref{fig:ball-h}).
Initial material parameters for the ball and plank were $E = \qty{e5}{\Pa}$, $\nu = 0.48$ and $E = \qty{e9}{\Pa}$, $\nu = 0.48$, respectively and the elasticity model used was NeoHookean. Note that we added a smoothing term to the optimization to increase smoothness in the material parameters.

\begin{figure}\centering\footnotesize
\parbox{.46\linewidth}{
\centering\includegraphics[trim={200 100 300 200},clip=True,width=\linewidth]{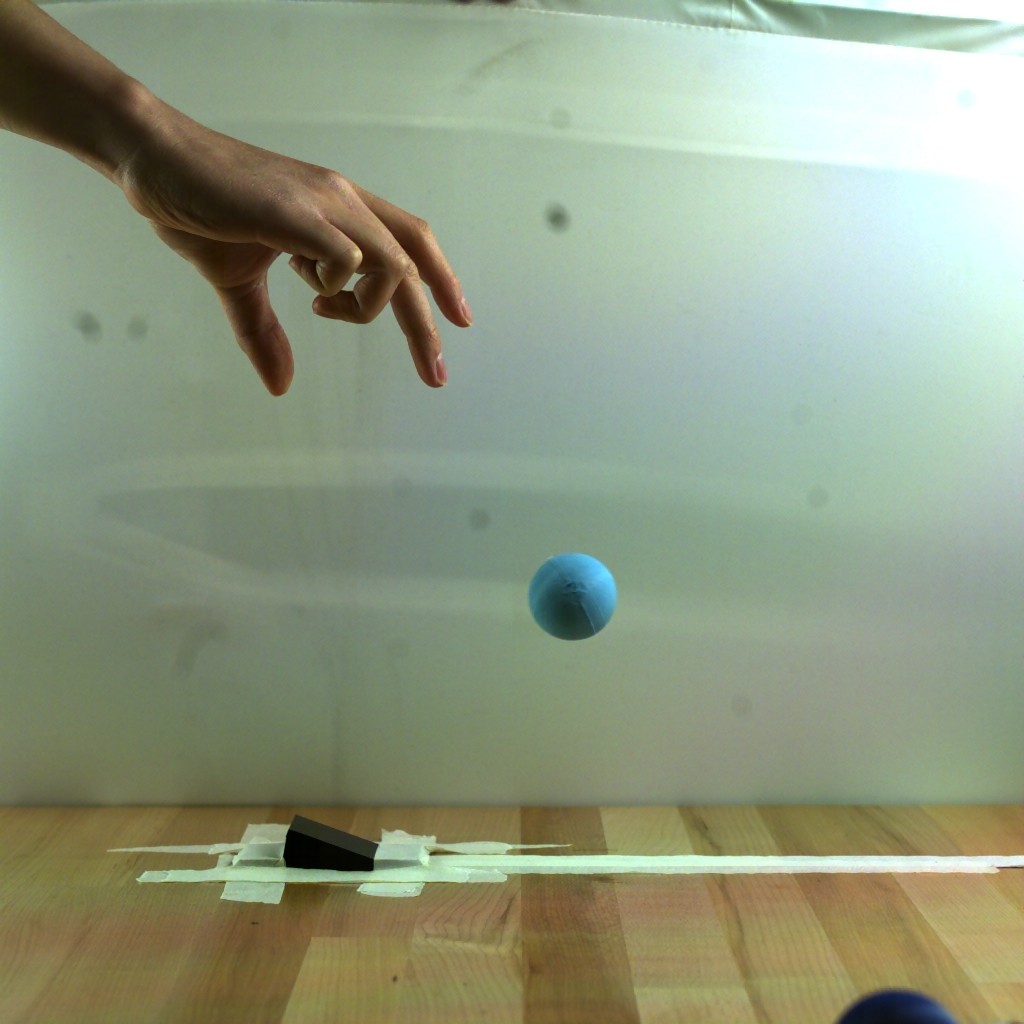}
Physical experiment.
}\hfill
\parbox{0.48\linewidth}{\centering
\includegraphics[width=\linewidth]{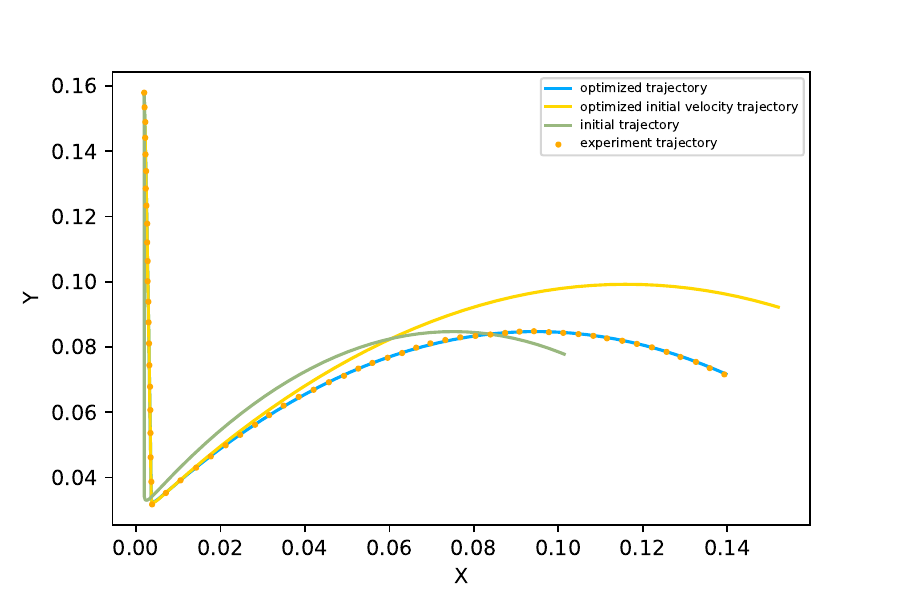}
$xy$ coordinates of the barycenter of the ball over time.
\includegraphics[width=\linewidth]{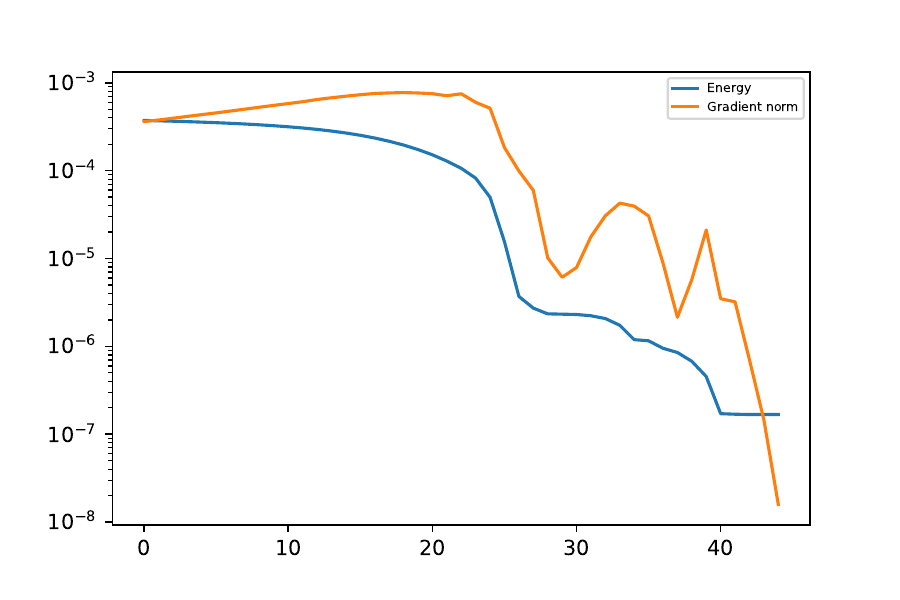}
Energy and gradient over the optimization iterations.
}
\parbox{.8\linewidth}{\centering
\includegraphics[width=0.75\linewidth]{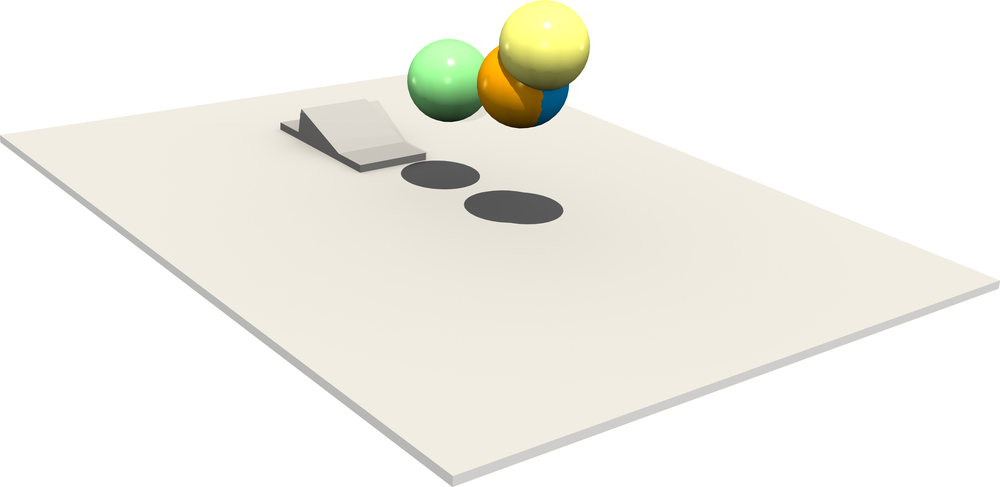}\\
Initial guess (green), initial velocity optimization (yellow), material optimization (blue), and experimental data (orange).
}\par
\caption{\textbf{Transient: Physical Experiment Bouncing Ball.} Optimize the material and initial velocity of the ball to match the observed physical result.}
\label{fig:bounce-ball}
\Description{}
\end{figure}
\paragraph{Transient: Physical Experiment Bouncing Ball}
We show that we can optimize for the initial velocity, material parameters, friction coefficient, and damping parameters of a silicone rubber ball bouncing on an incline, using trajectory data from a physical experiment. The real-world dynamics of the ball are captured using a high-speed camera and used to formulate a functional based on  \eqref{eq:obj:centermasstrajectory}, which penalizes differences between the observed and simulated barycenter of the ball. The material model used is NeoHookean and we match initial conditions by optimizing for them using the observed barycenters of the ball before it hits the ground.

\subsection{Comparisons}\label{sec:res:comparisons}

Finally, we compare our method with existing methods in terms of solution quality, contact handling, and efficiency. Due to stability issues, different time step sizes are chosen for different methods so that no visible artifacts appear in the forward simulations. See Table~\ref{Table:comparisons} for statistics. We also compare our method with finite difference and automatic differentiation on PolyFEM \cite{polyfem} in terms of efficiency.

\begin{table}[htb]
\caption{\textbf{Comparisons.} Columns from left to right are examples, methods, degree of freedom of the simulation, time step size, peak memory (\unit{\mega\byte}), running time of the simulation (\unit{\s}), and running time of computing gradients (\unit{\s}).}
\resizebox{.96\linewidth}{!}{
\begin{tabular}{llccccc}
\toprule
Example & Method & Dofs & dt & Memory & Solve time & Grad time \\
\multirow{3}{*}{Armadillo} & DiffPD & 36699 & $3\times 10^{-3}$ & 1246 & 37.9 & 131.2 \\
& GradSim & 36699 & $1.5\times 10^{-5}$ & 17164 & 167.2 & N/A \\
& \textbf{Ours} & 36699 & $6\times 10^{-3}$ & 2068 & 220.6 & 14.1 \\ \hline
\multirow{3}{*}{Hilbert Cube} & DiffPD & 4050 & $5\times 10^{-2}$ & 240 & 1.555 & 2.12 \\
& GradSim & 4050 & $5\times 10^{-4}$ & 1323 & 11.1 & 27.7 \\
& \textbf{Ours} & 4050 & $5\times 10^{-2}$ & 1599 & 73.2 & 1.73\\ \hline
\multirow{2}{*}{Billiards} & DiffPD & 978 & $2.5\times 10^{-3}$ & 226 & 11.3 & 10.5 \\
& \textbf{Ours} & 978 & $2.5\times 10^{-3}$ & 190 & 66.2 & 3.1\\
\bottomrule
\end{tabular}
}
\label{Table:comparisons}
\end{table}

\begin{figure}\centering\footnotesize
\includegraphics[width=\linewidth]{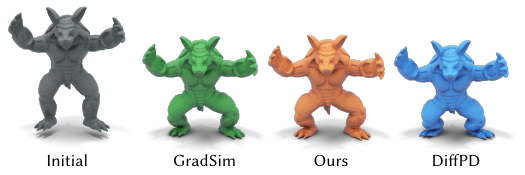}
\caption{\textbf{Transient: Armadillo.} Simulation of dropping an Armadillo onto the floor.}
\label{fig:armadillo}
\Description{}
\end{figure}

\paragraph{Transient: Armadillo}
We simulate dropping the Armadillo (using the same material parameters) onto a fixed plane (\cref{fig:armadillo}) and compute the material derivatives with our method, DiffPD~\cite{du2021diffpd} and GradSim~\cite{gradsim}. The results of GradSim and our method are similar, which is expected as both methods are based on a finite element formulation with a similar material model. However, the backward solve of GradSim encounters NAN and fails to compute the gradient, likely due to the instability from its semi-implicit time integration or the non-differentiable contact model (Its contact force is only $C^0$). DiffPD creates a result that is different from the two, likely due to the use of a different elastic model.

\begin{figure}\centering\footnotesize
\includegraphics[width=\linewidth]{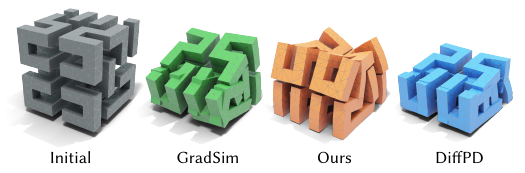}
\caption{\textbf{Transient: Hilbert Cube.} Simulation of dropping a Hilbert cube onto the floor.}
\label{fig:hilbert_cube}
\Description{}
\end{figure}

\paragraph{Transient: Hilbert Cube}
In this example, we simulate the drop of a Hilbert cube (Figure ~\ref{fig:hilbert_cube}), compute the material derivatives, and compare our method with GradSim and DiffPD. Although GradSim and DiffPD can resolve the planar contact, they do not support  self-collision, resulting in visible and physically implausible self-intersections. In contrast, the solution computed by our method has no self-intersections or inverted elements.

\begin{figure}\centering\footnotesize
\includegraphics[width=\linewidth]{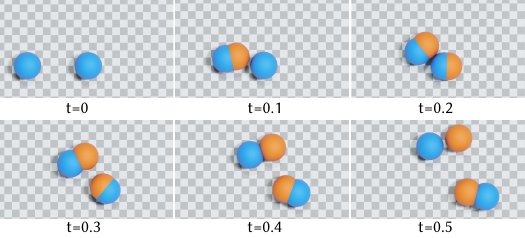}
\caption{\textbf{Transient: Billiards.} The ball on the left with initial velocity $(\cos(\frac{15}{180}\pi),\sin(\frac{15}{180}\pi))$ hits the ball on the right, simulated with our method (orange) and DiffPD (blue).}
\label{fig:billiards}
\Description{}
\end{figure}

\begin{figure}\centering\footnotesize
\includegraphics[width=\linewidth]{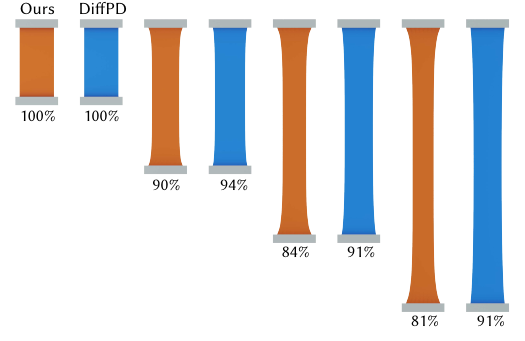}
\caption{\textbf{Static: Tensile Test.} Stretch a 3D bar up to 300\% strain with our method (orange) and DiffPD (blue). The thickness of the deformed bar is shown as a percentage with respect to the initial thickness.}
\label{fig:tensile}
\Description{}
\end{figure}

\paragraph{Static: Tensile Test}
We perform the tensile testing on a bar of size $0.16\unit{\meter}\times 0.08\unit{\meter}\times 0.08\unit{\meter}$, with Poisson's ratio $\nu=0.3$ and Young's modulus $E=10^3$\unit{\Pa}, using both our method and DiffPD. We refine the meshes used in both methods until the results become stable and show the converged results. Since there is no contact, our method is equivalent to the standard FEM with Neo-Hookean material. Since the material model used in DiffPD is an approximation of the hyper-elastic model designed for high efficiency, there is a noticeable difference between DiffPD and the standard model when the deformation is large (\cref{fig:tensile}). We favor using the Neo-Hookean material model, as we are interested in accurately capturing large physical deformations.

\paragraph{Transient: Billiards}
In this example we reproduce the billiards example in ~\cite{du2021diffpd} (\cref{fig:billiards}), and compute the material derivatives. Since GradSim does not support collisions between spheres (or between meshes), we restrict the comparison to DiffPD.

Although the same mesh is used in both methods, there is a significant difference in the contact handling: Our method detects the collision between the discrete meshes, while DiffPD uses the averaged sphere center and radius to detect the collision between spheres. While more efficient, the DiffPD solution is customized for this example, while our approach works on arbitrary geometries. Due to the difference in both the elastic model (\cref{fig:tensile}) and contact handling, the results are different. Our forward simulation is 6 times slower than DiffPD.

\paragraph{Finite Difference}
To evaluate the correctness and efficiency of our method, we compute the gradient using finite differences and compare it with our method. We use the central difference scheme, which requires solving the forward problem for $2n$ times if the parameter dimension is $n$. As a result, the finite difference is approximately twice as expensive as the forward solve, while the time of our method is negligible (\cref{Table:finite_diff}).

\begin{table}[htb]
\caption{\textbf{Finite Difference.} Columns from left to right are examples, dimension of the design parameters, running time of the simulation (\unit{\s}), running time of the adjoint method (\unit{\s}), and running time of the finite difference (\unit{\s}). The accuracy is the relative error between the finite difference and the adjoint method.}
\resizebox{.96\linewidth}{!}{
\begin{tabular}{lccccc}
\toprule
Example & Dim & Solve time & Grad time & FD time & Accuracy \\
\midrule
Shock Protection (Figure~\ref{fig:shock-protect}) & 24 & 1273 & 131.8 & 63502 & $1.12\times 10^{-2}$\\
Micro-Structure (Figure~\ref{fig:physical-microstructure}) & 2 & 42.1 & 0.089 & 172.1 & $2.25\times 10^{-6}$ \\
Sliding Bunny (Figure~\ref{fig:bunny}) & 1 & 544.6 & 1.74 & 1092 & $6.35\times 10^{-10}$ \\
\bottomrule
\end{tabular}
}
\label{Table:finite_diff}
\end{table}

\paragraph{Automatic Differentiation (AD)} While it is impossible to transform the linear solver to AD form for large problems (see Section~\ref{sec:related:compute-grad}), we could use AD to compute the terms needed in the adjoint method. To evaluate the difference in performance between AD and analytic derivation, we focus our investigation on the local assembly of the elastic force vector into AD form \cite{autodiff} to compute the $\vadj^T \dpar \pde^k$ in Equation~(\ref{eq:static-adjoint}). We solve the static NeoHookean PDE on a tetrahedral mesh with 4670 vertices and using linear FE bases. Our method of computing $\vadj^T \dpar \pde^k$ takes 0.0174 seconds, while AD takes 0.247 seconds. The forward nonlinear solve takes 2.85 seconds, and the backward adjoint solve takes 0.0212 seconds. Given this experiment, we opted to spend the additional effort in analytically deriving the adjoint terms to avoid this unnecessary additional computational cost.  In our setting with expensive implicit solves, the cost of computing the adjoint terms is a small overhead on the whole optimization, and computing the derivatives with AD makes the implementation simpler and makes it easier to switch the material models.
However, we found that for more complex contact models, not included in this paper, the cost of AD can still be significant, and in settings in which forward solves can be done explicitly or semi-implicitly, the computational costs are distributed differently. 
\section{Concluding Remarks}
\label{sec:concluding}
We introduced a generic, robust, and accurate framework for PDE-constrained optimization problems involving elastic deformations of multiple objects with contact and friction forces. Our framework supports customizable objective functions and allows for the optimization of functionals involving the geometry of the objects involved, material parameters, contact/friction parameters, and boundary/initial conditions. %

There are several limitations in our work. First, our derivation is limited to hyper-elastic and visco-elastic materials. We don't support simulating shells (cloth), plastic materials, fluid, etc. Second, rigid and articulated objects, which are widely used in robotics, are not supported. Although it can be approximated by very large stiffness in our framework, the simulation is much slower than rigid body simulations. Third, our forward simulation, though robust, is less efficient than previous works like \cite{du2021diffpd, gradsim} in simple scenes (Section \ref{sec:res:comparisons}).

We believe the benefits of our analytic derivation of the adjoint system (efficiency, generality, guarantee of convergence under refinement) outweigh its downsides (complexity of derivation, difficulty in implementation, and requirement of an explicit FE mesh). We plan to extend our approach to a wider set of PDE-constrained problems and to further optimize it for common use cases in material design and robotics. In particular, we would like to explore the following directions:
\begin{enumerate}
\item Add support for periodic boundary conditions, which are required for the design of micro-structure families \cite{Tozoni2020}.
\item Add support for rigid and articulated objects (i.e. allow the material stiffness to be infinite). We plan to incorporate the IPC formulation introduced in \cite{Ferguson:2021:RigidIPC} to improve performance in design problems involving rigid objects.
\item Many robotics problems involve the manipulation of plastic objects or interaction with fluids: adding support for additional physical models will widen the applicability of our simulator.
\item We designed our system to provide accurate modeling of elastic, contact, and friction forces, as the majority of PDE-constrained applications require accurate simulations faithfully reproducing the behavior observable in the real works. However, there are applications where this is not necessary, and in these cases, it would be possible to either use simpler elastic models or reduce the accuracy of the collision/friction forces by using proxy geometry. This is commonly done in graphics settings, and it would be interesting to add this option to our system to accelerate its performance.
\end{enumerate}

\begin{acks}
This work was supported in part through the NYU IT High Performance Computing resources, services, and staff expertise.
This work was also partially supported by the NSF CAREER award under
Grant No. 1652515, the NSF grants OAC-1835712, CHS-1908767, CHS-1901091, IIS-2313156, a Sloan Fellowship, and a gift from
Adobe Research.
\end{acks}

\bibliographystyle{ACM-Reference-Format}
\bibliography{99-biblio.bib}


\begin{thebibliography}{84}


\ifx \showCODEN    \undefined \def \showCODEN     #1{\unskip}     \fi
\ifx \showDOI      \undefined \def \showDOI       #1{#1}\fi
\ifx \showISBNx    \undefined \def \showISBNx     #1{\unskip}     \fi
\ifx \showISBNxiii \undefined \def \showISBNxiii  #1{\unskip}     \fi
\ifx \showISSN     \undefined \def \showISSN      #1{\unskip}     \fi
\ifx \showLCCN     \undefined \def \showLCCN      #1{\unskip}     \fi
\ifx \shownote     \undefined \def \shownote      #1{#1}          \fi
\ifx \showarticletitle \undefined \def \showarticletitle #1{#1}   \fi
\ifx \showURL      \undefined \def \showURL       {\relax}        \fi
\providecommand\bibfield[2]{#2}
\providecommand\bibinfo[2]{#2}
\providecommand\natexlab[1]{#1}
\providecommand\showeprint[2][]{arXiv:#2}

\bibitem[Alappat et~al\mbox{.}(2020)]%
        {pardiso-7.2a}
\bibfield{author}{\bibinfo{person}{Christie Alappat}, \bibinfo{person}{Achim
  Basermann}, \bibinfo{person}{Alan~R. Bishop}, \bibinfo{person}{Holger
  Fehske}, \bibinfo{person}{Georg Hager}, \bibinfo{person}{Olaf Schenk},
  \bibinfo{person}{Jonas Thies}, {and} \bibinfo{person}{Gerhard Wellein}.}
  \bibinfo{year}{2020}\natexlab{}.
\newblock \showarticletitle{A Recursive Algebraic Coloring Technique for
  Hardware-Efficient Symmetric Sparse Matrix-Vector Multiplication}.
\newblock \bibinfo{journal}{\emph{ACM Trans. Parallel Comput.}}
  \bibinfo{volume}{7}, \bibinfo{number}{3}, Article \bibinfo{articleno}{19}
  (\bibinfo{date}{June} \bibinfo{year}{2020}), \bibinfo{numpages}{37}~pages.
\newblock
\showISSN{2329-4949}
\urldef\tempurl%
\url{https://doi.org/10.1145/3399732}
\showURL{%
\tempurl}


\bibitem[Allaire et~al\mbox{.}(2021)]%
        {ALLAIRE20211}
\bibfield{author}{\bibinfo{person}{Grégoire Allaire}, \bibinfo{person}{Charles
  Dapogny}, {and} \bibinfo{person}{François Jouve}.}
  \bibinfo{year}{2021}\natexlab{}.
\newblock \showarticletitle{Chapter 1 - Shape and topology optimization}.
\newblock In \bibinfo{booktitle}{\emph{Geometric Partial Differential Equations
  - Part II}}, \bibfield{editor}{\bibinfo{person}{Andrea Bonito} {and}
  \bibinfo{person}{Ricardo~H. Nochetto}} (Eds.). \bibinfo{series}{Handbook of
  Numerical Analysis}, Vol.~\bibinfo{volume}{22}.
  \bibinfo{publisher}{Elsevier}, \bibinfo{pages}{1--132}.
\newblock
\showISSN{1570-8659}
\urldef\tempurl%
\url{https://doi.org/10.1016/bs.hna.2020.10.004}
\showDOI{\tempurl}


\bibitem[Alnaes et~al\mbox{.}(2015)]%
        {AlnaesEtal2015}
\bibfield{author}{\bibinfo{person}{M.~S. Alnaes}, \bibinfo{person}{J. Blechta},
  \bibinfo{person}{J. Hake}, \bibinfo{person}{A. Johansson},
  \bibinfo{person}{B. Kehlet}, \bibinfo{person}{A. Logg}, \bibinfo{person}{C.
  Richardson}, \bibinfo{person}{J. Ring}, \bibinfo{person}{M.~E. Rognes}, {and}
  \bibinfo{person}{G.~N. Wells}.} \bibinfo{year}{2015}\natexlab{}.
\newblock \showarticletitle{The {FEniCS} Project Version 1.5}.
\newblock \bibinfo{journal}{\emph{Archive of Numerical Software}}
  \bibinfo{volume}{3} (\bibinfo{year}{2015}).
\newblock
\urldef\tempurl%
\url{https://doi.org/10.11588/ans.2015.100.20553}
\showDOI{\tempurl}


\bibitem[B{\"a}cher et~al\mbox{.}(2021)]%
        {Bacher2021}
\bibfield{author}{\bibinfo{person}{Moritz B{\"a}cher}, \bibinfo{person}{Espen
  Knoop}, {and} \bibinfo{person}{Christian Schumacher}.}
  \bibinfo{year}{2021}\natexlab{}.
\newblock \showarticletitle{Design and Control of Soft Robots Using
  Differentiable Simulation}.
\newblock \bibinfo{journal}{\emph{Current Robotics Reports}}
  (\bibinfo{year}{2021}), \bibinfo{pages}{1--11}.
\newblock


\bibitem[Baque et~al\mbox{.}(2018)]%
        {Baque2018}
\bibfield{author}{\bibinfo{person}{Pierre Baque}, \bibinfo{person}{Edoardo
  Remelli}, \bibinfo{person}{Fran{\c{c}}ois Fleuret}, {and}
  \bibinfo{person}{Pascal Fua}.} \bibinfo{year}{2018}\natexlab{}.
\newblock \showarticletitle{Geodesic convolutional shape optimization}. In
  \bibinfo{booktitle}{\emph{International Conference on Machine Learning}}.
  PMLR, \bibinfo{pages}{472--481}.
\newblock


\bibitem[Belytschko et~al\mbox{.}(2000)]%
        {belytschko2000nonlinear}
\bibfield{author}{\bibinfo{person}{Ted Belytschko}, \bibinfo{person}{{Wing Kam}
  Liu}, {and} \bibinfo{person}{Brian Moran}.} \bibinfo{year}{2000}\natexlab{}.
\newblock \bibinfo{booktitle}{\emph{Nonlinear Finite Elements for Continua and
  Structures}}.
\newblock \bibinfo{publisher}{John Wiley \& Sons, Ltd}.
\newblock


\bibitem[Beremlijski et~al\mbox{.}(2014)]%
        {Beremlijski2014}
\bibfield{author}{\bibinfo{person}{P. Beremlijski}, \bibinfo{person}{J.
  Haslinger}, \bibinfo{person}{J. Outrata}, {and} \bibinfo{person}{R. Pathó}.}
  \bibinfo{year}{2014}\natexlab{}.
\newblock \showarticletitle{Shape {Optimization} in {Contact} {Problems} with
  {Coulomb} {Friction} and a {Solution}-{Dependent} {Friction} {Coefficient}}.
\newblock \bibinfo{journal}{\emph{SIAM Journal on Control and Optimization}}
  \bibinfo{volume}{52}, \bibinfo{number}{5} (\bibinfo{date}{Jan.}
  \bibinfo{year}{2014}), \bibinfo{pages}{3371--3400}.
\newblock
\showISSN{0363-0129}
\urldef\tempurl%
\url{https://doi.org/10.1137/130948070}
\showDOI{\tempurl}


\bibitem[Bern et~al\mbox{.}(2019)]%
        {Bern2019}
\bibfield{author}{\bibinfo{person}{James Bern}, \bibinfo{person}{Pol Banzet},
  \bibinfo{person}{Roi Poranne}, {and} \bibinfo{person}{Stelian Coros}.}
  \bibinfo{year}{2019}\natexlab{}.
\newblock \showarticletitle{Trajectory Optimization for Cable-Driven Soft Robot
  Locomotion}. In \bibinfo{booktitle}{\emph{Robotics: Science and Systems XV}},
  Vol.~\bibinfo{volume}{1}. \bibinfo{publisher}{Robotics: Science and Systems
  Foundation}.
\newblock
\urldef\tempurl%
\url{https://doi.org/10.15607/rss.2019.xv.052}
\showDOI{\tempurl}


\bibitem[Bern et~al\mbox{.}(2020)]%
        {Bern2020}
\bibfield{author}{\bibinfo{person}{James~M. Bern}, \bibinfo{person}{Yannick
  Schnider}, \bibinfo{person}{Pol Banzet}, \bibinfo{person}{Nitish Kumar},
  {and} \bibinfo{person}{Stelian Coros}.} \bibinfo{year}{2020}\natexlab{}.
\newblock \showarticletitle{Soft Robot Control With a Learned Differentiable
  Model}. In \bibinfo{booktitle}{\emph{2020 3rd IEEE International Conference
  on Soft Robotics (RoboSoft)}}. \bibinfo{publisher}{IEEE},
  \bibinfo{pages}{417--423}.
\newblock
\urldef\tempurl%
\url{https://doi.org/10.1109/robosoft48309.2020.9116011}
\showDOI{\tempurl}


\bibitem[Bischof and B{\"u}cker(2000)]%
        {bischof2000computing}
\bibfield{author}{\bibinfo{person}{C.~H. Bischof} {and} \bibinfo{person}{H.~M.
  B{\"u}cker}.} \bibinfo{year}{2000}\natexlab{}.
\newblock \showarticletitle{Computing Derivatives of Computer Programs}.
\newblock In \bibinfo{booktitle}{\emph{Modern Methods and Algorithms of Quantum
  Chemistry: Proceedings, Second Edition}},
  \bibfield{editor}{\bibinfo{person}{J.~Grotendorst}} (Ed.).
  \bibinfo{series}{NIC Series}, Vol.~\bibinfo{volume}{3}.
  \bibinfo{publisher}{NIC-Directors}, \bibinfo{address}{J{\"u}lich},
  \bibinfo{pages}{315--327}.
\newblock
\urldef\tempurl%
\url{http://hdl.handle.net/2128/6053}
\showURL{%
\tempurl}


\bibitem[Bollh{\"o}fer et~al\mbox{.}(2019)]%
        {pardiso-7.2c}
\bibfield{author}{\bibinfo{person}{Matthias Bollh{\"o}fer},
  \bibinfo{person}{Aryan Eftekhari}, \bibinfo{person}{Simon Scheidegger}, {and}
  \bibinfo{person}{Olaf Schenk}.} \bibinfo{year}{2019}\natexlab{}.
\newblock \showarticletitle{Large-scale Sparse Inverse Covariance Matrix
  Estimation}.
\newblock \bibinfo{journal}{\emph{SIAM Journal on Scientific Computing}}
  \bibinfo{volume}{41}, \bibinfo{number}{1} (\bibinfo{year}{2019}),
  \bibinfo{pages}{A380--A401}.
\newblock
\urldef\tempurl%
\url{https://doi.org/10.1137/17M1147615}
\showDOI{\tempurl}
\showeprint{https://doi.org/10.1137/17M1147615}


\bibitem[Bollh{\"o}fer et~al\mbox{.}(2020)]%
        {pardiso-7.2b}
\bibfield{author}{\bibinfo{person}{Matthias Bollh{\"o}fer},
  \bibinfo{person}{Olaf Schenk}, \bibinfo{person}{Radim Janalik},
  \bibinfo{person}{Steve Hamm}, {and} \bibinfo{person}{Kiran Gullapalli}.}
  \bibinfo{year}{2020}\natexlab{}.
\newblock \showarticletitle{State-of-the-Art Sparse Direct Solvers}.
\newblock  (\bibinfo{year}{2020}), \bibinfo{pages}{3--33}.
\newblock
\showISBNx{978-3-030-43736-7}
\urldef\tempurl%
\url{https://doi.org/10.1007/978-3-030-43736-7_1}
\showURL{%
\tempurl}


\bibitem[Bridson et~al\mbox{.}(2002)]%
        {bridson2002robust}
\bibfield{author}{\bibinfo{person}{Robert Bridson}, \bibinfo{person}{Ronald
  Fedkiw}, {and} \bibinfo{person}{John Anderson}.}
  \bibinfo{year}{2002}\natexlab{}.
\newblock \showarticletitle{Robust Treatment of Collisions, Contact and
  Friction for Cloth Animation}.
\newblock \bibinfo{journal}{\emph{ACM Trans. on Graph.}}  \bibinfo{volume}{21}
  (\bibinfo{date}{05} \bibinfo{year}{2002}).
\newblock


\bibitem[Brogliato(1999)]%
        {Brogliato99}
\bibfield{author}{\bibinfo{person}{Bernard Brogliato}.}
  \bibinfo{year}{1999}\natexlab{}.
\newblock \bibinfo{booktitle}{\emph{Nonsmooth Mechanics}}.
\newblock \bibinfo{publisher}{Springer-Verlag}.
\newblock


\bibitem[Brown et~al\mbox{.}(2018)]%
        {damping2018}
\bibfield{author}{\bibinfo{person}{George~E. Brown}, \bibinfo{person}{Matthew
  Overby}, \bibinfo{person}{Zahra Forootaninia}, {and} \bibinfo{person}{Rahul
  Narain}.} \bibinfo{year}{2018}\natexlab{}.
\newblock \showarticletitle{Accurate Dissipative Forces in Optimization
  Integrators}.
\newblock \bibinfo{journal}{\emph{ACM Trans. Graph.}} \bibinfo{volume}{37},
  \bibinfo{number}{6}, Article \bibinfo{articleno}{282} (\bibinfo{date}{dec}
  \bibinfo{year}{2018}), \bibinfo{numpages}{14}~pages.
\newblock
\showISSN{0730-0301}
\urldef\tempurl%
\url{https://doi.org/10.1145/3272127.3275011}
\showDOI{\tempurl}


\bibitem[Chang et~al\mbox{.}(2016)]%
        {chang2016compositional}
\bibfield{author}{\bibinfo{person}{Michael~B Chang}, \bibinfo{person}{Tomer
  Ullman}, \bibinfo{person}{Antonio Torralba}, {and} \bibinfo{person}{Joshua~B
  Tenenbaum}.} \bibinfo{year}{2016}\natexlab{}.
\newblock \showarticletitle{A compositional object-based approach to learning
  physical dynamics}.
\newblock \bibinfo{journal}{\emph{arXiv preprint arXiv:1612.00341}}
  (\bibinfo{year}{2016}).
\newblock


\bibitem[Chen et~al\mbox{.}(2020)]%
        {Chen2020}
\bibfield{author}{\bibinfo{person}{Bicheng Chen}, \bibinfo{person}{Nianfeng
  Wang}, \bibinfo{person}{Xianmin Zhang}, {and} \bibinfo{person}{Wei Chen}.}
  \bibinfo{year}{2020}\natexlab{}.
\newblock \showarticletitle{Design of dielectric elastomer actuators using
  topology optimization on electrodes}.
\newblock \bibinfo{journal}{\emph{Smart Mater. Struct.}} \bibinfo{volume}{29},
  \bibinfo{number}{7} (\bibinfo{date}{June} \bibinfo{year}{2020}),
  \bibinfo{pages}{075029}.
\newblock
\showISSN{0964-1726, 1361-665X}
\urldef\tempurl%
\url{https://doi.org/10.1088/1361-665x/ab8b2d}
\showDOI{\tempurl}


\bibitem[Daviet et~al\mbox{.}(2011)]%
        {gilles20111hybrid}
\bibfield{author}{\bibinfo{person}{Gilles Daviet}, \bibinfo{person}{Florence
  Bertails-Descoubes}, {and} \bibinfo{person}{Laurence Boissieux}.}
  \bibinfo{year}{2011}\natexlab{}.
\newblock \showarticletitle{A Hybrid Iterative Solver for Robustly Capturing
  Coulomb Friction in Hair Dynamics}.
\newblock \bibinfo{journal}{\emph{ACM Trans. on Graph.}}  \bibinfo{volume}{30}
  (\bibinfo{date}{12} \bibinfo{year}{2011}).
\newblock


\bibitem[de~Vaucorbeil et~al\mbox{.}(2019)]%
        {de2019material}
\bibfield{author}{\bibinfo{person}{Alban de Vaucorbeil},
  \bibinfo{person}{Vinh~Phu Nguyen}, \bibinfo{person}{Sina Sinaie}, {and}
  \bibinfo{person}{Jian~Ying Wu}.} \bibinfo{year}{2019}\natexlab{}.
\newblock \showarticletitle{Material point method after 25 years: theory,
  implementation and applications}.
\newblock \bibinfo{journal}{\emph{Submitted to Advances in Applied Mechanics}}
  (\bibinfo{year}{2019}), \bibinfo{pages}{1}.
\newblock


\bibitem[Desmorat(2007)]%
        {Desmorat2007}
\bibfield{author}{\bibinfo{person}{B. Desmorat}.}
  \bibinfo{year}{2007}\natexlab{}.
\newblock \showarticletitle{Structural rigidity optimization with frictionless
  unilateral contact}.
\newblock \bibinfo{journal}{\emph{International Journal of Solids and
  Structures}} \bibinfo{volume}{44}, \bibinfo{number}{3} (\bibinfo{date}{Feb.}
  \bibinfo{year}{2007}), \bibinfo{pages}{1132--1144}.
\newblock
\showISSN{0020-7683}
\urldef\tempurl%
\url{https://doi.org/10.1016/j.ijsolstr.2006.06.010}
\showDOI{\tempurl}


\bibitem[Dokken et~al\mbox{.}(2020)]%
        {dokken2020automatic}
\bibfield{author}{\bibinfo{person}{Jørgen~S. Dokken},
  \bibinfo{person}{Sebastian~K. Mitusch}, {and} \bibinfo{person}{Simon~W.
  Funke}.} \bibinfo{year}{2020}\natexlab{}.
\newblock \bibinfo{title}{Automatic shape derivatives for transient PDEs in
  FEniCS and Firedrake}.
\newblock
\newblock
\showeprint[arxiv]{2001.10058}~[math.OC]


\bibitem[Du et~al\mbox{.}(2021)]%
        {du2021diffpd}
\bibfield{author}{\bibinfo{person}{Tao Du}, \bibinfo{person}{Kui Wu},
  \bibinfo{person}{Pingchuan Ma}, \bibinfo{person}{Sebastien Wah},
  \bibinfo{person}{Andrew Spielberg}, \bibinfo{person}{Daniela Rus}, {and}
  \bibinfo{person}{Wojciech Matusik}.} \bibinfo{year}{2021}\natexlab{}.
\newblock \showarticletitle{DiffPD: Differentiable Projective Dynamics}.
\newblock \bibinfo{journal}{\emph{ACM Trans. Graph.}} \bibinfo{volume}{41},
  \bibinfo{number}{2}, Article \bibinfo{articleno}{13} (\bibinfo{date}{nov}
  \bibinfo{year}{2021}), \bibinfo{numpages}{21}~pages.
\newblock
\showISSN{0730-0301}
\urldef\tempurl%
\url{https://doi.org/10.1145/3490168}
\showDOI{\tempurl}


\bibitem[Eck et~al\mbox{.}(2005)]%
        {Eck2005}
\bibfield{author}{\bibinfo{person}{Christof Eck}, \bibinfo{person}{Jiri
  Jarusek}, \bibinfo{person}{Miroslav Krbec}, \bibinfo{person}{Jiri Jarusek},
  {and} \bibinfo{person}{Miroslav Krbec}.} \bibinfo{year}{2005}\natexlab{}.
\newblock \bibinfo{booktitle}{\emph{Unilateral {Contact} {Problems} :
  {Variational} {Methods} and {Existence} {Theorems}}}.
\newblock \bibinfo{publisher}{CRC Press}.
\newblock
\showISBNx{978-1-4200-2736-5}
\urldef\tempurl%
\url{https://doi.org/10.1201/9781420027365}
\showDOI{\tempurl}


\bibitem[Ferguson et~al\mbox{.}(2020)]%
        {ipc_toolkit}
\bibfield{author}{\bibinfo{person}{Zachary Ferguson} {et~al\mbox{.}}}
  \bibinfo{year}{2020}\natexlab{}.
\newblock \bibinfo{booktitle}{\emph{{IPC Toolkit}}}.
\newblock
\urldef\tempurl%
\url{https://ipc-sim.github.io/ipc-toolkit/}
\showURL{%
\tempurl}


\bibitem[Ferguson et~al\mbox{.}(2021)]%
        {Ferguson:2021:RigidIPC}
\bibfield{author}{\bibinfo{person}{Zachary Ferguson}, \bibinfo{person}{Minchen
  Li}, \bibinfo{person}{Teseo Schneider}, \bibinfo{person}{Francisca
  Gil-Ureta}, \bibinfo{person}{Timothy Langlois}, \bibinfo{person}{Chenfanfu
  Jiang}, \bibinfo{person}{Denis Zorin}, \bibinfo{person}{Danny~M. Kaufman},
  {and} \bibinfo{person}{Daniele Panozzo}.} \bibinfo{year}{2021}\natexlab{}.
\newblock \showarticletitle{Intersection-free Rigid Body Dynamics}.
\newblock \bibinfo{journal}{\emph{ACM Transactions on Graphics (SIGGRAPH)}}
  \bibinfo{volume}{40}, \bibinfo{number}{4}, Article \bibinfo{articleno}{183}
  (\bibinfo{year}{2021}).
\newblock


\bibitem[Gavriil et~al\mbox{.}(2020)]%
        {Gavriil2020}
\bibfield{author}{\bibinfo{person}{Konstantinos Gavriil},
  \bibinfo{person}{Ruslan Guseinov}, \bibinfo{person}{Jes\'{u}s P\'{e}rez},
  \bibinfo{person}{Davide Pellis}, \bibinfo{person}{Paul Henderson},
  \bibinfo{person}{Florian Rist}, \bibinfo{person}{Helmut Pottmann}, {and}
  \bibinfo{person}{Bernd Bickel}.} \bibinfo{year}{2020}\natexlab{}.
\newblock \showarticletitle{Computational Design of Cold Bent Glass
  Fa\c{c}Ades}.
\newblock \bibinfo{journal}{\emph{ACM Trans. Graph.}} \bibinfo{volume}{39},
  \bibinfo{number}{6}, Article \bibinfo{articleno}{208} (\bibinfo{date}{nov}
  \bibinfo{year}{2020}), \bibinfo{numpages}{16}~pages.
\newblock
\showISSN{0730-0301}
\urldef\tempurl%
\url{https://doi.org/10.1145/3414685.3417843}
\showDOI{\tempurl}


\bibitem[Geilinger et~al\mbox{.}(2020)]%
        {geilinger2020add}
\bibfield{author}{\bibinfo{person}{Moritz Geilinger}, \bibinfo{person}{David
  Hahn}, \bibinfo{person}{Jonas Zehnder}, \bibinfo{person}{Moritz B{\"a}cher},
  \bibinfo{person}{Bernhard Thomaszewski}, {and} \bibinfo{person}{Stelian
  Coros}.} \bibinfo{year}{2020}\natexlab{}.
\newblock \showarticletitle{ADD: analytically differentiable dynamics for
  multi-body systems with frictional contact}.
\newblock \bibinfo{journal}{\emph{ACM Transactions on Graphics (TOG)}}
  \bibinfo{volume}{39}, \bibinfo{number}{6} (\bibinfo{year}{2020}),
  \bibinfo{pages}{1--15}.
\newblock


\bibitem[Geuzaine and Remacle(2009)]%
        {GMSH}
\bibfield{author}{\bibinfo{person}{Christophe Geuzaine} {and}
  \bibinfo{person}{Jean-François Remacle}.} \bibinfo{year}{2009}\natexlab{}.
\newblock \showarticletitle{Gmsh: A 3-D finite element mesh generator with
  built-in pre- and post-processing facilities}.
\newblock \bibinfo{journal}{\emph{Internat. J. Numer. Methods Engrg.}}
  \bibinfo{volume}{79}, \bibinfo{number}{11} (\bibinfo{year}{2009}),
  \bibinfo{pages}{1309--1331}.
\newblock
\urldef\tempurl%
\url{https://doi.org/10.1002/nme.2579}
\showDOI{\tempurl}
\showeprint{https://onlinelibrary.wiley.com/doi/pdf/10.1002/nme.2579}


\bibitem[Griewank and Walther(2008)]%
        {griewank2008evaluating}
\bibfield{author}{\bibinfo{person}{Andreas Griewank} {and}
  \bibinfo{person}{Andrea Walther}.} \bibinfo{year}{2008}\natexlab{}.
\newblock \bibinfo{booktitle}{\emph{Evaluating derivatives: principles and
  techniques of algorithmic differentiation}}. Vol.~\bibinfo{volume}{105}.
\newblock \bibinfo{publisher}{Siam}.
\newblock
\urldef\tempurl%
\url{https://doi.org/10.1137/1.9780898717761}
\showDOI{\tempurl}


\bibitem[Hafner et~al\mbox{.}(2019)]%
        {Hafner2019}
\bibfield{author}{\bibinfo{person}{Christian Hafner},
  \bibinfo{person}{Christian Schumacher}, \bibinfo{person}{Espen Knoop},
  \bibinfo{person}{Thomas Auzinger}, \bibinfo{person}{Bernd Bickel}, {and}
  \bibinfo{person}{Moritz Bächer}.} \bibinfo{year}{2019}\natexlab{}.
\newblock \showarticletitle{X-CAD: Optimizing CAD Models with Extended Finite
  Elements}.
\newblock \bibinfo{journal}{\emph{ACM Trans. Graph.}} \bibinfo{volume}{38},
  \bibinfo{number}{6}, Article \bibinfo{articleno}{157} (\bibinfo{date}{Nov.}
  \bibinfo{year}{2019}), \bibinfo{numpages}{15}~pages.
\newblock
\showISSN{0730-0301, 1557-7368}
\urldef\tempurl%
\url{https://doi.org/10.1145/3355089.3356576}
\showDOI{\tempurl}


\bibitem[Hahn et~al\mbox{.}(2019)]%
        {Hahn2019}
\bibfield{author}{\bibinfo{person}{David Hahn}, \bibinfo{person}{Pol Banzet},
  \bibinfo{person}{James~M. Bern}, {and} \bibinfo{person}{Stelian Coros}.}
  \bibinfo{year}{2019}\natexlab{}.
\newblock \showarticletitle{Real2Sim: Visco-Elastic Parameter Estimation from
  Dynamic Motion}.
\newblock \bibinfo{journal}{\emph{ACM Trans. Graph.}} \bibinfo{volume}{38},
  \bibinfo{number}{6}, Article \bibinfo{articleno}{236} (\bibinfo{date}{Nov.}
  \bibinfo{year}{2019}), \bibinfo{numpages}{13}~pages.
\newblock
\showISSN{0730-0301, 1557-7368}
\urldef\tempurl%
\url{https://doi.org/10.1145/3355089.3356548}
\showDOI{\tempurl}


\bibitem[Harmon et~al\mbox{.}(2009)]%
        {harmon2009asynchronous}
\bibfield{author}{\bibinfo{person}{David Harmon}, \bibinfo{person}{Etienne
  Vouga}, \bibinfo{person}{Breannan Smith}, \bibinfo{person}{Rasmus Tamstorf},
  {and} \bibinfo{person}{Eitan Grinspun}.} \bibinfo{year}{2009}\natexlab{}.
\newblock \showarticletitle{Asynchronous contact mnumpageechanics}. In
  \bibinfo{booktitle}{\emph{ACM Trans. on Graph. (TOG)}},
  Vol.~\bibinfo{volume}{28}. ACM.
\newblock


\bibitem[Harmon et~al\mbox{.}(2008)]%
        {harmon2008RTSC}
\bibfield{author}{\bibinfo{person}{David Harmon}, \bibinfo{person}{Etienne
  Vouga}, \bibinfo{person}{Rasmus Tamstorf}, {and} \bibinfo{person}{Eitan
  Grinspun}.} \bibinfo{year}{2008}\natexlab{}.
\newblock \showarticletitle{Robust Treatment of Simultaneous Collisions}.
\newblock \bibinfo{journal}{\emph{SIGGRAPH (ACM Trans. on Graph.)}}
  \bibinfo{volume}{27}, \bibinfo{number}{3} (\bibinfo{year}{2008}).
\newblock


\bibitem[Haslinger et~al\mbox{.}(1986)]%
        {Haslinger1986b}
\bibfield{author}{\bibinfo{person}{Jaroslav Haslinger}, \bibinfo{person}{Pekka
  Neittaanmaki}, {and} \bibinfo{person}{Timo Tiihonen}.}
  \bibinfo{year}{1986}\natexlab{}.
\newblock \showarticletitle{Shape optimization in contact problems based on
  penalization of the state inequality}.
\newblock \bibinfo{journal}{\emph{Aplikace matematiky}} \bibinfo{volume}{31},
  \bibinfo{number}{1} (\bibinfo{year}{1986}), \bibinfo{pages}{54--77}.
\newblock
\showISSN{0862-7940}
\urldef\tempurl%
\url{https://eudml.org/doc/15435}
\showURL{%
\tempurl}


\bibitem[Heiden et~al\mbox{.}(2021)]%
        {heiden2021disect}
\bibfield{author}{\bibinfo{person}{Eric Heiden}, \bibinfo{person}{Miles
  Macklin}, \bibinfo{person}{Yashraj~S Narang}, \bibinfo{person}{Dieter Fox},
  \bibinfo{person}{Animesh Garg}, {and} \bibinfo{person}{Fabio Ramos}.}
  \bibinfo{year}{2021}\natexlab{}.
\newblock \showarticletitle{{DiSECt: A Differentiable Simulation Engine for
  Autonomous Robotic Cutting}}. In \bibinfo{booktitle}{\emph{Proceedings of
  Robotics: Science and Systems}}. \bibinfo{address}{Virtual}.
\newblock
\urldef\tempurl%
\url{https://doi.org/10.15607/RSS.2021.XVII.067}
\showDOI{\tempurl}


\bibitem[Heiden et~al\mbox{.}(2020)]%
        {heiden2020neuralsim}
\bibfield{author}{\bibinfo{person}{Eric Heiden}, \bibinfo{person}{David
  Millard}, \bibinfo{person}{Erwin Coumans}, \bibinfo{person}{Yizhou Sheng},
  {and} \bibinfo{person}{Gaurav~S Sukhatme}.} \bibinfo{year}{2020}\natexlab{}.
\newblock \showarticletitle{NeuralSim: Augmenting Differentiable Simulators
  with Neural Networks}.
\newblock \bibinfo{journal}{\emph{arXiv preprint arXiv:2011.04217}}
  (\bibinfo{year}{2020}).
\newblock


\bibitem[Herskovits et~al\mbox{.}(2000)]%
        {Herskovits2000}
\bibfield{author}{\bibinfo{person}{J. Herskovits}, \bibinfo{person}{A.
  Leontiev}, \bibinfo{person}{G. Dias}, {and} \bibinfo{person}{G. Santos}.}
  \bibinfo{year}{2000}\natexlab{}.
\newblock \showarticletitle{Contact shape optimization: a bilevel programming
  approach}.
\newblock \bibinfo{journal}{\emph{Structural and Multidisciplinary
  Optimization}} \bibinfo{volume}{20}, \bibinfo{number}{3}
  (\bibinfo{date}{Nov.} \bibinfo{year}{2000}), \bibinfo{pages}{214--221}.
\newblock
\showISSN{1615-147X, 1615-1488}
\urldef\tempurl%
\url{https://doi.org/10.1007/s001580050149}
\showDOI{\tempurl}


\bibitem[Hoshyari et~al\mbox{.}(2019)]%
        {Hoshyari2019}
\bibfield{author}{\bibinfo{person}{Shayan Hoshyari}, \bibinfo{person}{Hongyi
  Xu}, \bibinfo{person}{Espen Knoop}, \bibinfo{person}{Stelian Coros}, {and}
  \bibinfo{person}{Moritz Bächer}.} \bibinfo{year}{2019}\natexlab{}.
\newblock \showarticletitle{Vibration-minimizing motion retargeting for robotic
  characters}.
\newblock \bibinfo{journal}{\emph{ACM Trans. Graph.}} \bibinfo{volume}{38},
  \bibinfo{number}{4} (\bibinfo{date}{July} \bibinfo{year}{2019}),
  \bibinfo{pages}{1--14}.
\newblock
\showISSN{0730-0301, 1557-7368}
\urldef\tempurl%
\url{https://doi.org/10.1145/3306346.3323034}
\showDOI{\tempurl}


\bibitem[Hsu et~al\mbox{.}(2022)]%
        {sagfree}
\bibfield{author}{\bibinfo{person}{Jerry Hsu}, \bibinfo{person}{Nghia Truong},
  \bibinfo{person}{Cem Yuksel}, {and} \bibinfo{person}{Kui Wu}.}
  \bibinfo{year}{2022}\natexlab{}.
\newblock \showarticletitle{A General Two-Stage Initialization for Sag-Free
  Deformable Simulations}.
\newblock \bibinfo{journal}{\emph{ACM Trans. Graph.}} \bibinfo{volume}{41},
  \bibinfo{number}{4}, Article \bibinfo{articleno}{64} (\bibinfo{date}{jul}
  \bibinfo{year}{2022}), \bibinfo{numpages}{13}~pages.
\newblock
\showISSN{0730-0301}
\urldef\tempurl%
\url{https://doi.org/10.1145/3528223.3530165}
\showDOI{\tempurl}


\bibitem[Hu et~al\mbox{.}(2019a)]%
        {hu2019difftaichi}
\bibfield{author}{\bibinfo{person}{Yuanming Hu}, \bibinfo{person}{Luke
  Anderson}, \bibinfo{person}{Tzu-Mao Li}, \bibinfo{person}{Qi Sun},
  \bibinfo{person}{Nathan Carr}, \bibinfo{person}{Jonathan Ragan-Kelley}, {and}
  \bibinfo{person}{Fredo Durand}.} \bibinfo{year}{2019}\natexlab{a}.
\newblock \showarticletitle{DiffTaichi: Differentiable Programming for Physical
  Simulation}. In \bibinfo{booktitle}{\emph{International Conference on
  Learning Representations}}.
\newblock


\bibitem[Hu et~al\mbox{.}(2019b)]%
        {hu2019chainqueen}
\bibfield{author}{\bibinfo{person}{Yuanming Hu}, \bibinfo{person}{Jiancheng
  Liu}, \bibinfo{person}{Andrew Spielberg}, \bibinfo{person}{Joshua~B
  Tenenbaum}, \bibinfo{person}{William~T Freeman}, \bibinfo{person}{Jiajun Wu},
  \bibinfo{person}{Daniela Rus}, {and} \bibinfo{person}{Wojciech Matusik}.}
  \bibinfo{year}{2019}\natexlab{b}.
\newblock \showarticletitle{Chainqueen: A real-time differentiable physical
  simulator for soft robotics}. In \bibinfo{booktitle}{\emph{2019 International
  conference on robotics and automation (ICRA)}}. IEEE,
  \bibinfo{pages}{6265--6271}.
\newblock


\bibitem[Hu et~al\mbox{.}(2020)]%
        {fTetWild}
\bibfield{author}{\bibinfo{person}{Yixin Hu}, \bibinfo{person}{Teseo
  Schneider}, \bibinfo{person}{Bolun Wang}, \bibinfo{person}{Denis Zorin},
  {and} \bibinfo{person}{Daniele Panozzo}.} \bibinfo{year}{2020}\natexlab{}.
\newblock \showarticletitle{Fast Tetrahedral Meshing in the Wild}.
\newblock \bibinfo{journal}{\emph{ACM Trans. Graph.}} \bibinfo{volume}{39},
  \bibinfo{number}{4}, Article \bibinfo{articleno}{117} (\bibinfo{date}{July}
  \bibinfo{year}{2020}), \bibinfo{numpages}{18}~pages.
\newblock
\showISSN{0730-0301}
\urldef\tempurl%
\url{https://doi.org/10.1145/3386569.3392385}
\showDOI{\tempurl}


\bibitem[Jakob(2010)]%
        {autodiff}
\bibfield{author}{\bibinfo{person}{Wenzel Jakob}.}
  \bibinfo{year}{2010}\natexlab{}.
\newblock \bibinfo{title}{Mitsuba renderer}.
\newblock
\newblock
\newblock
\shownote{http://www.mitsuba-renderer.org}.


\bibitem[Jatavallabhula et~al\mbox{.}(2021)]%
        {gradsim}
\bibfield{author}{\bibinfo{person}{Krishna~Murthy Jatavallabhula},
  \bibinfo{person}{Miles Macklin}, \bibinfo{person}{Florian Golemo},
  \bibinfo{person}{Vikram Voleti}, \bibinfo{person}{Linda Petrini},
  \bibinfo{person}{Martin Weiss}, \bibinfo{person}{Breandan Considine},
  \bibinfo{person}{Jerome Parent-Levesque}, \bibinfo{person}{Kevin Xie},
  \bibinfo{person}{Kenny Erleben}, \bibinfo{person}{Liam Paull},
  \bibinfo{person}{Florian Shkurti}, \bibinfo{person}{Derek Nowrouzezahrai},
  {and} \bibinfo{person}{Sanja Fidler}.} \bibinfo{year}{2021}\natexlab{}.
\newblock \showarticletitle{gradSim: Differentiable simulation for system
  identification and visuomotor control}.
\newblock \bibinfo{journal}{\emph{International Conference on Learning
  Representations (ICLR)}} (\bibinfo{year}{2021}).
\newblock
\urldef\tempurl%
\url{https://openreview.net/forum?id=c_E8kFWfhp0}
\showURL{%
\tempurl}


\bibitem[Jiang et~al\mbox{.}(2020)]%
        {Jiang:2020}
\bibfield{author}{\bibinfo{person}{Zhongshi Jiang}, \bibinfo{person}{Teseo
  Schneider}, \bibinfo{person}{Denis Zorin}, {and} \bibinfo{person}{Daniele
  Panozzo}.} \bibinfo{year}{2020}\natexlab{}.
\newblock \showarticletitle{Bijective Projection in a Shell}.
\newblock \bibinfo{journal}{\emph{ACM Trans. Graph.}} \bibinfo{volume}{39},
  \bibinfo{number}{6}, Article \bibinfo{articleno}{247} (\bibinfo{date}{nov}
  \bibinfo{year}{2020}), \bibinfo{numpages}{18}~pages.
\newblock
\showISSN{0730-0301}
\urldef\tempurl%
\url{https://doi.org/10.1145/3414685.3417769}
\showDOI{\tempurl}


\bibitem[Kikuchi and Oden(1988)]%
        {Kikuchi88}
\bibfield{author}{\bibinfo{person}{Noboru Kikuchi} {and}
  \bibinfo{person}{John~Tinsley Oden}.} \bibinfo{year}{1988}\natexlab{}.
\newblock \bibinfo{booktitle}{\emph{{Contact Problems in Elasticity: A Study of
  Variational Inequalities and Finite Element Methods}}}. \bibinfo{series}{SIAM
  Studies in App. and Numer. Math.}, Vol.~\bibinfo{volume}{8}.
\newblock \bibinfo{publisher}{Society for Industrial and Applied Mathematics}.
\newblock


\bibitem[Knupp(2001)]%
        {Knupp:2001}
\bibfield{author}{\bibinfo{person}{Patrick~M. Knupp}.}
  \bibinfo{year}{2001}\natexlab{}.
\newblock \showarticletitle{Algebraic Mesh Quality Metrics}.
\newblock \bibinfo{journal}{\emph{SIAM Journal on Scientific Computing}}
  \bibinfo{volume}{23}, \bibinfo{number}{1} (\bibinfo{year}{2001}),
  \bibinfo{pages}{193--218}.
\newblock
\urldef\tempurl%
\url{https://doi.org/10.1137/S1064827500371499}
\showDOI{\tempurl}
\showeprint{https://doi.org/10.1137/S1064827500371499}


\bibitem[Li et~al\mbox{.}(2020)]%
        {Li2020IPC}
\bibfield{author}{\bibinfo{person}{Minchen Li}, \bibinfo{person}{Zachary
  Ferguson}, \bibinfo{person}{Teseo Schneider}, \bibinfo{person}{Timothy
  Langlois}, \bibinfo{person}{Denis Zorin}, \bibinfo{person}{Daniele Panozzo},
  \bibinfo{person}{Chenfanfu Jiang}, {and} \bibinfo{person}{Danny~M. Kaufman}.}
  \bibinfo{year}{2020}\natexlab{}.
\newblock \showarticletitle{Incremental Potential Contact: Intersection- and
  Inversion-free Large Deformation Dynamics}.
\newblock \bibinfo{journal}{\emph{ACM Transactions on Graphics}}
  \bibinfo{volume}{39}, \bibinfo{number}{4} (\bibinfo{year}{2020}).
\newblock


\bibitem[Li et~al\mbox{.}(2023a)]%
        {Li:2022:ConvergentIPC}
\bibfield{author}{\bibinfo{person}{Minchen Li}, \bibinfo{person}{Zachary
  Ferguson}, \bibinfo{person}{Teseo Schneider}, \bibinfo{person}{Timothy
  Langlois}, \bibinfo{person}{Denis Zorin}, \bibinfo{person}{Daniele Panozzo},
  \bibinfo{person}{Chenfanfu Jiang}, {and} \bibinfo{person}{Danny~M. Kaufman}.}
  \bibinfo{year}{2023}\natexlab{a}.
\newblock \bibinfo{title}{Convergent Incremental Potential Contact}.
\newblock
\newblock
\showeprint[arxiv]{2307.15908}~[math.NA]


\bibitem[Li et~al\mbox{.}(2022)]%
        {li2022diffcloth}
\bibfield{author}{\bibinfo{person}{Yifei Li}, \bibinfo{person}{Tao Du},
  \bibinfo{person}{Kui Wu}, \bibinfo{person}{Jie Xu}, {and}
  \bibinfo{person}{Wojciech Matusik}.} \bibinfo{year}{2022}\natexlab{}.
\newblock \showarticletitle{DiffCloth: Differentiable Cloth Simulation with Dry
  Frictional Contact}.
\newblock \bibinfo{journal}{\emph{ACM Trans. Graph.}} (\bibinfo{date}{mar}
  \bibinfo{year}{2022}).
\newblock
\showISSN{0730-0301}
\urldef\tempurl%
\url{https://doi.org/10.1145/3527660}
\showDOI{\tempurl}
\newblock
\shownote{Just Accepted}.


\bibitem[Li et~al\mbox{.}(2023b)]%
        {2023SPHRigid}
\bibfield{author}{\bibinfo{person}{Zhehao Li}, \bibinfo{person}{Qingyu Xu},
  \bibinfo{person}{Xiaohan Ye}, \bibinfo{person}{Bo Ren}, {and}
  \bibinfo{person}{Ligang Liu}.} \bibinfo{year}{2023}\natexlab{b}.
\newblock \showarticletitle{DiffFR: Differentiable SPH-Based Fluid-Rigid
  Coupling for Rigid Body Control}.
\newblock \bibinfo{journal}{\emph{ACM Trans. Graph.}} \bibinfo{volume}{42},
  \bibinfo{number}{6}, Article \bibinfo{articleno}{179} (\bibinfo{date}{dec}
  \bibinfo{year}{2023}), \bibinfo{numpages}{17}~pages.
\newblock
\showISSN{0730-0301}
\urldef\tempurl%
\url{https://doi.org/10.1145/3618318}
\showDOI{\tempurl}


\bibitem[Liang et~al\mbox{.}(2019)]%
        {liang2019differentiable}
\bibfield{author}{\bibinfo{person}{Junbang Liang}, \bibinfo{person}{Ming Lin},
  {and} \bibinfo{person}{Vladlen Koltun}.} \bibinfo{year}{2019}\natexlab{}.
\newblock \showarticletitle{Differentiable cloth simulation for inverse
  problems}.
\newblock \bibinfo{journal}{\emph{Neural Information Processing Systems}}
  (\bibinfo{year}{2019}).
\newblock


\bibitem[Ly et~al\mbox{.}(2018)]%
        {elasticshell2018}
\bibfield{author}{\bibinfo{person}{Micka\"{e}l Ly}, \bibinfo{person}{Romain
  Casati}, \bibinfo{person}{Florence Bertails-Descoubes},
  \bibinfo{person}{M\'{e}lina Skouras}, {and} \bibinfo{person}{Laurence
  Boissieux}.} \bibinfo{year}{2018}\natexlab{}.
\newblock \showarticletitle{Inverse Elastic Shell Design with Contact and
  Friction}.
\newblock \bibinfo{journal}{\emph{ACM Trans. Graph.}} \bibinfo{volume}{37},
  \bibinfo{number}{6}, Article \bibinfo{articleno}{201} (\bibinfo{date}{dec}
  \bibinfo{year}{2018}), \bibinfo{numpages}{16}~pages.
\newblock
\showISSN{0730-0301}
\urldef\tempurl%
\url{https://doi.org/10.1145/3272127.3275036}
\showDOI{\tempurl}


\bibitem[Maloisel et~al\mbox{.}(2021)]%
        {Maloisel2021}
\bibfield{author}{\bibinfo{person}{Guirec Maloisel}, \bibinfo{person}{Espen
  Knoop}, \bibinfo{person}{Christian Schumacher}, {and} \bibinfo{person}{Moritz
  Bacher}.} \bibinfo{year}{2021}\natexlab{}.
\newblock \showarticletitle{Automated Routing of Muscle Fibers for Soft
  Robots}.
\newblock \bibinfo{journal}{\emph{IEEE Trans. Robot.}} \bibinfo{volume}{37},
  \bibinfo{number}{3} (\bibinfo{date}{June} \bibinfo{year}{2021}),
  \bibinfo{pages}{996--1008}.
\newblock
\showISSN{1552-3098, 1941-0468}
\urldef\tempurl%
\url{https://doi.org/10.1109/tro.2020.3043654}
\showDOI{\tempurl}


\bibitem[Margossian(2019)]%
        {margossian2019review}
\bibfield{author}{\bibinfo{person}{Charles~C Margossian}.}
  \bibinfo{year}{2019}\natexlab{}.
\newblock \showarticletitle{A review of automatic differentiation and its
  efficient implementation}.
\newblock \bibinfo{journal}{\emph{Wiley Interdisciplinary Reviews: Data Mining
  and Knowledge Discovery}} \bibinfo{volume}{9}, \bibinfo{number}{4}
  (\bibinfo{year}{2019}), \bibinfo{pages}{e1305}.
\newblock
\urldef\tempurl%
\url{https://doi.org/10.1002/widm.1305}
\showDOI{\tempurl}


\bibitem[Maury et~al\mbox{.}(2017)]%
        {Maury2017}
\bibfield{author}{\bibinfo{person}{Aymeric Maury}, \bibinfo{person}{Gr\'egoire
  Allaire}, {and} \bibinfo{person}{Fran\c{c}ois Jouve}.}
  \bibinfo{year}{2017}\natexlab{}.
\newblock \bibinfo{title}{Shape optimisation with the level set method for
  contact problems in linearised elasticity}.  (\bibinfo{date}{Jan.}
  \bibinfo{year}{2017}).
\newblock
\urldef\tempurl%
\url{https://hal.archives-ouvertes.fr/hal-01435325}
\showURL{%
\tempurl}


\bibitem[McNamara et~al\mbox{.}(2004)]%
        {McNamara:2004:fluid}
\bibfield{author}{\bibinfo{person}{Antoine McNamara}, \bibinfo{person}{Adrien
  Treuille}, \bibinfo{person}{Zoran Popovi\'{c}}, {and} \bibinfo{person}{Jos
  Stam}.} \bibinfo{year}{2004}\natexlab{}.
\newblock \showarticletitle{Fluid Control Using the Adjoint Method}.
\newblock \bibinfo{journal}{\emph{ACM Transactions on Graphics / SIGGRAPH
  2004}} \bibinfo{volume}{23}, \bibinfo{number}{3} (\bibinfo{date}{Aug.}
  \bibinfo{year}{2004}).
\newblock


\bibitem[Mitusch et~al\mbox{.}(2019)]%
        {Mitusch2019}
\bibfield{author}{\bibinfo{person}{Sebastian~K. Mitusch},
  \bibinfo{person}{Simon~W. Funke}, {and} \bibinfo{person}{Jørgen~S. Dokken}.}
  \bibinfo{year}{2019}\natexlab{}.
\newblock \showarticletitle{dolfin-adjoint 2018.1: automated adjoints for
  FEniCS and Firedrake}.
\newblock \bibinfo{journal}{\emph{Journal of Open Source Software}}
  \bibinfo{volume}{4}, \bibinfo{number}{38} (\bibinfo{year}{2019}),
  \bibinfo{pages}{1292}.
\newblock
\urldef\tempurl%
\url{https://doi.org/10.21105/joss.01292}
\showDOI{\tempurl}


\bibitem[Moses et~al\mbox{.}(2022)]%
        {Moses:2022}
\bibfield{author}{\bibinfo{person}{William~S. Moses}, \bibinfo{person}{Sri
  Hari~Krishna Narayanan}, \bibinfo{person}{Ludger Paehler},
  \bibinfo{person}{Valentin Churavy}, \bibinfo{person}{Michel Schanen},
  \bibinfo{person}{Jan H\"{u}ckelheim}, \bibinfo{person}{Johannes Doerfert},
  {and} \bibinfo{person}{Paul Hovland}.} \bibinfo{year}{2022}\natexlab{}.
\newblock \showarticletitle{Scalable Automatic Differentiation of Multiple
  Parallel Paradigms through Compiler Augmentation}. In
  \bibinfo{booktitle}{\emph{Proceedings of the International Conference on High
  Performance Computing, Networking, Storage and Analysis}} (Dallas, Texas)
  \emph{(\bibinfo{series}{SC '22})}. \bibinfo{publisher}{IEEE Press}, Article
  \bibinfo{articleno}{60}, \bibinfo{numpages}{18}~pages.
\newblock
\showISBNx{9784665454445}


\bibitem[Naumann(2012)]%
        {naumann2012art}
\bibfield{author}{\bibinfo{person}{Uwe Naumann}.}
  \bibinfo{year}{2012}\natexlab{}.
\newblock \bibinfo{booktitle}{\emph{The art of differentiating computer
  programs: an introduction to algorithmic differentiation}}.
  Vol.~\bibinfo{volume}{24}.
\newblock \bibinfo{publisher}{SIAM}.
\newblock
\urldef\tempurl%
\url{https://doi.org/10.1137/1.9781611972078}
\showDOI{\tempurl}


\bibitem[Otaduy et~al\mbox{.}(2009)]%
        {otaduy2009implicit}
\bibfield{author}{\bibinfo{person}{Miguel Otaduy}, \bibinfo{person}{Rasmus
  Tamstorf}, \bibinfo{person}{Denis Steinemann}, {and} \bibinfo{person}{Markus
  Gross}.} \bibinfo{year}{2009}\natexlab{}.
\newblock \showarticletitle{Implicit Contact Handling for Deformable Objects}.
\newblock \bibinfo{journal}{\emph{Comp. Graph. Forum}}  \bibinfo{volume}{28}
  (\bibinfo{date}{04} \bibinfo{year}{2009}).
\newblock


\bibitem[Panetta et~al\mbox{.}(2017)]%
        {panetta2017}
\bibfield{author}{\bibinfo{person}{Julian Panetta}, \bibinfo{person}{Abtin
  Rahimian}, {and} \bibinfo{person}{Denis Zorin}.}
  \bibinfo{year}{2017}\natexlab{}.
\newblock \showarticletitle{Worst-case stress relief for microstructures}.
\newblock \bibinfo{journal}{\emph{ACM Transactions on Graphics}}
  \bibinfo{volume}{36}, \bibinfo{number}{4} (\bibinfo{year}{2017}).
\newblock
\showISSN{0730-0301}
\urldef\tempurl%
\url{https://doi.org/10.1145/3072959.3073649}
\showDOI{\tempurl}


\bibitem[Panetta et~al\mbox{.}(2015)]%
        {Panetta2015}
\bibfield{author}{\bibinfo{person}{Julian Panetta}, \bibinfo{person}{Qingnan
  Zhou}, \bibinfo{person}{Luigi Malomo}, \bibinfo{person}{Nico Pietroni},
  \bibinfo{person}{Paolo Cignoni}, {and} \bibinfo{person}{Denis Zorin}.}
  \bibinfo{year}{2015}\natexlab{}.
\newblock \showarticletitle{Elastic Textures for Additive Fabrication}.
\newblock \bibinfo{journal}{\emph{ACM Trans. Graph.}} \bibinfo{volume}{34},
  \bibinfo{number}{4}, Article \bibinfo{articleno}{135} (\bibinfo{date}{July}
  \bibinfo{year}{2015}), \bibinfo{numpages}{12}~pages.
\newblock
\showISSN{0730-0301}


\bibitem[Qiao et~al\mbox{.}(2020)]%
        {qiao2020scalable}
\bibfield{author}{\bibinfo{person}{Yi-Ling Qiao}, \bibinfo{person}{Junbang
  Liang}, \bibinfo{person}{Vladlen Koltun}, {and} \bibinfo{person}{Ming Lin}.}
  \bibinfo{year}{2020}\natexlab{}.
\newblock \showarticletitle{Scalable Differentiable Physics for Learning and
  Control}. In \bibinfo{booktitle}{\emph{International Conference on Machine
  Learning}}. PMLR, \bibinfo{pages}{7847--7856}.
\newblock


\bibitem[Rabinovich et~al\mbox{.}(2017)]%
        {rabinovich2017scalable}
\bibfield{author}{\bibinfo{person}{Michael Rabinovich}, \bibinfo{person}{Roi
  Poranne}, \bibinfo{person}{Daniele Panozzo}, {and} \bibinfo{person}{Olga
  Sorkine-Hornung}.} \bibinfo{year}{2017}\natexlab{}.
\newblock \showarticletitle{Scalable locally injective mappings}.
\newblock \bibinfo{journal}{\emph{ACM Transactions on Graphics (TOG)}}
  \bibinfo{volume}{36}, \bibinfo{number}{4} (\bibinfo{year}{2017}),
  \bibinfo{pages}{1}.
\newblock


\bibitem[Rojas et~al\mbox{.}(2021)]%
        {rojas2021differentiable}
\bibfield{author}{\bibinfo{person}{Junior Rojas}, \bibinfo{person}{Eftychios
  Sifakis}, {and} \bibinfo{person}{Ladislav Kavan}.}
  \bibinfo{year}{2021}\natexlab{}.
\newblock \showarticletitle{Differentiable Implicit Soft-Body Physics}.
\newblock \bibinfo{journal}{\emph{arXiv preprint arXiv:2102.05791}}
  (\bibinfo{year}{2021}).
\newblock


\bibitem[Schenck and Fox(2018)]%
        {pmlr-v87-schenck18a}
\bibfield{author}{\bibinfo{person}{Connor Schenck} {and}
  \bibinfo{person}{Dieter Fox}.} \bibinfo{year}{2018}\natexlab{}.
\newblock \showarticletitle{SPNets: Differentiable Fluid Dynamics for Deep
  Neural Networks}. In \bibinfo{booktitle}{\emph{Proceedings of The 2nd
  Conference on Robot Learning}} \emph{(\bibinfo{series}{Proceedings of Machine
  Learning Research}, Vol.~\bibinfo{volume}{87})},
  \bibfield{editor}{\bibinfo{person}{Aude Billard}, \bibinfo{person}{Anca
  Dragan}, \bibinfo{person}{Jan Peters}, {and} \bibinfo{person}{Jun Morimoto}}
  (Eds.). \bibinfo{publisher}{PMLR}, \bibinfo{pages}{317--335}.
\newblock
\urldef\tempurl%
\url{https://proceedings.mlr.press/v87/schenck18a.html}
\showURL{%
\tempurl}


\bibitem[Schneider et~al\mbox{.}(2019)]%
        {polyfem}
\bibfield{author}{\bibinfo{person}{Teseo Schneider}, \bibinfo{person}{Jérémie
  Dumas}, \bibinfo{person}{Xifeng Gao}, \bibinfo{person}{Denis Zorin}, {and}
  \bibinfo{person}{Daniele Panozzo}.} \bibinfo{year}{2019}\natexlab{}.
\newblock \bibinfo{title}{{PolyFEM}}.
\newblock \bibinfo{howpublished}{\url{https://polyfem.github.io/}}.
\newblock


\bibitem[Schumacher et~al\mbox{.}(2020)]%
        {Schumacher2020}
\bibfield{author}{\bibinfo{person}{Christian Schumacher},
  \bibinfo{person}{Espen Knoop}, {and} \bibinfo{person}{Moritz Bacher}.}
  \bibinfo{year}{2020}\natexlab{}.
\newblock \showarticletitle{Simulation-Ready Characterization of Soft Robotic
  Materials}.
\newblock \bibinfo{journal}{\emph{IEEE Robot. Autom. Lett.}}
  \bibinfo{volume}{5}, \bibinfo{number}{3} (\bibinfo{date}{July}
  \bibinfo{year}{2020}), \bibinfo{pages}{3775--3782}.
\newblock
\showISSN{2377-3766, 2377-3774}
\urldef\tempurl%
\url{https://doi.org/10.1109/lra.2020.2982058}
\showDOI{\tempurl}


\bibitem[Schumacher et~al\mbox{.}(2018)]%
        {Schumacher2018}
\bibfield{author}{\bibinfo{person}{Christian Schumacher},
  \bibinfo{person}{Jonas Zehnder}, {and} \bibinfo{person}{Moritz B\"{a}cher}.}
  \bibinfo{year}{2018}\natexlab{}.
\newblock \showarticletitle{Set-in-Stone: Worst-Case Optimization of Structures
  Weak in Tension}.
\newblock \bibinfo{journal}{\emph{ACM Trans. Graph.}} \bibinfo{volume}{37},
  \bibinfo{number}{6}, Article \bibinfo{articleno}{252} (\bibinfo{date}{dec}
  \bibinfo{year}{2018}), \bibinfo{numpages}{13}~pages.
\newblock
\showISSN{0730-0301}
\urldef\tempurl%
\url{https://doi.org/10.1145/3272127.3275085}
\showDOI{\tempurl}


\bibitem[Shan et~al\mbox{.}(2015)]%
        {bistable2015}
\bibfield{author}{\bibinfo{person}{Sicong Shan}, \bibinfo{person}{Sung Kang},
  \bibinfo{person}{Jordan Raney}, \bibinfo{person}{Pai Wang},
  \bibinfo{person}{Lichen Fang}, \bibinfo{person}{Francisco Candido},
  \bibinfo{person}{Jennifer Lewis}, {and} \bibinfo{person}{Katia Bertoldi}.}
  \bibinfo{year}{2015}\natexlab{}.
\newblock \showarticletitle{Multistable Architected Materials for Trapping
  Elastic Strain Energy}.
\newblock \bibinfo{journal}{\emph{Advanced materials (Deerfield Beach, Fla.)}}
  \bibinfo{volume}{27} (\bibinfo{date}{06} \bibinfo{year}{2015}).
\newblock
\urldef\tempurl%
\url{https://doi.org/10.1002/adma.201501708}
\showDOI{\tempurl}


\bibitem[Sharma and Maute(2018)]%
        {Sharma2018}
\bibfield{author}{\bibinfo{person}{Ashesh Sharma} {and} \bibinfo{person}{Kurt
  Maute}.} \bibinfo{year}{2018}\natexlab{}.
\newblock \showarticletitle{Stress-based topology optimization using spatial
  gradient stabilized XFEM}.
\newblock \bibinfo{journal}{\emph{Structural and Multidisciplinary
  Optimization}} \bibinfo{volume}{57}, \bibinfo{number}{1}
  (\bibinfo{year}{2018}), \bibinfo{pages}{17--38}.
\newblock


\bibitem[Skouras et~al\mbox{.}(2013)]%
        {Skouras2013}
\bibfield{author}{\bibinfo{person}{M\'{e}lina Skouras},
  \bibinfo{person}{Bernhard Thomaszewski}, \bibinfo{person}{Stelian Coros},
  \bibinfo{person}{Bernd Bickel}, {and} \bibinfo{person}{Markus Gross}.}
  \bibinfo{year}{2013}\natexlab{}.
\newblock \showarticletitle{Computational Design of Actuated Deformable
  Characters}.
\newblock \bibinfo{journal}{\emph{ACM Trans. Graph.}} \bibinfo{volume}{32},
  \bibinfo{number}{4}, Article \bibinfo{articleno}{82} (\bibinfo{date}{jul}
  \bibinfo{year}{2013}), \bibinfo{numpages}{10}~pages.
\newblock
\showISSN{0730-0301}
\urldef\tempurl%
\url{https://doi.org/10.1145/2461912.2461979}
\showDOI{\tempurl}


\bibitem[Stewart(2001)]%
        {Stewart01}
\bibfield{author}{\bibinfo{person}{David~E Stewart}.}
  \bibinfo{year}{2001}\natexlab{}.
\newblock \showarticletitle{{Finite-dimensional contact mechanics}}.
\newblock \bibinfo{journal}{\emph{Phil. Trans. R. Soc. Lond. A}}
  \bibinfo{volume}{359} (\bibinfo{year}{2001}).
\newblock


\bibitem[Stupkiewicz et~al\mbox{.}(2010)]%
        {Stupkiewicz2010}
\bibfield{author}{\bibinfo{person}{Stanisław Stupkiewicz},
  \bibinfo{person}{Jakub Lengiewicz}, {and} \bibinfo{person}{Jože Korelc}.}
  \bibinfo{year}{2010}\natexlab{}.
\newblock \showarticletitle{Sensitivity analysis for frictional contact
  problems in the augmented {Lagrangian} formulation}.
\newblock \bibinfo{journal}{\emph{Computer Methods in Applied Mechanics and
  Engineering}} \bibinfo{volume}{199}, \bibinfo{number}{33}
  (\bibinfo{date}{July} \bibinfo{year}{2010}), \bibinfo{pages}{2165--2176}.
\newblock
\showISSN{0045-7825}
\urldef\tempurl%
\url{https://doi.org/10.1016/j.cma.2010.03.021}
\showDOI{\tempurl}


\bibitem[Tapia et~al\mbox{.}(2020)]%
        {Tapia2020}
\bibfield{author}{\bibinfo{person}{Javier Tapia}, \bibinfo{person}{Espen
  Knoop}, \bibinfo{person}{Mojmir Mutný}, \bibinfo{person}{Miguel~A. Otaduy},
  {and} \bibinfo{person}{Moritz Bächer}.} \bibinfo{year}{2020}\natexlab{}.
\newblock \showarticletitle{MakeSense: Automated Sensor Design for
  Proprioceptive Soft Robots}.
\newblock \bibinfo{journal}{\emph{Soft Rob.}} \bibinfo{volume}{7},
  \bibinfo{number}{3} (\bibinfo{date}{June} \bibinfo{year}{2020}),
  \bibinfo{pages}{332--345}.
\newblock
\showISSN{2169-5172, 2169-5180}
\urldef\tempurl%
\url{https://doi.org/10.1089/soro.2018.0162}
\showDOI{\tempurl}


\bibitem[Tozoni et~al\mbox{.}(2020)]%
        {Tozoni2020}
\bibfield{author}{\bibinfo{person}{Davi~Colli Tozoni},
  \bibinfo{person}{J\'{e}r\'{e}mie Dumas}, \bibinfo{person}{Zhongshi Jiang},
  \bibinfo{person}{Julian Panetta}, \bibinfo{person}{Daniele Panozzo}, {and}
  \bibinfo{person}{Denis Zorin}.} \bibinfo{year}{2020}\natexlab{}.
\newblock \showarticletitle{A Low-Parametric Rhombic Microstructure Family for
  Irregular Lattices}.
\newblock \bibinfo{journal}{\emph{ACM Trans. Graph.}} \bibinfo{volume}{39},
  \bibinfo{number}{4}, Article \bibinfo{articleno}{101} (\bibinfo{date}{jul}
  \bibinfo{year}{2020}), \bibinfo{numpages}{20}~pages.
\newblock
\showISSN{0730-0301}
\urldef\tempurl%
\url{https://doi.org/10.1145/3386569.3392451}
\showDOI{\tempurl}


\bibitem[Tozoni et~al\mbox{.}(2021)]%
        {Tozoni2021}
\bibfield{author}{\bibinfo{person}{Davi~Colli Tozoni}, \bibinfo{person}{Yunfan
  Zhou}, {and} \bibinfo{person}{Denis Zorin}.} \bibinfo{year}{2021}\natexlab{}.
\newblock \showarticletitle{Optimizing Contact-Based Assemblies}.
\newblock \bibinfo{journal}{\emph{ACM Trans. Graph.}} \bibinfo{volume}{40},
  \bibinfo{number}{6}, Article \bibinfo{articleno}{269} (\bibinfo{date}{dec}
  \bibinfo{year}{2021}), \bibinfo{numpages}{19}~pages.
\newblock
\showISSN{0730-0301}
\urldef\tempurl%
\url{https://doi.org/10.1145/3478513.3480552}
\showDOI{\tempurl}


\bibitem[{van Keulen} et~al\mbox{.}(2005)]%
        {VANKEULEN20053213}
\bibfield{author}{\bibinfo{person}{F. {van Keulen}}, \bibinfo{person}{R.T.
  Haftka}, {and} \bibinfo{person}{N.H. Kim}.} \bibinfo{year}{2005}\natexlab{}.
\newblock \showarticletitle{Review of options for structural design sensitivity
  analysis. Part 1: Linear systems}.
\newblock \bibinfo{journal}{\emph{Computer Methods in Applied Mechanics and
  Engineering}} \bibinfo{volume}{194}, \bibinfo{number}{30}
  (\bibinfo{year}{2005}), \bibinfo{pages}{3213--3243}.
\newblock
\showISSN{0045-7825}
\urldef\tempurl%
\url{https://doi.org/10.1016/j.cma.2005.02.002}
\showDOI{\tempurl}
\newblock
\shownote{Structural and Design Optimization}.


\bibitem[Verschoor and Jalba(2019)]%
        {verschoor2019efficient}
\bibfield{author}{\bibinfo{person}{Mickeal Verschoor} {and}
  \bibinfo{person}{Andrei~C Jalba}.} \bibinfo{year}{2019}\natexlab{}.
\newblock \showarticletitle{Efficient and accurate collision response for
  elastically deformable models}.
\newblock \bibinfo{journal}{\emph{ACM Trans. on Graph. (TOG)}}
  \bibinfo{volume}{38}, \bibinfo{number}{2} (\bibinfo{year}{2019}).
\newblock


\bibitem[Wieschollek(2016)]%
        {cppoptlib}
\bibfield{author}{\bibinfo{person}{Patrick Wieschollek}.}
  \bibinfo{year}{2016}\natexlab{}.
\newblock \bibinfo{title}{CppOptimizationLibrary}.
\newblock
  \bibinfo{howpublished}{\url{https://github.com/PatWie/CppNumericalSolvers}}.
\newblock


\bibitem[Wriggers(1995)]%
        {wriggers1995finite}
\bibfield{author}{\bibinfo{person}{Peter Wriggers}.}
  \bibinfo{year}{1995}\natexlab{}.
\newblock \showarticletitle{Finite Element Algorithms for Contact Problems}.
\newblock \bibinfo{journal}{\emph{Archives of Comp. Meth. in Eng.}}
  \bibinfo{volume}{2} (\bibinfo{date}{12} \bibinfo{year}{1995}).
\newblock


\bibitem[Xu et~al\mbox{.}(2022)]%
        {nvidiawarp}
\bibfield{author}{\bibinfo{person}{Jie Xu}, \bibinfo{person}{Viktor
  Makoviychuk}, \bibinfo{person}{Yashraj Narang}, \bibinfo{person}{Fabio
  Ramos}, \bibinfo{person}{Wojciech Matusik}, \bibinfo{person}{Animesh Garg},
  {and} \bibinfo{person}{Miles Macklin}.} \bibinfo{year}{2022}\natexlab{}.
\newblock \bibinfo{title}{Accelerated Policy Learning with Parallel
  Differentiable Simulation}.
\newblock
\newblock
\urldef\tempurl%
\url{https://doi.org/10.48550/ARXIV.2204.07137}
\showDOI{\tempurl}


\bibitem[Zhang et~al\mbox{.}(2016)]%
        {Zhang2016}
\bibfield{author}{\bibinfo{person}{Xiaoting Zhang}, \bibinfo{person}{Xinyi Le},
  \bibinfo{person}{Zihao Wu}, \bibinfo{person}{Emily Whiting}, {and}
  \bibinfo{person}{Charlie~C.L. Wang}.} \bibinfo{year}{2016}\natexlab{}.
\newblock \showarticletitle{Data-Driven Bending Elasticity Design by Shell
  Thickness}.
\newblock \bibinfo{journal}{\emph{Computer Graphics Forum (Proceedings of
  Symposium on Geometry Processing)}} \bibinfo{volume}{35}, \bibinfo{number}{5}
  (\bibinfo{year}{2016}), \bibinfo{pages}{157--166}.
\newblock


\end{thebibliography}

\appendix

\section{Time-dependent problems}
\label{appsec:timedependent}

In this section, we show how to compute the derivative and adjoint equation for the time-dependent case.
We do this in general form, only assuming that the force terms depend on solution and optimization parameters but not explicitly on time.

\paragraph{Problem setup}
We assume all quantities involved in the adjoint equations and shape derivatives for the static case known from the main text. In this appendix, we derive how to update these to obtain the adjoint equation for the time-dependent PDE.

We consider the following time-dependent system, discretized in space.
\[
\dot{\vu} = \vv;\; M(\prm) \dot{\vv} =  \pde(\vu,\prm);\; \vu(0) = \ic^u(\prm);\; \vv(0) = \ic^v(\prm)
\]
where $M(\prm)$ is the mass matrix, which may also depend on parameters $p$.

We  assume that the discretization in time uses a BDF scheme of order $m$:
\[\dot{\vu} \approx \frac{1}{\beta \dt}(\vu^i + \sum_{j=1}^{\min(i,m)}\alpha^i_j \vu^{i-j}).\]

In general, $\alpha^i_j$ does not depend on $i$, except
at the first $m-1$ steps, when a higher-order scheme needs to be initialized with lower-order steps; more specifically,
$\alpha^i_j$ is $j$-th coefficient of BDFi, for $1 \leq i < m$,
and $j$-th coefficient of BDFm otherwise.

We assume that $ \pde(\vu,\prm)$ does not directly depend on the
velocities $\vv$; if a dependence on velocities is needed,
as we see below, it can be expressed directly in terms of $u$.

The discrete system has the form
\begin{equation}
\begin{split}
\vu^i + \sum_{j=1}^{\min(i,m)}\alpha^i_j \vu^{i-j} &= \beta_i \dt \vv^{i},\\
M \bigg(\vv^i + \sum_{j=1}^{\min(i,m)}\alpha^i_j \vv^{i-j}\bigg) &= \beta_i \dt \pde^i(\vu^i, \vu^{i-1},\prm) = \hht^i.
\end{split}
\label{eq:forward}
\end{equation}
where $M$  is the mass matrix. This is the form in which the system is solved in~\cite{Li2020IPC}.

For  time-dependent problems, we consider functionals of the form
\[
J(\vu,\prm) = \int_{t=0}^T \objt(\vu,t,\prm) dt,
\]
where $\objt(\vu,t,\prm)$ is a spatial functional, e.g., integral over
the solid $\Omega(t)$ or its surface, of some pointwise quantity depending on the solution and/or its derivatives pointwise.
In discretized form, this functional is

\[
J(\vu,\prm) = \sum_{i=0}^N w_i \objt_{i}(\vu^i,\prm) = \sum_{i=0}^N \htobj^i,
\]
where $w_i$ are quadrature weights (e.g., all $\dt$ in the simplest case), and N is the number of time steps.

\paragraph{Remark on notation} We omit most of the explicit arguments in functions $h$ and $J$
used in the expressions, to make the formulas more readable. The following is implied:
$ \pde(\vu^i, \vu^{i-1}, \prm,t_i) = \pde_i(\vu^i,\vu^{i-1}, \prm) = \pde_i$ and similarly for $J_i$.

\paragraph{Summary}
Computing the derivative $\fdpar J$ requires the following
components
\begin{itemize}
    \item Derivatives $\dpu J_i$,  $\dpu \pde_i$, $\dpp J_i$ and
    $\dpp \pde_i$. See Sections 7 to 9 in main text for corresponding formulas.
    \item Derivatives $\dpp \ic^u$ and $\dpp \ic^v$, derivatives of the initial conditions. See Section 5.4 in main text.
\end{itemize}

To compute the parametric derivative of $J$, the steps are as follows:
\begin{itemize}
    \item Solve the forward system \eqref{eq:forward}, and store
    the resulting solutions $\vu^i, \vv^i$, $i=0\ldots N$ at every step.
    \item Initialize adjoint variables $\vadj_N,\vnu_N$ from   \eqref{eq:adjoint-init-euler} (general BDF: \eqref{eq:adjoint-init}).
    \item Perform backward time stepping using   \eqref{eq:time-adjoint-euler} (general BDF: \eqref{eq:time-adjoint}).
    \item At every step, evaluate derivative of the
    mass matrix $\fdpar M$, if applicable, and use formula \eqref{eq:deriv-from-adjoint-euler} (general BDF: \eqref{eq:deriv-from-adjoint}) to update $\fdpar J$.
\end{itemize}

\subsection{Implicit Euler}
\label{appsec:euler-adjoint}

\paragraph{Discrete Lagrangian}
We use the Lagrangian-based approach (C\'{e}a's method) to derive the  adjoint equation.  The overall idea is to
write the Lagrangian $\cL$ for the functional $J$ viewing
the equations for $\dot{\vv}$ and $\dot{\vu}$ as constraints
with Lagrange multipliers $\vadj$ and $\vmu$.
For the solution $(\vu,\vv)$  for any optimization parameter values, the constraints are satisfied,
$\fdpar J = \fdpar \cL$, as the constraint terms identically vanish. The goal of introducing the adjoint variables is
to eliminate the direct dependence of  $\fdpar J$ on the displacement and velocity derivatives: $\fdpar \vu^i$ or $\fdpar \vv^i$.

To achieve our objective, we expand the derivative $\fdpar \cL$, and isolate the terms multiplying $\fdpar \vu$ and $\fdpar \vv$. By setting the sum of each of these two sets of terms to zero (which corresponds to our adjoint equations), we can find $\vadj$ and $\vmu$ so that the derivative of the functional $\fdpar J$ does not directly depend on $\fdpar \vu^i$ or $\fdpar \vv^i$.

The time-stepping for implicit Euler/BDF1 has the following simple form:
\begin{equation}
\begin{split}
\vu^i - \vu^{i-1} &= \dt\vv^{i},\\
M (\vv^i  -\vv^{i-1}) &= \dt \pde^i = \hht^i.
\end{split}
\label{eq:forward-euler}
\end{equation}

We introduce adjoint variables $\adj_i$ and $\mu_i$
(we use subscripts for the adjoint variables to indicate the time step,  as these are often transposed in the formulas to make the formulas more readable).

In the derivation below, we drop most dependencies on variables, assuming $\htobj^i = \htobj^i(\vu^i,\prm)$,
$\ic^u = \ic^u(\prm)$, $\ic^v = \ic^v(\prm)$,
$\hht^i =  \hht(\vu^i, \vu^{i-1}, \prm)$ and $M = M(\prm)$.

The Lagrangian $\cL$ has the form

\[
\footnotesize
\begin{aligned}
\cL &= \sum_{i=0}^N \htobj^i & \mbox{\{objective terms\}} \\
&+\vadj_0^T (\vv^0 - \ic^v) + \vmu_0^T(\vu^0 - \ic^u) &\mbox{\{initial condition terms\}}\\
&+\sum_{i=1}^N \lmT_i
(M (\vv^i  -\vv^{i-1}) - \hht^i) + \vmu^T_i (
\vu^i - \vu^{i-1} - \dt\vv^{i}).&\mbox{\{PDE terms\}}\\
\end{aligned}
\]
Rearranging terms, and shifting summation index for $\vu^{i-1}$,
\[
\small
\begin{aligned}
\cL &=
\vadj_0^T (\vv^0 - \ic^v) + \vmu_0^T(\vu^0 - \ic^u) +
 \htobj^0\\
 &+
\sum_{i=1}^N
 \htobj^i +\lmT_i
(M (\vv^i  -\vv^{i-1}) - \hht^i) + \vmu^T_i (
\vu^i - \vu^{i-1} - \dt\vv^{i})\\
&=
\vadj_0^T (\vv^0 - \ic^v) + \vmu_0^T(\vu^0 - \ic^u) +
 \htobj^0\\
 &+
\sum_{i=1}^N
 \htobj^i +\lmT_i
(M\vv^i - \hht^i) + \vmu^T_i (
\vu^i - \dt\vv^{i})
-
\sum_{i=0}^{N-1}
\lmT_{i+1} M\vv^{i} + \vmu^T_{i+1}  \vu^{i}.
\end{aligned}
\]
Combining two sums back together and separating $N$-th term from the first, we get

\[
\small
\begin{split}
\cL &=
\vadj_0^T (\vv^0 - \ic^v) + \vmu_0^T(\vu^0 - \ic^u) +
 \htobj^0 -\lmT_1 M\vv^0 - \vmu^T_1  \vu^{0}\\
 &+
\sum_{i=1}^{N-1}
 \htobj^i +\lmT_i
(M\vv^i - \hht^i) + \vmu^T_i (
\vu^i - \dt\vv^{i})
- \lmT_{i+1} M\vv^{i} - \vmu^T_{i+1}  \vu^{i}
\\
&+ \htobj^N +\lmT_N
(M\vv^N - \hht^N) + \vmu^T_N (
\vu^N - \dt\vv^{N}).
\end{split}
\]
Collecting $\vu^i$ and $\vv^i$ terms:

\[
\small
\begin{split}
\cL &=
 \htobj^0 -\vadj_0^T \ic^v - \vmu_0^T \ic^u  +
(\vadj_0^T - \lmT_1 M)\vv^0  + (\vmu_0^T  - \vmu^T_1) \vu^0   \\
 &+
\sum_{i=1}^{N-1}
 \htobj^i -  \lmT_i \hht^i +
 (\vmu^T_i   - \vmu^T_{i+1}) \vu^{i}
+ ((\lmT_i   - \lmT_{i+1} )M  - \vmu^T_i \dt)\vv^{i}
\\
&+ \htobj^N  - \lmT_N \hht^N
+ \vmu^T_N  \vu^N  + (\lmT_N M  - \vmu^T_N \dt)\vv^N.
\end{split}
\]

Differentiating with respect to $\prm$:
\[
\scriptsize
\begin{split}
\fdpar \cL &=
 \dpp \htobj^0
 -\vadj_0^T \dpp \ic^v - \vmu_0^T \dpp \ic^u
 - \lmT_1 \dpp M\vv^0 +
(\vadj_0^T - \lmT_1 M) \fdpar\vv^0
+\\&+ (  \dpui \htobj^0 + \vmu_0^T  - \vmu^T_1) \fdpar \vu^0   +\\
 &+
\sum_{i=1}^{N-1}
 \dpp \htobj^i  -  \lmT_i \dpp \hht^i  + (\lmT_i - \lmT_{i+1}) \dpp M \vv^{i}+
 \\
&+ \sum_{i=1}^{N-1}
 (\vmu^T_i   - \vmu^T_{i+1} + \dpui \htobj^i  -  \lmT_i \dpui \hht^i )\fdpar \vu^i - \lmT_i \dpuim \hht^i \, \fdpar \vu^{i-1}
+\\&+ ((\lmT_i   - \lmT_{i+1} )M  - \vmu^T_i \dt) \fdpar\vv^{i}
+\\
&
+ \dpp \htobj^N  - \lmT_N \dpp \hht^N
+ (\dpui \htobj^N  - \lmT_N \dpun \hht^N + \vmu^T_N) \fdpar \vu^N +\\&- \lmT_N \dpunm \hht^N \fdpar \vu^{N-1} + (\lmT_N M  - \vmu^T_N \dt) \fdpar\vv^N + \lmT_N \dpp M \vv^N.
\end{split}
\]
Reorganizing to have all terms for each $\fdpar \vu^i$ together:
\[
\scriptsize
\begin{split}
\fdpar \cL &=
 \dpp \htobj^0
 -\vadj_0^T \dpp \ic^v - \vmu_0^T \dpp \ic^u
 - \lmT_1 \dpp M\vv^0 +
(\vadj_0^T - \lmT_1 M) \fdpar\vv^0
+\\&+ (  \dpui \htobj^0 + \vmu_0^T  - \vmu^T_1 - \lmT_{1} \dpuz \hht^{1}) \fdpar \vu^0   \\
 &+
\sum_{i=1}^{N-1}
 \dpp \htobj^i  -  \lmT_i \dpp \hht^i  + (\lmT_i - \lmT_{i+1}) \dpp M \vv^{i}
 \\
&+ \sum_{i=1}^{N-1}
 (\vmu^T_i   - \vmu^T_{i+1} + \dpui \htobj^i  -  \lmT_i \dpui \hht^i - \lmT_{i+1} \dpui \hht^{i+1})\fdpar \vu^i
+\\&+ ((\lmT_i   - \lmT_{i+1} )M  - \vmu^T_i \dt) \fdpar\vv^{i}
\\
&
+ \dpp \htobj^N  - \lmT_N \dpp \hht^N
+ (\dpui \htobj^N  - \lmT_N \dpun \hht^N + \vmu^T_N) \fdpar \vu^N +\\&+ (\lmT_N M  - \vmu^T_N \dt) \fdpar\vv^N + \lmT_N \dpp M \vv^N.
\end{split}
\]

Introducing $\vmu = M^T \vnu$, we obtain the following adjoint equations from the terms multiplying $\fdpar \vv^i$
and $\fdpar \vu^i$ in the summation:

\begin{equation}
\boxed{
\begin{aligned}
\vadj_i   - \vadj_{i+1}   &=   \dt \vnu_i,\\
M^T(\vnu_i   - \vnu_{i+1})  &=    (\dpui \hht^i)^T \vadj_i + (\dpui \hht^{i+1})^T \vadj_{i+1}  - (\dpui \htobj^i)^T.
\end{aligned}
\label{eq:time-adjoint-euler}
}
\end{equation}
For the initial conditions we get from the terms multiplying
$\fdpar \vv^N$ and  $\fdpar \vu^N$:

\begin{equation}
\begin{aligned}
&\vadj_N = \dt \vnu_N,\\
&M^T\vnu_N=(\dpui \hht^N)^T\vadj_N -(\dpui \htobj^N)^T.
\end{aligned}
\end{equation}

By introducing $\vadj_{N+1}$ and $\vnu_{N+1}$, the initial conditions can be simplified as
\begin{equation}
\boxed{
\begin{aligned}
&\vadj_{N+1} = 0,\\
&\vnu_{N+1} = 0.
\end{aligned}
}
\label{eq:adjoint-init-euler}
\end{equation}

For $\vadj_0, \vnu_0$ we have $M^T \vnu_0 = -(\dpuz \htobj^0)^T + M^T \vnu_1 + \lmT_{1} \dpuz \hht^{1}$ and $\vadj_0 = M^T \vadj_1$.

Finally, the expression for $\fdpar J$ is obtained by dropping all terms with  $\fdpar \vv^i$ and  $\fdpar \vu^i$, as these are set to zero by our choice of equations for the adjoint, and retaining the rest:

\begin{equation}
\boxed{
\begin{aligned}
\fdpar J &=
 \dpp \htobj^0   -\vadj_0^T \dpp \ic^v - \vmu_0^T \dpp \ic^u
 - \lmT_1 \dpp M\vv^0, \\
 &+
\sum_{i=1}^{N}
\dpp \htobj^i  -  \lmT_i \dpp \hht^i + \dt \vnu_i^T \dpp M \vv^{i}.
\end{aligned}
}
\label{eq:deriv-from-adjoint-euler}
\end{equation}

\subsection{General BDF time integration}
\label{appsec:adjoint-bdf}

\paragraph{Discrete Lagrangian}
For the general case, we split the Lagrangian $\cL(\vu,\vv,\vadj,\vmu,\prm)$  into three  parts:
$J(\vu,\prm)$ itself,  the part $\cL_c$ containing the Lagrange  multipliers for the time steps $i=1\ldots N$,  and the part for initial conditions
$\cL_{in}$
\[
\cL(\vu,\vv,\vadj,\vmu,\prm) = J(\vu,\prm) + \cL_c(\vu,\vv,\vadj,\vmu,\prm) + \cL_{in}
(\vu^0,\vv^0,\vadj^0,\vmu^0,\prm),
\]
where
\[
\cL_{in} = \vadj_0^T (\vv^0 - \ic^v) + \vmu_0^T(\vu^0 - \ic^u).
\]

We start with $\cL_c$. Remember that $\cL_c$ depends on $h^i$, which has inputs $x$, $u^i$ and $u^{i-1}$ (due to friction):
\[
\small
\cL_c = \sum_{i=1}^N \lmT_i M \bigg(\vv^i + \sum_{j=1}^{\min(i,m)}\alpha^i_j\vv^{i-j} - \hht^i \bigg) + \vmu^T_i \bigg( \vu^i + \sum_{j=1}^{\min(i,m)}\alpha^i_j \vu^{i-j} - \beta_i \dt\vv^{i}\bigg).
\]

We rearrange the double summations in this expression, so that
each term depends only on $\vu^i$ and $\vv^i$, as the adjoint
equations will be obtained by setting coefficients of
$\fdpar \vu^i$ and $\fdpar\vv^i$ to zero after differentiation.

If we have a sum of the form
\[ \sum_{i=1}^N \sum_{j=1}^{\min(i,m)} \alpha_j^i c_i^T z_{i-j},\]
we can change the summation order: let $r = i-j$,
\tiny
\[
\sum_{i=1}^N \sum_{r=\max(0,i-m)}^{i-1}  \alpha_{i-r}^i c_i^T z_r
=
\sum_{r=0}^{N-1} \sum_{i=r+1}^{\min(r+m,N)} \alpha_{i-r}^i c_i^T z_r =
\sum_{r=0}^{N-1} \sum_{j=1}^{\min(m,N-r)} \alpha_{j}^{j+r}  c_{r+j}^T z_r,
\]
\normalsize
where we introduced back $j = i-r$ in the last equation.
Finally, renaming $r$ to $i$, we obtain the form for which each term
contains $z_i$ only:
\begin{equation}
\small
\sum_{i=0}^{N-1} \bigg(\sum_{j=1}^{\min(m,N-i)} \alpha_{j}^{i+j} c_{i+j}^T \bigg) z_i.
\label{eq:switchsum}
\end{equation}

Returning to the Lagrangian, we regroup the terms in  $\cL$ as:
\[
\tiny
\cL_c =
\sum_{i=1}^N \left( \lmT_i \left(M\vv^i - \hht^i\right) +
 \vmu^T_i  \left( \vu^i  - \beta_i \dt\vv^{i}\right)\right) +
 \sum_{i=1}^N \sum_{j=1}^{\min(i,m)}
\left( \lmT_i M \alpha^i_j\vv^{i-j} + \vmu^T_i \alpha^i_j \vu^{i-j} \right).
\]
Using \eqref{eq:switchsum}, we get
\begin{align*}
\tiny
\cL_c =
\sum_{i=1}^N \left( \lmT_i \left(M\vv^i - \hht^i\right) +
 \vmu^T_i  \left( \vu^i  - \beta_i \dt\vv^{i}\right)\right) +
 \sum_{i=0}^{N-1} \sum_{j=1}^{\min(m,N-i)}
\left( \lmT_{i+j} M \alpha^{i+j}_j\vv^i + \vmu^T_{i+j} \alpha^{i+j}_j \vu^i \right).
\end{align*}
Collecting the terms for $\vu^i$ and $\vv^i$:
\[
\scriptsize
\begin{split}
\cL_c &= \\
&\sum_{i=1}^{N-1}
- \lmT_i \hht^i +
  \left( \left( \lmT_i
+ \sum_{j=1}^{\min(m,N-i)}
\alpha^{i+j}_j \lmT_{i+j}\right) M
 - \beta_i \dt \vmu^T_i
\right)\vv^i +\\&+
 \left(\vmu^T_i    +
 \sum_{j=1}^{\min(m,N-i)}
 \alpha^{i+j}_j \vmu^T_{i+j} \right) \vu^i
 +
 \\
&\sum_{j=1}^m
 \lmT_{j} M \alpha^{j}_j\vv^0 + \vmu^T_{j} \alpha^{j}_j \vu^0
+ \\
&  -  \lmT_N \hht^N + (\lmT_N M - \beta_N  \dt \vmu^T_N)\vv^N +
 \vmu^T_N  \vu^N.
\end{split}
\]
We split this expression again, into $\cL^{mid}_c + \cL^0_c + \cL^N_c$,
corresponding to three lines of the equation; these terms contribute
to  the time-dependent adjoint equations and boundary conditions.
In this form, it is straightforward to differentiate with respect
to $q$:

\[
\scriptsize
\begin{split}
\fdpar \cL_c^{mid} &= \\
&\sum_{i=1}^{N-1}
- \lmT_i \dpp \hht^i
+  \left( \lmT_i
+ \sum_{j=1}^{\min(m,N-i)}
\alpha^{i+j}_j \lmT_{i+j}\right) \fdpar M
\vv^i +
\\
&\sum_{i=1}^{N-1}
  \left( \bigg( \lmT_i
+ \sum_{j=1}^{\min(m,N-i)}
\alpha^{i+j}_j \lmT_{i+j}\bigg) M
 - \beta_i \dt \vmu^T_i
\right) \fdpar\vv^i +\\&+
 \left(\vmu^T_i    +
 \sum_{j=1}^{\min(m,N-i)}
 \alpha^{i+j}_j \vmu^T_{i+j}
 - \lmT_i \dpui \hht^i -\lmT_{i+1} \, \dpui \hht^{i+1}
 \right) \fdpar \vu^i +
 \\
& -\lmT_{1} \, \dpuz \hht^{1} \, \fdpar \vu^{0}
\\
\fdpar \cL_c^0 &=
\sum_{j=1}^m
\alpha^{j}_j \left( \lmT_{j} \fdpar M \vv^0 +
\lmT_{j} M \fdpar\vv^0 +  \vmu^T_{j}  \fdpar \vu^0 \right)
\\
\fdpar \cL_c^N &=
 -  \lmT_N \dpp \hht^N -  \lmT_N \dpun \hht^N \fdpar \vu^N + \lmT_N \fdpar M \vv^N +\\&+  (\lmT_N M - \beta_N  \dt \vmu^T_N) \fdpar\vv^N +
 \vmu^T_N  \fdpar \vu^N.
 \end{split}
\]
\normalsize

Similarly, we split
\tiny
\[
\fdpar J = \sum_{i=1}^{N-1} \dpp \htobj^{i} +
\sum_{i=1}^{N-1} \dpu \htobj^{i} \fdpar \vu^i + \big(\dpp \htobj^0+  \dpu \htobj^{0} \fdpar \vu^0\big)  + \big(\dpp \htobj^{N} +  \dpu \htobj^{N} \fdpar \vu^N\big)
= \fdpar J^{mid} + \fdpar J^0 + \fdpar J^N.
\]
\normalsize

\paragraph{Adjoint equations}
Equating the coefficients of $\fdpar \vu^i$ and $\fdpar\vv^i$, $i=1\ldots N-1$ to zero  in $\fdpar \cL_c^{mid} + \fdpar J^{mid}$, we obtain the
adjoint equations:
\begin{equation}
\scriptsize
\begin{split}
& M^T \bigg( \vadj_i   + \sum_{j=1}^{\min(m,N-i)} \alpha^{i+j}_j \vadj_{i+j}\bigg)   = \beta_i \dt \vmu_i,\\
&\vmu_i    +
 \sum_{j=1}^{\min(m,N-i)}
 \alpha^{i+j}_j \vmu_{i+j}
 = (\dpui \hht^i)^T \vadj_i  + (\dpui \hht^{i+1})^T \vadj_{i+1} -  (\dpu \htobj^{i})^T,
\end{split}
\label{eq:time-adjoint-mu}
\end{equation}
for $i = 1 \ldots N-1$.

Making a substitution $\vmu = M^T \hat{\vnu}$, we obtain~\eqref{eq:time-adjoint}.

We obtain the adjoint equation in time in the form very similar
to the forward equations \eqref{eq:forward}.
The most important difference is that the integration is "back in time",
i.e., the finite difference formula for time derivative is applied to $i,\ldots i+m$.
This means that the system is integrated backwards,
starting with $(\vadj_N, \vmu_N)$. Second, there is a slight
difference in the coefficients of the scheme used. Specifically,
the starting iterations do \emph{not} use the lower-order BDF formulas,
rather truncations of the same order BDF formula. At the same time,
the end iterations, for small $i$, will use lower order coefficients,
even though this is not needed.
The reason for preferring this (although this seemingly damages the accuracy
of the integration of the adjoint) is consistency with the functional
discretization: as this fact is a consequence
of deriving the adjoint from the time discretization, if
we compute the functional using the same discretization,
finite differences for the functional will be closer to the adjoint.

\paragraph{Initial conditions for adjoint}
The initial conditions follow from setting coefficients of $\fdpar \vu^N$
and $\fdpar\vv^N$ to zero in $\fdpar \cL_c^N  \fdpar J^N$,
i.e.
\[
(\dpu \htobj^N)^T - \dpu \hht^N \vadj_N  +  \vmu_N  = 0;\;
 M^T \vadj_N - \beta_N  \dt \vmu_N  = 0.
\]
Substituting $\vmu = M^T\vnu$, we get
\[
(\dpu \htobj^N)^T - \dpun \hht^N \vadj_N  +  M^T \vnu_N  = 0;\;
  \vadj_N - \beta_N  \dt \vnu_N  = 0,
\]
and a linear system for  for $\vnu_N$:
\[
(\dpu \htobj^N)^T  +  (M^T - \beta_N \dt \dpun \hht^N) \vnu_N = 0.
\]
Solving these
\begin{equation}
    (M^T - \beta_N \dt \dpun \hht^N) \vnu_N = - (\dpu \htobj^N)^T,\;
      \vadj_N =  \beta_N  \dt \vnu_N.
\end{equation}

By introducing $\vadj_{N+1},\ \vnu_{N+1}$, the initial condition can be simplified as~\eqref{eq:adjoint-init}.

Finally, the adjoint equations  \eqref{eq:time-adjoint}
only allow to solve down to to $i = 1$; the equations for
$\vadj_0, \vmu_0$ are derived from $\fdpar J^0 + \fdpar \cL_{in} + \fdpar \cL_c^0 $;  adding terms containing $\fdpar \vu^0$ and $\fdpar\vv^0$, we get:
\[
\left(  \dpu \htobj^{0}  + \vmu_0^T
+\sum_{j=1}^m
\alpha^{j}_j   \vmu^T_{j} - \vadj_1^T \dpuz \hht^{1}  \right)  \fdpar \vu^0 +
\left( \vadj_0^T  +\sum_{j=1}^m
\alpha^{j}_j
\lmT_{j} M   \right) \fdpar\vv^0,
\]
which yields direct expressions \eqref{eq:zero-adjoint} for $\vadj_0$ and $\vmu_0$,
based on $\vadj_i, \vmu_i$ for $i=1\ldots m$.

\paragraph{Computing the derivative of $J$ from the forward and adjoint solutions}
Finally, once the adjoint variables are obtained, we can compute \eqref{eq:deriv-from-adjoint},
by collecting all terms not containing $\fdpar \vu^i$
and $\fdpar\vv^i$.

Partial derivatives $\dpp \hht, \dpu \hht$ and $\dpp \htobj_i$, $\dpu \htobj_i$
are exactly the same as used in the construction of the system for
static adjoint and computation of the functional. The differences, specific
to time discretization, are:
\begin{itemize}
\item Mass matrix derivative $\fdpar M$. See Section~\ref{appsec:mass-matrix-deriv}.
\item Partial derivatives  of the initial conditions with respect to parameters  $\dpp \ic^v$ and $\dpp \ic^u$, for positions and velocities (See Appendix~\ref{appsec:initcond-derivatives}).
Typically, a 3D position and velocity for the whole object (or angular velocity for the object rotating as a rigid body) are used as parameters, so these are trivial to compute.
\end{itemize}

\subsection{Mass matrix derivative}
\label{appsec:mass-matrix-deriv}
Consider our Mass Matrix as follows:
\begin{align*}
    M_{Ds+i, Dt+i} = \sum_{e \in E(s) \cap E(t)} \int_{\fprms^e(\hat{K}_e)} \rho(\q) \, \ubas^{\loc{e}{s}}(x) \,\, \ubas^{\loc{e}{t}}(x) \,\, dx,
\end{align*}
where $i \in \{1, .. , D\}$ and $D$ equals $2$ or $3$ (representing dimension). This means mass matrix $M$ has $D \cdot n$ rows and columns.
Here, we use notation to define the local index of a node $\loc{e}{\ell}$ with respect to elements $e$ containing it.

Then, we can obtain the shape derivative with respect to perturbation $\theta$ ($x^\epsilon = x + \epsilon \theta(x)$) by computing the Gateaux Derivative below for each element:
\footnotesize
\begin{align*}
    \partial_q M_{Ds+i, Dt+i} &= \frac{d}{d\epsilon} \Big|_{\epsilon=0} \sum_{e \in E(s) \cap E(t)} \int_{\fprms^e(\hat{K}_e)^\epsilon} \rho(\q) \, \ubas^{\loc{e}{s}}(x^\epsilon) \,\, \ubas^{\loc{e}{t}}(x^\epsilon) \,\, dx^\epsilon \\
    &= \sum_{e \in E(s) \cap E(t)} \int_{\fprms^e(\hat{K}_e)} \dpar \rho(\q) \, \ubas^{\loc{e}{s}}(x) \,\, \ubas^{\loc{e}{t}}(x) +\\&+ \rho(\q) \, \ubas^{\loc{e}{s}}(x) \,\, \ubas^{\loc{e}{t}}(x) \,\, \nabla \cdot \theta(x) \,\, dx \\
    &= \sum_{e \in E(s) \cap E(t)} \sum_{l \in \Loc{e}} \int_{\fprms^e(\hat{K}_e)} \dpar \rho(\q) \, \ubas^{\loc{e}{s}}(x) \,\, \ubas^{\loc{e}{t}}(x) \, +\\&+ \, \rho(\q) \, \ubas^{\loc{e}{s}} \,\, \ubas^{\loc{e}{t}} \,\, \nabla \vbas^{l} \,\, dx  \cdot \theta^l. \\
\end{align*}
\normalsize

\subsection{Initial condition derivatives}
\label{appsec:initcond-derivatives}

We need to compute partial derivatives  of the initial conditions with respect to optimization parameters $\prm$, $\dpp \ic^v$ and $\dpp \ic^u$, for positions and velocities.
Notice that both $\dpp \ic^v$ and $\dpp \ic^u$ are discrete vector fields on domain $\Omega_\prms$. Consider we have one vector value $\prm^m$ per node of the domain.

If $(\ic^v)_s = \prm^m$, where $(\ic^v)_s$ is initial condition at node $s$, the derivative with respect to $q_m$ is simply the identity matrix ($\dppm (\ic^v)_s = I$). At the same time, it is the zero matrix w.r.t. any other $q^{m^*}$, with $m^* \neq m$. Same thing goes for $\ic^u$.
\section{Parametric derivatives of forces}
\label{appsec:general-forces}

In this section, we derive general expressions for gradient-dependent volume forces.

In a general form,  the contribution to the PDE can be written as 
\[
\fh^v(u,w,Q) = \int_{\Omega_\fprms} f^v( (\nabla u(x),\fprm(x) ): \nabla w\, dx.
\]

\subsection{Gradient-dependent volume forces}

\paragraph{Shape derivatives}
we omit the dependence on $\prm$, and use $\nabla f$ to denote 
$\partial_{\nabla u} f$.

Define $\Omega_\epsilon = \Omega_{\prms + \vper \epsilon}$, 
$\Omega = \Omega_{\prms}$ and  $x^\epsilon = x + \epsilon\per$.
Let $u^\epsilon$ be the solution on domain $\Omega_\epsilon$.
Then computing the G\^{a}teaux derivative of $\fh^v$
we get: 
\begin{equation}
\small
\begin{split}
    \frac{d}{d\epsilon} \Big|_{\epsilon=0} \fh^{f}  = & \frac{d}{d\epsilon} \Big|_{\epsilon=0} \int_{\Omega_\epsilon} f(\nabla_{x^\epsilon} u^\epsilon) : \nabla_{x^\epsilon} w^\epsilon \, \, dx^\epsilon \\
    = & \frac{d}{d\epsilon} \Big|_{\epsilon=0} \int_{\Omega} f((\nabla u) F_\epsilon^{-1}) : (\nabla w) F_\epsilon^{-1} \,\,  \det F_\epsilon\, \, dx \\
    = & \int_{\Omega} \frac{d}{d\epsilon} \Big|_{\epsilon=0} \left(f((\nabla u) F_\epsilon^{-1})F_\epsilon^{-T} : \nabla  w \,\,  \det F_\epsilon \right)\, \, dx \\
    = & \int_{\Omega} - f(\nabla u) \nabla \per^T : \nabla w + \boxed{(\nabla f(\nabla u) : \nabla \delta u) : \nabla w } + \\ -& (\nabla f(\nabla u) : (\nabla u\nabla \per)) : \nabla w + \left( f(\nabla u) : \nabla w \,  \right) \nabla \cdot \per\, \, dx.
\end{split}
\label{eq:grad-vol-PDEdiff}
\end{equation}

Thus we have for the shape derivative contribution: 
\[
\scriptsize
\boxed{
B^f(\per,p) = \int_{\Omega} - f(\nabla u) \nabla \per^T : \nabla p - (\nabla f(\nabla u) : (\nabla u \nabla \per)) : \nabla p + \left( f(\nabla u) : \nabla p \,  \right) \nabla \cdot \per\, \, dx.
}
\]

$B^f(\per,p)$ is linear in $\per$ and $p$, and we convert it to 
a matrix form by substituting basis functions for $\per$ and $p$:
\tiny
\begin{align*}
[&B^f]_{Da+i, Db+j} = \sum_{e \in E(a) \cap E(b)} \, \sum_{k,l \in {1..D}} \delta_{i,j} \int_{\fprms^e(\hat{K}_e)} - [\nabla \vbas^{\loc{e}{b}}]_k \, f_{kl} \, [\nabla \ubas^{\loc{e}{a}}]_l  \,\, dx \,\, +\\
&+ \sum_{e \in E(a) \cap E(b)} \, \sum_{k,l,m \in {1..D}} \int_{\fprms^e(\hat{K}_e)} - \nabla f(\nabla u)_{iklm} [\nabla \vbas^{\loc{e}{b}}]_l [\nabla u]_{jm} [\nabla \varphi^{\loc{e}{a}}]_k  \,\, dx \,\, +\\
&+ \sum_{e \in E(a) \cap E(b)} \, \sum_{k \in {1..D}} \int_{\fprms^e(\hat{K}_e)} f(\nabla u)_{ik} [\nabla \ubas^{\loc{e}{a}}]_k [\nabla \vbas^{\loc{e}{b}}]_j \,\, dx.
\end{align*}
\normalsize
the sum is over elements $e$ containing both $a$ and $b$. Again, we use notation $\loc{e}{\ell}$ to define the local index of a node with respect to the elements $e$ containing it.

 The contribution to the left-hand side of the adjoint equation is
 \[
 \boxed{
A^f(\psi,p) = \int_{\Omega} (\nabla f(\nabla u ) : \nabla \psi) : \nabla p \, dx,
}
 \]
 which is the boxed term from \eqref{eq:grad-vol-PDEdiff}, 
 corresponding to $w^T \dpu h \, \delta u$, with replacements
 $w := p$ and $\delta u := \psi$.
Discretizing according to our FE basis:
\scriptsize
\begin{align*}
    [A^f]_{Da+i, Db+j} &= \sum_{e \in E(a) \cap E(b)} \, \sum_{k,l \in {1..D}} \int_{\fprms^e(\hat{K}_e)} (\nabla f)_{ikjl} [\nabla \ubas^{\loc{e}{b}}]_l [\nabla \ubas^{\loc{e}{a}}]_k \,\, dx,
\end{align*}
\normalsize

where we sum over all elements $\Omega_e =  \fprms(\hat{K})$, with $\hat{K}$ being the reference element.
 
 \paragraph{Non-shape volumetric parameter derivatives} 
We assume that the force depends on $\fprm = \fprm(x)$, a function of the point in $\Omega_\fprms$, defined by its values $\prm$ at the same nodes as the solution, and interpolated using the same basis $\phi$.

The perturbed parameter function $\fprm$ is defined as
\begin{align*}
    \fprm^\epsilon(x) = \fprm(x) + \epsilon \per(x),
\end{align*}
where $\per(x)$ represents the perturabtion, assumed to be given in the same basis as $\fprm$ and solution. 

\begin{equation}
\begin{split}
    \frac{d}{d\epsilon} \Big|_{\epsilon=0} \fh^v  = & \frac{d}{dt} \Big|_{\epsilon=0} \int_{\Omega_\epsilon} f(\nabla u^\epsilon, \fprm^\epsilon) : \nabla w^\epsilon \, \, dx^\epsilon \\
    = & \int_{\Omega} (\nabla_1 f : \nabla \delta u) : \nabla w \, + \, (\nabla_2 f \cdot \frac{d}{d\epsilon}\fprm^\epsilon) : \nabla w \,\, dx \\
    = & \int_{\Omega} \boxed{(\nabla_1 f: \nabla \delta u) : \nabla w} \, + \, (\nabla_2 f \cdot \per) : \nabla w \,\, dx.
\end{split}
\label{eq:forceparam-grad-vol-PDEdiff}
\end{equation}
Thus, the shape derivative contribution is:
\[
\boxed{
B^f(\per,p) = \int_{\Omega} (\nabla_2 f \cdot \per) : \nabla p \,\, dx.
}
\]

Discretizing:
\scriptsize
\begin{align*}
    [B^f]_{D_s a+i, D_s b+j} &= \sum_{e \in E(a) \cap E(b)} \, \sum_{k \in {1..D_d}} \int_{\fprms^e(\hat{K}_e)} [\nabla_2 f]_{ikj} \vbas^{\loc{e}{b}} [\nabla \ubas^{\loc{e}{a}}]_k \,\, dx.
\end{align*}
\normalsize

 The contribution to the left-hand side of the adjoint equation is
 \[
 \boxed{
A^f(\psi,p) = \int_{\Omega} (\nabla_1 f : \nabla \psi) : \nabla p \, dx,
}
 \]
 which is the boxed term from \eqref{eq:forceparam-grad-vol-PDEdiff}, depending on $\delta u$ with replacements $w := p$ and $\delta u := \psi$.

Discretizing according to our FE basis
\tiny
\begin{align*}
    [A^f]_{D_s a+i, D_s b+j} &= \sum_{e \in E(a) \cap E(b)} \, \sum_{k,l \in {1..D_d}} \int_{\fprms^e(\hat{K}_e)} (\nabla_1 f)_{ikjl} [\nabla \ubas^{\loc{e}{b}}]_l [\nabla \ubas^{\loc{e}{a}}]_k \,\, dx.
\end{align*}
\normalsize

\section{General form of objective derivatives} 
\label{appsec:objectives}

For each objective $J$, the derivations below include vectors $R^o$ and $S^o$, corresponding to $\partial_u J$ and $\partial_q J$, which are necessary to compute the adjoint solution and the final shape derivative.

\subsection{Objectives depending on gradient of solution and shape}
Consider an objective that depends on both the solution of the PDE and the domain :
\begin{align}
    J(\nabla u, \Omega) = \int_{\Omega} j(\nabla u, x) dx.
\end{align}

Computing the Gateaux derivative, while considering perturbation of the domain  $x^\epsilon: = x + \epsilon\per$: 
\small
\begin{align*}
    \frac{d}{d\epsilon} \Big|_{t=0} J  &= \frac{d}{d\epsilon} \Big|_{\epsilon=0} \int_{\Omega_\epsilon} j(\nabla u^\epsilon, x^\epsilon) \,\, dx^\epsilon \\
    &= \frac{d}{d\epsilon} \Big|_{\epsilon=0} \int_{\Omega} j((\nabla u)F_\epsilon^{-1}, x+\epsilon\per) \det(F_\epsilon) \, \, dx \\
    & = \int_{\Omega} \frac{d}{d\epsilon} \Big|_{\epsilon=0} \left( j((\nabla u)F_\epsilon^{-1}, x+\epsilon\per) \det(F_\epsilon) \right) \, \, dx \\
    &= \int_{\Omega} \boxed{\nabla_1 j : \nabla \delta u} \,   - \nabla_1 j : \nabla u\nabla \per + \nabla_2 j\cdot \per + j(\nabla u, x) \nabla \cdot \per \, \, dx.
\end{align*}
\normalsize

We can select the parts not depending on $\delta u$ to be part of $S$:
\begin{align}
    S^o(\per) = \int_{\Omega} - \nabla_1 j : \nabla u\nabla \per + \nabla_2 j\cdot \per + j(\nabla u,x) \nabla \cdot \per \,\, dx.
\end{align}
Discretizing according to our FE basis:
\small
\begin{align*}
    [S^o]_{Da+i} &= \sum_{e \in E(a)} \, \int_{\fprms^{e}(\hat{K}_{e})} - (\nabla u^T)_{ij} \left( \nabla_1 j \right)_{jk}  [\nabla \vbas^{\loc{e}{a}}]_k  +\\&+ \frac{\partial j}{\partial x_i} \vbas^{\loc{e}{a}} + j(u,x) [\nabla \vbas^{\loc{e}{a}}]_i \,\, dx.
\end{align*} 
\normalsize

And,
\begin{align}
    R^o(\psi) = \int_{\Omega} \nabla_1 j : \nabla \psi \,\, dx,
\end{align}
which can be discretized as follows:
\begin{align*}
    [R^o]_{Da+i} &= \sum_{e \in E(a)} \, \int_{\fprms^{e}(\hat{K}_{e})}  (\nabla_1 j)_{ij} \,\, [\nabla \ubas^{\loc{e}{a}}]_j  \,\, dx.
\end{align*}

\subsection{Objectives depending on solution and shape}

Consider an objective that depends on both the solution of the PDE and the domain:
\begin{align}
    J(u, \Omega) = \int_{\Omega} j(u, x) dx.
\end{align}

Computing the Gateaux derivative, while considering perturbation of the domain  $x^\epsilon: = x + \epsilon\per$: 
\begin{align*}
    \frac{d}{d\epsilon} \Big|_{\epsilon=0} J  &= \frac{d}{d\epsilon} \Big|_{\epsilon=0} \int_{\Omega_\epsilon} j(u^\epsilon, x^\epsilon) \,\, dx^\epsilon \\
    &= \int_{\Omega} \boxed{\nabla_1 j \cdot \delta  u} + \nabla_2 j \cdot \per +   j \nabla \cdot \per \,\,dx.
\end{align*}

We can select the parts depending on $\delta u$, which will be the rhs of our adjoint PDE (represented by vector $R$), while the remaining part is a term that should be added directly to the shape derivative (vector $S$).

So,
\begin{align}
    S^o(\per) = \int_{\Omega} \nabla_2 j \cdot \per + j(u,x) \nabla \cdot \per \,\, dx.
\end{align}
Discretizing according to our FE basis:
\begin{align*}
    [S]_{Da+i} &= \sum_{e \in E(a)} \, \int_{\fprms^{e}(\hat{K}_{e})} \frac{\partial j}{\partial x_i} \vbas^{\loc{e}{a}} + j(u,x) [\nabla \vbas^{\loc{e}{a}}]_i \,\, dx.
\end{align*} 

And,
\begin{align}
    R^o(\psi) = \int_{\Omega} \nabla_1 j \cdot \psi \,\, dx,
\end{align}
which can be discretized as follows:
\begin{align*}
    [R]_{Da+i} &= \sum_{e \in E(a)} \, \int_{\fprms^{e}(\hat{K}_{e})}  \frac{\partial j}{\partial u_i} \,\, \ubas^{\loc{e}{a}}  \,\, dx.
\end{align*}

\section{Contact and Friction Area Term}
\label{appsec:contact-area}
In our contact and friction formulas, we use $A_k$ as a weight for our forces, which measures the area of our contact pair $k$. In the formulation, it corresponds to the sum of surface areas associated with each primitive. In 3D, it is 1/3 of the sum of areas of incident triangles for vertices and edges, and the area of triangles. For a triangle $T = (t_0, t_1, t_2)$, where $t_i$ corresponds to the position of each triangle's vertex, the corresponding triangle area will be:
\[
A_\Delta(T) = A_\Delta(t_0, t_1, t_2) = \tfrac{1}{2}\|(t_1 - t_0) \times (t_2 - t_0)\|.
\]

If $k$ corresponds to a point-triangle contact pair between point $p$ and triangle $T$, and $\text{Incid(p)}$ has the incident triangles of $p$, we have:
\[
A_k = \sum_{T \in \text{Incid}(p)} \tfrac{1}{3}  A_\Delta(T).
\]

In this context, $\dparms A_k$ corresponds to the gradient of the area term, which can be computed as a sum of the $\dparms A_\Delta$ terms:
\[
\dparms A_\Delta(T) = \frac{(t_1 - t_0) \times (t_2 - t_0)}{2\|(t_1 - t_0) \times (t_2 - t_0)\|} \dparms ((t_1 - t_0) \times (t_2 - t_0)).
\]

\section{Friction Derivative Terms}
\label{appsec:friction-derivative}
For friction, we have:
\[
B^f(\adj,\per) = \sum_k \dparms F^f_ks \per \cdot p \, A_k + F^f_k \cdot p \,\,  \dparms A_k  \,\, A_k,
\]
and 
\[
A^f(\psi,p)= \sum_k \dpu F_k \psi \cdot p \, \, A_k,
\]

which reduces to computing the derivative of each $F^f_k$ term with respect to $x$, $u^i$ and $u^{i-1}$.

\scriptsize
\begin{align}
\begin{split}
    \partial_{x_\ell} F^f_k &= - \friccoeff_{k_1,k_2}\,  \, T_k f_\eta(\lVert \tau_k \rVert) \frac{\tau_k}{\lVert \tau_k \rVert} \times\\
    &\left( \kappa \frac{N_k}{\lVert N_k \rVert} \cdot ( b'' (\partial_{\mathbf{x^d}} d_k) (\partial_{\mathbf{x^d}} d_k)^T + b' \partial_{\mathbf{x^d}}(\partial_{\mathbf{x^d}} d_k)) \,\, M^*  \right)_\ell   + \\
    &- \friccoeff_{k_1,k_2}\, N_k \left(\partial_{\mathbf{x^p}} T_k \, M^*\right)_\ell f_\eta(\lVert \tau_k \rVert) \frac{\tau_k}{\lVert \tau_k \rVert} +\\
    &- \friccoeff_{k_1,k_2}\, N_k T_k \, \frac{\tau_k}{\lVert \tau_k \rVert} \left( f_\eta' \frac{\tau_k}{\lVert \tau_k \rVert} \cdot ((\partial_{\mathbf{x^p}} T_k\, M^*)_\ell^T (\vu^i - \vu^{i-1}))\right) +\\
    &- \friccoeff_{k_1,k_2}\, N_k T_k f_\eta(\lVert \tau_k \rVert)  \left( \left( \frac{I_2}{\lVert \tau_k \rVert} - \frac{\tau_k \tau_k^T}{\lVert \tau_k \rVert ^3} \right) (\partial_{\mathbf{x^p}} T_k \, M^*)_\ell^T  (\vu^i - \vu^{i-1}) \right).
\end{split}
\end{align}

\begin{align}
\begin{split}
    \partial_{u^i_\ell} F^f_k  &= - \friccoeff_{k_1,k_2}\, N_k T_k \, \frac{\tau_k}{\lVert \tau_k \rVert} \left( f_\eta' \frac{\tau_k}{\lVert \tau_k \rVert} \cdot \left(T_k^T\right)_\ell \right)  +\\
    &- \friccoeff_{k_1,k_2}\, N_k T_k f_\eta(\lVert \tau_k \rVert)  \left( \left( \frac{I_2}{\lVert \tau_k \rVert} - \frac{\tau_k \tau_k^T}{\lVert \tau_k \rVert ^3} \right) \left(T_k^T\right)_\ell  \right).
\end{split}
\end{align}

\begin{align}
\begin{split}
    \partial_{u^{i-1}_\ell} F^f_k &= - \friccoeff_{k_1,k_2}\,  \, T_k f_\eta(\lVert \tau_k \rVert) \frac{\tau_k}{\lVert \tau_k \rVert} \times\\
    &\left( \kappa \frac{N_k}{\lVert N_k \rVert} \cdot ( b'' \nabla d_k \nabla d_k^T + b' \nabla^2 d_k)  \right)_\ell  + \\
    &- \friccoeff_{k_1,k_2}\, N_k \left(\partial_{\mathbf{x^p}} T_k \right)_\ell f_\eta(\lVert \tau_k \rVert) \frac{\tau_k}{\lVert \tau_k \rVert}  + \\
    &- \friccoeff_{k_1,k_2}\, N_k T_k \, \frac{\tau_k}{\lVert \tau_k \rVert} \left( f_\eta' \frac{\tau_k}{\lVert \tau_k \rVert} \cdot \left(\partial_{\mathbf{x^p}} T_k\right)_\ell^T (\vu^i - \vu^{i-1}))\right)  +\\
    &- \friccoeff_{k_1,k_2}\, N_k T_k f_\eta(\lVert \tau_k \rVert)  \left( \left( \frac{I_2}{\lVert \tau_k \rVert} - \frac{\tau_k \tau_k^T}{\lVert \tau_k \rVert ^3} \right) \left( \partial_{\mathbf{x^p}} T_k\right)_\ell^T (\vu^i - \vu^{i-1}) \right).
\end{split}
\end{align}
\normalsize

\section{Regularization Derivatives}
\label{appsec:regularization-derivatives}
\paragraph{Scale-invariant smoothing}
The regularization in \eqref{eq:obj:smoothing} is used for shape optimization, when the optimization parameter is $q_m = v_i$, a vertex of the shape. That said, the derivative with respect to each vertex $v_i$ is:

\begin{align*}
\partial_{v_i} J^{\text{smooth}} &= p \|s_i\|^{p-2} s_i^T (\partial_{v_i} s_i) + \sum_{j\in N(i)\cap B} p \|s_j\|^{p-2} s_j^T (\partial_{v_i} s_j).
\end{align*}

And we have that
\scriptsize
\begin{align*}
   \partial_{v_i} s_i = \frac{| N(i)\cap B | I}{\sum_{j\in N(i)\cap B}\|v_i - v_j\|} - \frac{\left( \sum_{j\in N(i)\cap B} \frac{v_i - v_j}{\|v_i-v_j\|} \right) \left( \sum_{j\in N(i)\cap B} (v_i - v_j) \right)^T}{\left(\sum_{j\in N(i)\cap B}\|v_i - v_j\|\right)^2}.
\end{align*}
\normalsize

And, for $\partial_{v_i} s_j$, where $v_i$ is one of the neighbors of $v_j$:
\scriptsize
\begin{align*}
   \partial_{v_i} s_j = - \frac{I}{\sum_{k\in N(j)\cap B}\|v_j - v_k\|} + \frac{\left(\frac{v_j - v_i}{\|v_j-v_i\|} \right) \left( \sum_{k\in N(j)\cap B} (v_j - v_k) \right)^T}{\left(\sum_{k\in N(j)\cap B}\|v_j - v_k\|\right)^2}.
\end{align*}
\normalsize

\paragraph{Material parameter spatial smoothing}
For derivatives of \eqref{eq:obj:matsmoothing} with respect to material parameters $\lambda_i, \mu_i$ we have
$$
\partial_{\lambda_i} J^{\lambda, \mu \ \text{smooth}} = 2 \sum_{t' \in Adj(t_i)} \left( \frac{\lambda_{t_i}}{\lambda_{t'}} - 1 \right) + \left(1 - \frac{\lambda_{t'}}{\lambda_{t_i}} \right) \frac{\lambda_{t'}}{\lambda_{t_i}^2},
$$
and
$$
\partial_{\mu_i} J^{\lambda, \mu \ \text{smooth}} = 2 \sum_{t' \in Adj(t_i)} \left( \frac{\mu_{t_i}}{\mu_{t'}} - 1 \right) + \left(1 - \frac{\mu_{t'}}{\mu_{t_i}} \right) \frac{\mu_{t'}}{\mu_{t_i}^2}.
$$
\section{Difference between "Optimize-then-Discretize" and "Discretize-then-Optimize"}
\label{app:optimizediscretize}

There is a well-established theory showing that the equations derived through the Optimize-then-Discretize are the correct equations for optimality. This is, in general, not guaranteed for the "Discretize-then-Optimize" approach; the easiest approach is to ensure that for a choice of discretization methods, the results of both approaches are identical (which is what we do, although further analysis is needed to make any rigorous claims).

Specifically for shape optimization, "Optimize-then-Discretize" makes it possible to derive the gradients in the physical domain: “shape derivative calculus” \cite{ALLAIRE20211} allows one to compute shape derivatives with respect to changes in the shape of the domain on which PDE is solved using physical domain variables in which the PDEs have the standard form, e.g. for Poisson or elasticity, are expressed in terms of constant differentiation operators, e.g., the 2D Poisson equation in the weak form:
\begin{equation}\label{eq:poisson}
\int_{\Omega(q)} a(x,y) \nabla z \cdot \nabla w dxdy = 
\int_{\Omega(q)} f(x,y) w dx dy
\end{equation}
with $(x,y)$ coordinates on the physical domain $\Omega(q)$, where $z(x,y)$ is the unknown function, $a(x,y)$ is a material parameter, $f(x,y)$ is the source term, $q$ is the optimization shape parameters. 

In a typical FEM  discretization, the "Discretize-then-Optimize approach" requires converting the equations to a fixed reference domain and then need to substitute into the equation, leading to a variable coefficient equation with an explicit dependence on shape parameters. 
In this case, the unknown is defined on the reference domain $\Omega_{ref}$ with coordinates  $(u,v)$, mapped to the physical domain $\Omega(q)$ via the geometry map $x = x(u,v), y = y(u,v)$.  If we denote the inverse of this map 
$u = u(x,y)$, $v = v(x,y)$, then the left-hand side of Equation~(\ref{eq:poisson}) on the reference domain becomes
$$
\resizebox{\hsize}{!}{
$\int_{\Omega_{ref}} a(x,y) (\|\nabla u\|^2 z_x + \nabla u \cdot \nabla v z_y) w_x\ + (\|\nabla v\|^2 z_y + \nabla u \cdot \nabla v  z_x) w_y |\det J(x,y)| du dv$}
$$

with $\nabla$ denoting derivatives w.r.t. $x,y$, and $J(x,y)$ denoting the geometric map Jacobian matrix $\frac{\partial (x,y)}{\partial (u,v)}$. Here we spell out the expressions more explicitly instead of more concise matrix notation, to elucidate the increase in complexity. Please refer to Equation (\ref{eq:grad-vol-PDEdiff}) for the complete derivation of shape derivatives following this way.

As typically the geometry map $x(u,v),y(u,v)$, rather than its inverse, is given explicitly in terms of shape parameters $q$ (as a linear function of $q$, if it is e.g., represented in a FEM basis), derivatives of $u,v$ w.r.t. $x,y$ need to be expressed in terms of derivatives of $x,y$ w.r.t. $u,v$, i.e., $\nabla u$, $\nabla v$ are the rows of the inverse of $J(x,y)$. In other words, a simple constant coefficient equation on a variable domain becomes a complex variable coefficient equation on a fixed domain, with coefficients depending on the geometric map in a complex nonlinear way.
As a next step, these equations need to be discretized by substituting FEM expressions for $x(u,v;q)$,$y(u,v;q)$,$z(u,v)$, $f(u,v)$ in FEM basis, with the standard Galerkin procedure yielding stiffness matrix and right-hand side coefficients. 
Finally, the derivatives of the resulting coefficients with respect to $q$ need to be computed. 

Note that the derivatives with respect to material parameters (e.g., coefficients of $a(x,y)$ in a FEM basis) unlike shape derivatives have similar complexity in either form. This is also true for elasticity equations: shape derivatives in the Discretize-then-Optimize setting are even more elaborate, but material parameter derivatives are relatively simple.

While, in the end, most equations required for shape derivative adjoints are very close to the forward equations (as shown in Equation~(\ref{eq:time-adjoint}) in the paper, the coefficient matrices are the same as in the forward solves), and can be computed relatively concisely, it is nontrivial to see this from differentiating the coefficients obtained from the equations above with respect to $q$. We are not aware of any tool that can automatically do this conversion, nor of any manual attempt ever done to compute shape derivatives in this way.

\end{document}